\documentclass[12pt,a4]{scrartcl}
\usepackage[]{graphicx,color}
\usepackage{amssymb}                    % mathfrak + amsfonts + spezielle Symbole
\usepackage{amsmath}

\usepackage{bm}							% bold Greek letters in math mode
\usepackage[position=t,caption=false,justification=centering]{subfig}
\usepackage{hyperref}
\usepackage{mathtools}
\usepackage[makeroom]{cancel}
\usepackage{stmaryrd} % double square brackets
\usepackage{tabularx}
\usepackage{multirow}
\usepackage{authblk}
\renewcommand{\vec}[1]{{\mbox{\boldmath $ #1 $}}}
\newcommand{\DD}{\mathrm{D}}
\allowdisplaybreaks
\usepackage{authblk}
\usepackage{natbib}						% implements both author-year and numbered references
\usepackage[title,titletoc]{appendix}
\newcommand{\Fr}{\mbox{Fr}}	 % Froude number
\newcommand{\Rey}{\mbox{Re}} % Reynolds number
\newcommand{\We}{\mbox{We}}  % Weber number
\newcommand{\Eo}{\mbox{Eo}}  % E\"otv\"os number
\graphicspath{{./figures/}{./pictures/}}
\newcommand*{\email}[1]{%
	\normalsize\href{mailto:#1}{#1}\par
}

\title{\vspace{-2cm}A numerical framework for linear stability analysis of two-phase stratified pipe flows}

\author[]{Ilya Barmak\footnote{Corresponding author. E-mail address: \email{ilyab@tauex.tau.ac.il}}}
\author[]{Alexander Gelfgat}
\author[]{Neima Brauner}
\affil[]{\normalsize School of Mechanical Engineering, Tel Aviv University, Tel Aviv 6997801, Israel
}%
\date{}

\begin{document}

\maketitle

\vspace{-2cm}

\section*{Abstract}

A numerical framework for rigorous linear stability analysis of two-phase stratified flows of two immiscible fluids in horizontal circular pipes is presented. For the first time, three-dimensional disturbances, including those at the interface between two fluids, are considered. The proposed numerical framework is based on a finite--volume method and allows solving the problem numerically in bipolar cylindrical coordinates. In these coordinates, both the pipe wall and the unperturbed interface (of a constant curvature, e.g., plane interface, as considered in this work) coincide with the coordinate surfaces. Thereby, the no-slip as well as the interfacial boundary conditions can be imposed easily. It also enables investigation of the local behavior of the flow field and shear stresses in the vicinity of the triple points, where the interface contacts the pipe wall. The results obtained in the bipolar coordinates are verified by an independent numerical solution based on the problem formulation in Cartesian coordinates, where the pipe wall is treated by the immersed boundary method. Two representative examples of gas-liquid and liquid-liquid flows are included to demonstrate the applicability of the proposed numerical technique  for analyzing the flow stability.

\section{Introduction}

In many industrial processes involving two-phase liquid-liquid and gas-liquid pressure-driven flows (e.g., gas-condensate pipeline transport), the desirable flow configuration is the stratified smooth flow pattern, in which heavy and light fluids form two continuous layers, lower and upper, respectively. The stratified smooth flow pattern, however, is feasible only for certain operational conditions for which the flow is stable and the fluid-fluid interface remains smooth. The critical conditions corresponding to the onset of interfacial perturbations have been explored experimentally \citep[e.g.,][]{Charles65,Yu69,Kao72,Barnea80,Kokal89} and theoretically. To avoid complexity of rigorous stability analysis in circular pipe geometry, many theoretical studies were based on simplified mechanistic models, mainly in the framework of the Two-Fluid (TF) model \citep[e.g.,][]{Andritsos89,Brauner91,Barnea93,Ullmann06,Kushnir17}. 

In other studies, the comprehensive linear stability analysis was applied to two-phase flow between two infinite plates. This simplified configuration allows one to account for all possible infinitesimal perturbations \citep[e.g.,][]{Yih67,Charru94,ONaraigh14,Kaffel15,Barmak16a,Barmak19}. This approach yielded better understanding of mechanisms involved in the onset of instability of stratified flows. However, the simplification of geometry allows for only qualitative comparison of the stability characteristics with experimental data obtained in pipe flow. 

The realistic cross-sectional geometry of a rectangular duct was used recently in stability studies of horizontal and inclined two-phase stratified flows \citep{Gelfgat20a,Gelfgat21}. A qualitative agreement of stability characteristics with the flows in the two-plate geometry (TP) was established. At the same time, the quantitative differences were found to be substantial. The effect of circular pipe walls on the onset of instability in two-phase stratified flows has never been studied, and it is still unknown whether the effect is at least qualitatively similar to that for either the TP or square duct geometry. An additional complication found in two-phase pipe flows is that in the vicinity of triple points \citep{Goldstein21b}, i.e., the contact points of the fluid--fluid interface with the pipe wall, the shear stresses may differ strongly from those in the flow bulk \citep[e.g.,][]{Brauner96,Goldstein21a}. 

It is well-known that the single-phase Poiseuille flow in a circular pipe is linearly stable for all Reynolds numbers \citep[e.g.][]{Davey69}, while Poiseuille flow between two plates becomes unstable at Re=5772 \citep{Orszag71}. Since the transition to turbulence takes place at the Reynolds numbers slightly above 2000, the linear stability results are irrelevant for predicting transitions in these flows. Contrarily, the linear stability analysis describes the primary instability in the two-phase flows correctly, as was recently shown by comparison of experimental and numerical results for square ducts in \citep{Nezihovski22}. Therefore, one can expect that linear stability analysis conducted for a two-phase flow in a circular pipe will also describe the primary transition correctly. To the best of our knowledge, no attempt has been made to carry out a rigorous stability analysis of stratified two-phase flows in circular pipes, which is the main objective of this study. 

In the present study, we formulate and solve the linear stability problem in bipolar cylindrical coordinates \citep{Korn00}, where both a constant curvature interface and the pipe wall coincide with the coordinate surfaces. In these coordinates, the boundary conditions can be imposed relatively easily and there is an analytical solution available for the steady fully developed flow \citep[e.g.,][]{Goldstein15}. We also apply an alternative approach, treating the pipe wall by the immersed boundary method (IBM) in Cartesian coordinates, in which the plane interface coincides with the plane of constant vertical coordinate. Since there are no independent results on stability of two-phase stratified pipe flow to compare with, the results obtained by two approaches cross verify each other.  

In the following, we formulate the problem in the vector form and in the bipolar coordinates. We present in detail the finite-volume method in bipolar coordinates, which appears to be non-trivial. For two representative examples of gas-liquid and liquid-liquid flows, we perform a comprehensive stability analysis and discuss the stability characteristics.

\section{Problem formulation}

Two-phase stratified flow in a horizontal circular pipe, which is driven by pressure gradient imposed in the direction of the pipe axis, is a particular case of a general two-layer flow of two immiscible fluids in a channel of constant cross section. For such a flow, we formulate governing equations both in a vector form and in specially tailored curvilinear orthogonal coordinates that fit the boundaries of the problem, i.e., both the channel walls and the unperturbed interface coincide with coordinate lines (Appendix\ \ref{sec: Appendix_gov}).

\subsection{Governing equations in vector form}

The two-phase flow is described by velocity $\vec{u}^{(k)}$ and pressure $p^{(k)}$ fields that satisfy the dimensionless continuity and momentum equations defined in each layer $k=1,2$ (1 - heavy fluid, 2 - light fluid):
\begin{equation} \label{Eq: Continuity_vector}
	\nabla \cdot \vec{u}^{(k)} = 0,
\end{equation}
\begin{equation} \label{Eq: N-S_vector_form}
	\frac{\partial \vec{u}^{(k)}}{\partial t} + \left(\vec{u}^{(k)}\cdot\nabla\right)\vec{u}^{(k)}
	= -\frac{\rho_1}{\rho_{12} \rho_k}\nabla p^{(k)} + \frac{1}{\Rey} \frac{\rho_1}{\rho_{12} \rho_k} \frac{\mu_{12} \mu_k}{\mu_1} \Delta \vec{u}^{(k)} + \frac{1}{\Fr} \vec{e_g},
\end{equation}
where the velocity scale is the mixture velocity $\displaystyle U_m = U_{1S} + U_{2S}$.  $U_{1S}$ and $U_{2S}$ represent the superficial velocities of each phase, i.e., the flow rate of each phase divided by the cross-sectional area of the channel. The time and pressure are scaled by $\displaystyle D/U_m$ ($D$ - characteristic channel size, e.g., pipe diameter) and $\rho_2 U_m^2$, respectively. The dimensionless governing parameters are the density and viscosity ratios, $\displaystyle\rho_{12} = \rho_1/\rho_2$ and $\displaystyle\mu_{12} = \mu_1/\mu_2$, respectively, the Reynolds number $\displaystyle\Rey=\rho_2 D U_m/\mu_2$ and the Froude number $\displaystyle\Fr = U_m^2/(gD)$, where $g$ is the gravitational acceleration. $\displaystyle\vec{e_g}$ is a unit vector in the direction of gravity.

\subsection{Flow geometry} \label{Sec: Geometry}

\begin{figure}
	\setcounter{subfigure}{0}
	\subfloat[][]
	{\def\svgwidth{0.35\textwidth}
		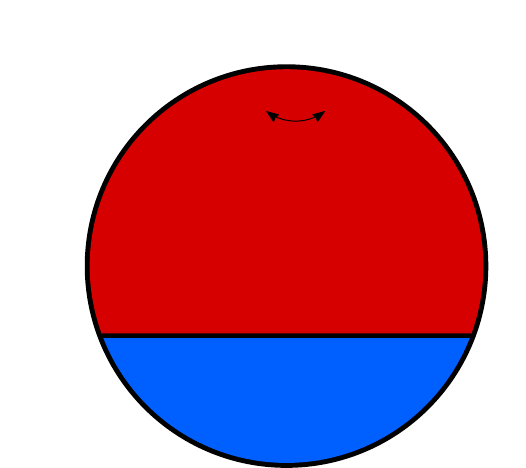}
	\qquad
	\subfloat[][]
	{\def\svgwidth{0.6\textwidth}
		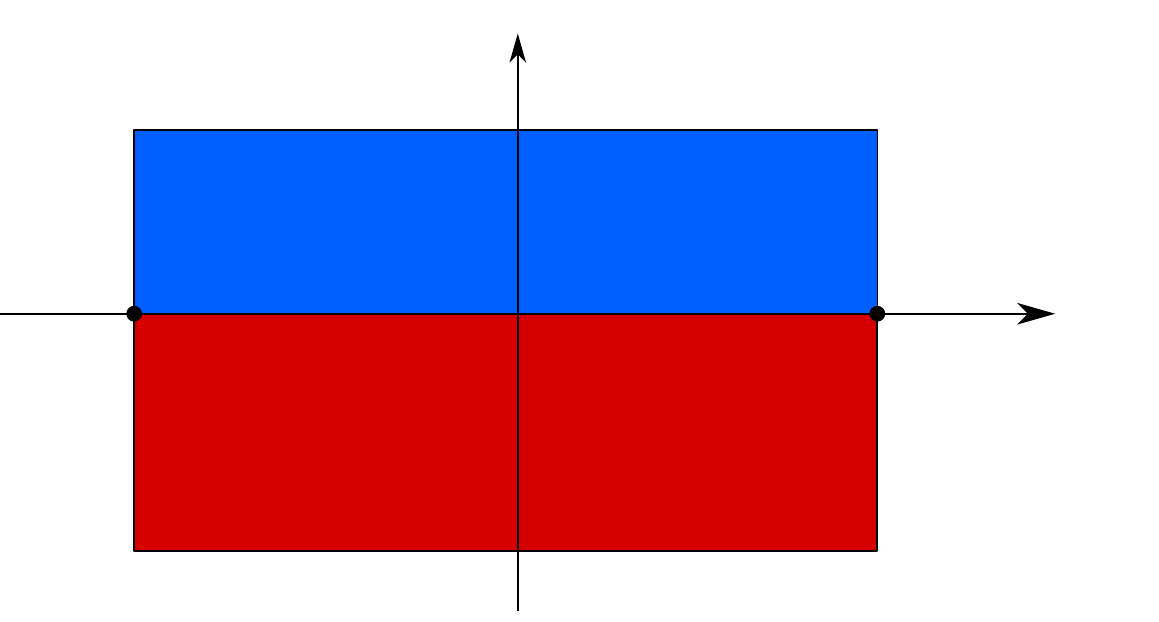}	
	\caption{Schematics of stratified two-phase pipe flow (plane interface, $\displaystyle\phi^*=\pi$). Cross-section of a circular pipe in (a) Cartesian and (b) bipolar coordinates.}
	\label{Fig: flow_geometry}
\end{figure}

For the considered two-phase stratified flow in a circular pipe of diameter $D$, the convenient coordinates are the bipolar cylindrical coordinates $(\xi,\phi,z)$ \citep[e.g.,][]{Korn00,Goldstein15}, in which two lines of foci align with the contact lines of two phases and pipe wall, while the z-axis coincides with the pipe axis. In the pipe cross section (Fig.\ \ref{Fig: flow_geometry}), $\xi$ is defined as the natural logarithm of the ratio of distances from a point to the foci F$_1$ and F$_2$ and $\phi$ as the angle formed by the two foci and the point. For example, for a point P on the pipe wall, $\displaystyle\xi_P=\ln(|\text{PF}_2|/|\text{PF}_1|)$ and $\displaystyle\phi_P=\phi_0=\angle\text{F}_1\text{PF}_2$. Thus, in the bipolar coordinates, the pipe wall corresponds to isolines of coordinate $\phi$. The upper section of the pipe wall bounds the light phase (shown in red in Fig.\ \ref{Fig: flow_geometry}) and is represented by $\phi_0$, while the bottom of the pipe, bounding the heavy phase (shown in blue in Fig.\ \ref{Fig: flow_geometry}), is represented by $\phi_0+\pi$. The interface that separates two fluids is a surface defined as $\phi_\eta = \phi^* + \eta\left(\xi,z\right)$.  

In this work, we consider gravity-dominated systems with E\"otv\"os number, $\displaystyle\Eo = (\rho_1 - \rho_2) g D^2/\sigma \gg 1$, where $\sigma$ is the surface tension coefficient. The smooth stratified flow in such systems is characterized by a plane interface \citep{Gorelik99}, so that $\phi^*=\pi$ (as shown in Fig.\ \ref{Fig: flow_geometry}). The heavy fluid occupies an infinite strip of $\displaystyle\bigl(-\infty<\xi<+\infty,\pi<\phi<\phi_0+\pi\bigr)$ with the cross-sectional area of $\displaystyle A_1 = 0.25 D^2 \big(\phi_0 - 0.5\sin2\phi_0\big)$, while the light fluid -- an infinite strip of $\displaystyle\bigl(-\infty<\xi<+\infty,\phi_0<\phi<\pi\bigr)$ with the area of $\displaystyle A_2 = A - A_1$, where $\displaystyle A = \pi D^2/4$ is the cross-sectional area of the pipe. Then, one of the dimensionless parameters defining two-phase stratified flow is the heavy phase holdup, defined as $\displaystyle h = A_1/A$. In the following, we normalize all the length scales by the pipe diameter, $D$. 

In Cartesian coordinates, the pipe center is located at $x,y=(0,0)$, and the foci of the bipolar coordinate system are located in the triple points, $\displaystyle \text{F}_1(-0.5\sin\phi_0,-0.5\cos\phi_0)$ and $\displaystyle \text{F}_2(0.5\sin\phi_0,-0.5\cos\phi_0)$. The relations of the Cartesian coordinates to the bipolar coordinates are 
\begin{subequations}
	\begin{align}
	x &= \frac{1}{2} \frac{\sin\phi_0 \sinh\xi}{\cosh\xi - \cos\phi},
	\\
	y &= \frac{1}{2} \left(\frac{\sin\phi \sin\phi_0}{\cosh\xi - \cos\phi}
	- \cos\phi_0\right). 
	\end{align}
\end{subequations}

The reciprocal relations from the Cartesian to the bipolar coordinates are
\begin{subequations}
	\begin{align}
	\xi &= \frac{1}{2} \ln\frac{\bigl(x + 0.5\sin\phi_0\bigr)^2 + y^2}{\bigl(x - 0.5\sin\phi_0\bigr)^2 + y^2},
	\\
	\phi &= \pi-2\arctan\frac{\sin\phi_0 y}{0.25 \sin^2\phi_0 - x^2 - y^2 + \bigl[\bigl(0.25 \sin^2\phi_0 - x^2 - y^2\bigr)^2 + \sin^2\phi_0 y^2\bigr]^{0.5}}
	\end{align}
\end{subequations}

The Lam\'e coefficients (scale factors) for the bipolar coordinates are
\begin{equation} \label{Eq: Lame_bipolar}
	H_\xi = H_\phi 
	= \frac{1}{2} \frac{\sin\phi_0}{\cosh \xi - \cos \phi}.
\end{equation}

It is worth noting that the cylindrical coordinate system, being a natural choice for computation of a single-phase flow in a circular pipe, is not convenient for the modeling of a two-phase stratified pipe flow. In addition to possible cross flows across the cylindrical axis, in two-phase flows, the fluid—fluid interface can intersect the axis, which is problematic due to its singularity in numerical formulation. The second problem in using cylindrical coordinates is the treatment of the interface. In case it is assumed to be plane in the unperturbed state \citep[valid for gravity-dominated systems, e.g.,][]{Gorelik99}, the interface depends on two cross-sectional coordinates and the formulation and implementation of the non-trivial interfacial boundary conditions becomes intricate. Moreover, as we will show below, the interface can be located at any height, depending on the flow rates of two fluids. In the following, we provide a formulation in bipolar cylindrical coordinates that appear to be the most suitable.

\subsection{Governing equations in bipolar coordinates}

The governing equations for three-dimensional time-dependent flow in the bipolar cylindrical coordinates can be obtained from the corresponding equations in general orthogonal coordinates,	Eqs.\ \ref{Eq: Continuity_orthogonal} and \ref{Eq: N-S_orthogonal} (see Appendix\ \ref{sec: Appendix_gov}), by introducing the Lam\'e coefficients (Eq.\ \ref{Eq: Lame_bipolar}):
\begin{equation} \label{Eq: Continuity_bipolar}
	\frac{1}{H_\xi}\frac{\partial u_\xi^{(k)}}{\partial \xi}
	+ \frac{1}{H_\phi} \frac{\partial u_\phi^{(k)}}{\partial \phi}
	- 2 \frac{\sinh\xi}{\sin\phi_0} u_\xi^{(k)}
	- 2 \frac{\sin\phi}{\sin\phi_0} u_\phi^{(k)}
	+ \frac{\partial u_z}{\partial z} = 0,
\end{equation}
\begin{subequations} \label{Eq: N-S_bipolar}
	\begin{align}
		\begin{aligned}
			&\frac{\partial u_\xi^{(k)}}{\partial t}
			+ \frac{u_\xi^{(k)}}{H_\xi}\frac{\partial u_\xi^{(k)}}{\partial \xi}
			+ \frac{u_\phi^{(k)}}{H_\phi}\frac{\partial u_\xi^{(k)}}{\partial \phi}			
			+ u_z^{(k)} \frac{\partial u_\xi^{(k)}}{\partial z}
			- 2 \frac{\sin\phi}{\sin\phi_0} 
			u_\xi^{(k)} u_\phi^{(k)}
			+ 2 \frac{\sinh\xi}{\sin\phi_0} 
			\left(u_\phi^{(k)}\right)^2
			\\
			&= - \frac{1}{\rho_k} 
			\frac{1}{H_\xi}
			\frac{\partial p^{(k)}}{\partial \xi}
			+ \frac{1}{\Rey} \frac{\rho_1}{\rho_{12} \rho_k} \frac{\mu_{12} \mu_k}{\mu_1}
			\Biggl[
			\frac{1}{H_\xi H_\phi}
			\frac{\partial}{\partial \xi} 
			\biggl(\frac{H_\phi}{H_\xi}\frac{\partial u_\xi^{(k)}}{\partial \xi}\biggr)
			+ \frac{1}{H_\xi H_\phi}
			\frac{\partial}{\partial \phi}
			\biggl(\frac{H_\xi}{H_\phi} 
			\frac{\partial u_\xi^{(k)}}{\partial \phi}\biggr)
			\\
			&+ \frac{\partial^2 u_\xi^{(k)}}{\partial z^2}
			- 4 \frac{\sin\phi}{\sin\phi_0}
			\frac{1}{H_\xi}\frac{\partial u_\phi^{(k)}}{\partial \xi} 	
			+ 4 \frac{\sinh\xi}{\sin\phi_0}
			\frac{1}{H_\phi} \frac{\partial u_\phi^{(k)}}{\partial \phi}
			+ 4 \frac{\cos^2\phi - \cosh^2\xi}{\sin^2\phi_0} 
			u_\xi^{(k)}
			\Biggr]
			\\
			&+ \frac{1}{\Fr} \frac{\sinh\xi \sin\phi}{\bigl(\cosh\xi - \cos\phi\bigr)},
		\end{aligned}
	\end{align}
	\begin{align}
		\begin{aligned}
			\frac{\partial u_\phi^{(k)}}{\partial t}
			&+ \frac{u_\xi^{(k)}}{H_\xi}\frac{\partial u_\phi^{(k)}}{\partial \xi}
			+ \frac{u_\phi^{(k)}}{H_\phi}\frac{\partial u_\phi^{(k)}}{\partial \phi}
			+ u_z^{(k)} \frac{\partial u_\phi^{(k)}}{\partial z}
			- 2 \frac{\sinh\xi}{\sin\phi_0} u_\xi^{(k)} u_\phi^{(k)}
			+ 2 \frac{\sin\phi}{\sin\phi_0} \left(u_\xi^{(k)}\right)^2
			\\
			=& - \frac{1}{\rho_k} 
			\frac{1}{H_\phi}
			\frac{\partial p^{(k)}}{\partial \phi}
			+ \frac{1}{\Rey} \frac{\rho_1}{\rho_{12} \rho_k} \frac{\mu_{12} \mu_k}{\mu_1}
			\Biggl[
			\frac{1}{H_\xi H_\phi}
			\frac{\partial}{\partial \xi}
			\biggl(\frac{H_\phi}{H_\xi}\frac{\partial u_\phi^{(k)}}{\partial \xi}\biggr)
			\\
			&+ \frac{1}{H_\xi H_\phi}
			\frac{\partial}{\partial \phi}	\biggl(
			\frac{H_\xi}{H_\phi}
			\frac{\partial u_\phi^{(k)}}{\partial \phi}\biggr)	
			+ \frac{\partial^2 u_\phi^{(k)}}{\partial z^2}
			+ 4 \frac{\sin\phi}{\sin\phi_0}
			\frac{1}{H_\xi} 
			\frac{\partial u_\xi^{(k)}}{\partial \xi}
			- 4 \frac{\sinh\xi}{\sin\phi_0}
			\frac{1}{H_\phi}
			\frac{\partial u_\xi^{(k)}}{\partial \phi}
			\\
			&+ 4 \frac{\cos^2\phi - \cosh^2\xi}{\sin^2\phi_0}
			u_\phi^{(k)}
			\Biggr]
			+ \frac{1}{\Fr} 
				\frac{\bigl(1 - \cosh\xi\cos\phi\bigr)}{\bigl(\cosh\xi-\cos\phi\bigr)},
		\end{aligned}
	\end{align}
	\begin{align}
		\begin{aligned}
			\frac{\partial u_z^{(k)}}{\partial t}
			&+ \frac{u_\xi^{(k)}}{H_\xi}\frac{\partial u_z^{(k)}}{\partial \xi} 
			+ \frac{u_\phi^{(k)}}{H_\phi}\frac{\partial u_z^{(k)}}{\partial \phi}			
			+ u_z^{(k)} \frac{\partial u_z^{(k)}}{\partial z} 
			= - \frac{1}{\rho_k} 
			\frac{\partial p^{(k)}}{\partial z}
			\\
			&+ \frac{1}{\Rey} \frac{\rho_1}{\rho_{12} \rho_k} \frac{\mu_{12} \mu_k}{\mu_1} 
			\Biggl[
			\frac{1}{H_\xi H_\phi}
			\frac{\partial}{\partial \xi}
			\left(\frac{H_\phi}{H_\xi} \frac{\partial u_z^{(k)}}{\partial \xi}\right)
			+ \frac{1}{H_\xi H_\phi} 
			\frac{\partial}{\partial \phi}
			\left(\frac{H_\xi}{H_\phi} \frac{\partial u_z^{(k)}}{\partial \phi}\right)
			\\
			&+ \frac{\partial^2 u_z^{(k)}}{\partial z^2}\Biggr].
		\end{aligned}
	\end{align}
\end{subequations}

Note that $\displaystyle H_\xi d\xi$ and $\displaystyle H_\phi d\phi$ have the meaning of physical lengths so that they are important for the purpose of the finite--volume integration on the computational grid and should not be canceled out (see sec.\ \ref{Sec: Numerics}).

\subsection{Boundary conditions} \label{Sec: BC}

The governing equations for the velocity field are coupled with the no-slip conditions at the pipe wall, including at the triple points ($\xi\to\pm\infty$)
\begin{equation} \label{Eq: BC_no-slip}
	\vec{u}^{(1)}(\phi=\phi_0 + \pi)
	= \vec{u}^{(2)}(\phi=\phi_0) 
	= 0
\end{equation}
and the boundary conditions at the interface between the two fluids. Here we present only the final expressions for the interfacial boundary conditions, while the detailed description can be found in Appendix\ \ref{sec: Appendix_BC}. The interface, which is considered plane in an unperturbed state, is described as
\begin{equation}
	\phi_\eta = \pi + \eta\left(\xi,z\right).
\end{equation}

The absence of mass transfer across the interface is accounted by the kinematic boundary condition
\begin{equation}	\label{Eq: BC_kinematic}	
	u_\phi^{(k)}(\phi=\phi_\eta) = H_\phi \frac{\partial \eta}{\partial t}
	+ u_\xi^{(k)}(\phi=\phi_\eta) \frac{H_\phi}{H_\xi} \frac{\partial \eta}{\partial \xi}
	+ u_z^{(k)}(\phi=\phi_\eta) H_\phi \frac{\partial \eta}{\partial z}.
\end{equation}

The velocity field is continuous across the interface, therefore
\begin{equation}
	\vec{u}^{(1)}(\phi=\phi_\eta) 
	= \vec{u}^{(2)}(\phi=\phi_\eta).
\end{equation} 

The interfacial boundary conditions require continuity of the tangential components of the viscous stress tensor, i.e., $\displaystyle\tau_{\xi \phi}$ in the plane $\displaystyle(\xi,\phi)$ and $\displaystyle\tau_{\phi z}$ in the plane $\displaystyle(\phi,z)$, while the discontinuity in the normal component, which is the sum of the normal viscous stress $\displaystyle\sigma_n$ and the pressure, is balanced by the surface tension
\begin{subequations}
	\begin{align}
		\tau_{\xi \phi}^{(1)}(\phi=\phi_\eta) 
		&= \tau_{\xi \phi}^{(2)}(\phi=\phi_\eta),
		\\
		\tau_{\phi z}^{(1)}(\phi=\phi_\eta) 
		&= \tau_{\phi z}^{(2)}(\phi=\phi_\eta),
		\\		
		\biggl\llbracket -p + \sigma_n \biggr\rrbracket_{\phi=\phi_\eta} 
		&= - \frac{1}{\We}\nabla \cdot \vec{n},	
	\end{align}
\end{subequations}
where $\vec{n}$ is a unit normal vector to the interface and double square brackets denote a jump in the corresponding quantity across the interface. The Weber number is defined as $\displaystyle\We=\rho_2 D U_m^2/\sigma$. Expressions for the stress tensor components are given in Appendix\ \ref{sec: Appendix_BC}.

\section{Base flow}

The base flow is assumed to be steady, laminar, and fully-developed flow. The only non-zero component is in the pipe axis direction, and it varies only in the cross section $\displaystyle\vec{U^{(k)}}=\left(0,0,U_z^{(k)}(\xi,\phi)\right)$. Then the base flow momentum equations are obtained from Eqs.\ \ref{Eq: N-S_bipolar}a-c and read

\begin{subequations} \label{Eq: Base_flow_eqs}
	\begin{align}
		\frac{1}{H_\phi H_\xi}
		\left[\frac{\partial}{\partial \xi}
		\left(\frac{H_\phi}{H_\xi}
		\frac{\partial U_z^{(k)}}{\partial \xi}\right)
		+\frac{\partial}{\partial \phi}
		\left(\frac{H_\xi}{H_\phi}
		\frac{\partial U_z^{(k)}}{\partial \phi}\right)\right]
		&= \frac{\mu_1}{\mu_{12} \mu_k}
		\frac{\partial P^{(k)}}{\partial z},						
		\\
		- \frac{\partial P^{(k)}}{\partial \xi} 
		+ \frac{\rho_k}{\Fr} H_\xi 
		\frac{\sinh\xi \sin\phi}{\cosh\xi - \cos\phi}
		&= 0,
		\\
		- \frac{\partial P^{(k)}}{\partial \phi} 
		+ \frac{\rho_k}{\Fr} H_\phi 
			\frac{\bigl(1 - \cosh\xi\cos\phi\bigr)}{\cosh\xi-\cos\phi}
		&= 0.
	\end{align}
\end{subequations}

Note that differentiation of Eqs.\ \ref{Eq: Base_flow_eqs}b,c in the axial $z$ direction, i.e., $\displaystyle \frac{\partial}{\partial z}$(Eqs.\ \ref{Eq: Base_flow_eqs}b,c), yields $\displaystyle \frac{\partial}{\partial \phi}\left(\frac{\partial P^{(k)}}{\partial z}\right)=0$ and $\displaystyle \frac{\partial}{\partial \xi}\left(\frac{\partial P^{(k)}}{\partial z}\right)=0$, therefore the base-flow pressure gradient in the axial direction is the same for the two fluids $\displaystyle\left(\frac{\partial P_1}{\partial z}=\frac{\partial P_2}{\partial z}=\frac{\partial P}{\partial z}\right)$.

The base flow velocity is subject to no-slip conditions at the pipe wall
\begin{equation}
	U_z^{(1)} \left(\phi=\phi_0+\pi\right) = 0, \qquad
	U_z^{(2)} \left(\phi=\phi_0\right) = 0, 
\end{equation}
and continuous across the fluid--fluid interface
\begin{equation}
	U_z^{(1)}(\phi=\pi) = U_z^{(2)}(\phi=\pi),
\end{equation}
while the tangential shear stress of the base flow is continuous as well
\begin{equation}
	\mu_{12} \bigg(\frac{\partial U_z^{(1)}}{\partial \phi}\bigg)_{\phi=\pi} 
	= \bigg(\frac{\partial U_z^{(2)}}{\partial \phi}\bigg)_{\phi=\pi}. 	
\end{equation}
For a plane unperturbed interface, continuity of the normal stress results in continuity of the pressure across the interface
\begin{equation}
	P^{(1)}(\phi=\pi) = P^{(2)}(\phi=\pi).
\end{equation}
Integration of the base-flow velocity profiles over the lower- and upper-layer cross-sectional areas yields a relative input of each fluid into the prescribed total volumetric flow rate, which can be written as a ratio of the corresponding superficial velocity to the mixture velocity
\begin{subequations} \label{Eq: BF_flow_rates}
	\begin{align}
		\int_{\pi}^{\phi_0 + \pi}\int_{-\infty}^{+\infty} U_z^{(1)} H_\xi H_\phi d\xi d\phi 
		&= \frac{U_{1S}}{U_{1S} + U_{2S}},
		\\
		\int_{\phi_0}^{\pi}\int_{-\infty}^{+\infty} U_z^{(2)} H_\xi H_\phi d\xi d\phi 
		&= \frac{U_{2S}}{U_{1S} + U_{2S}}.
	\end{align}
\end{subequations}

\section{Linear stability}

In the following, we study linear stability of above plane-parallel flow with respect to infinitesimal perturbations. The velocity and pressure of the perturbed flow are
\begin{align}
	\vec{u^{(k)}} &= \vec{U^{(k)}}\bigl(\xi,\phi\bigr)
	+ \vec{u'^{(k)}}\bigl(\xi,\phi,z\bigr),
	\\
	p^{(k)} &= P^{(k)}\bigl(\xi,\phi\bigr) 
	+ p'^{(k)}\bigl(\xi,\phi,z\bigr),
\end{align}
while the perturbed interface is described as a surface $\displaystyle \phi = \pi + \eta\left(\xi,z,t\right)$, where $\eta$ denotes the deformation of the interface from the undisturbed plane state. The infinitesimal perturbations are
\begin{equation} \label{Eq: Perturbation}
	\begin{pmatrix}
		\vec{u'^{(k)}} \\
		p'^{(k)} \\
		\eta
	\end{pmatrix}
	=
	\begin{pmatrix}
		\vec{\tilde{u}^{(k)}}(\xi,\phi)
		= \left[\tilde{u}_\xi^{(k)}(\xi,\phi),\tilde{u}_\phi^{(k)}(\xi,\phi),\tilde{u}_z^{(k)}(\xi,\phi)\right] 
		\\
		\tilde{p}^{(k)}(\xi,\phi)
		\\
		\tilde{\eta}(\xi)
	\end{pmatrix}
	e^{\left(i\alpha z + \lambda t\right)},
\end{equation}
where $\alpha$ is the wavenumber of the perturbation. 

To derive the linearized form of the boundary conditions \ref{Eq: BC_shear_stress} at the unperturbed interface $\phi=\pi$, we use the Taylor expansion of a scalar function $f$ of the radius vector $\vec{r}$ around $\vec{r_0}$ in bipolar cylindrical coordinates
\begin{align} \label{Eq: Taylor}
	&\begin{aligned}
		f(\vec{r})
		&= f(\vec{r_0})
		+ \nabla f\biggr|_{\vec{r_0}} 
		\cdot \left(\vec{r} -\vec{r_0}\right)
		+ \mathcal{O}(\left(\vec{r} -\vec{r_0}\right)^2)
		= f(\vec{r_0})
		+ \biggl(\frac{1}{H_\xi}\frac{\partial f}{\partial\xi}\biggr|_{\vec{r_0}} \vec{e_\xi}
		\\
		&+ \frac{1}{H_\phi}\frac{\partial f}{\partial\phi}\biggr|_{\vec{r_0}} \vec{e_\phi}
		+ \frac{\partial f}{\partial z}\biggr|_{\vec{r_0}} \vec{e_z}\biggr)
		\cdot \bigg(0\vec{e_\xi}
		+ \eta\left(\xi,z,t\right) H_\phi \vec{e_\phi}
		+ 0\vec{e_z}\bigg)
		+ \mathcal{O}(\eta^2)
		\\
		&= f\bigl(\xi,\pi,z\bigr)
		+ \frac{1}{H_\phi}\frac{\partial f}{\partial\phi} \biggr|_{\phi=\pi} H_\phi \eta\left(\xi,z,t\right)
		+ \mathcal{O}(\eta^2),
	\end{aligned}
\end{align}
where $\displaystyle \vec{r} = \bigl(\xi,\pi + \eta\left(\xi,z,t\right),z\bigr)$,
$\displaystyle \vec{r_0} = \bigl(\xi,\pi,z\bigr)$, and $\displaystyle \vec{r}-\vec{r_0} = \eta\left(\xi,z,t\right) H_\phi \vec{e_\phi}$ for infinitesimally small $\eta$. The function $f$ denotes either the axial velocity $u_z^{(k)}$ or the pressure $p^{(k)}$ in the derivations below.

\subsection{Linearized governing equations}

Linearizing the continuity \ref{Eq: Continuity_bipolar} and momentum \ref{Eq: N-S_bipolar} equations for the perturbed flow and taking into account the corresponding equations for the base flow (Eq.\ \ref{Eq: Base_flow_eqs}), we get equations formulated for the perturbation amplitude (Eq.\ \ref{Eq: Perturbation})
\begin{equation} \label{Eq: Stability_continuity}
	\frac{1}{H_\xi}\frac{\partial \tilde{u}_\xi^{(k)}}{\partial \xi}
	+ \frac{1}{H_\phi} \frac{\partial \tilde{u}_\phi^{(k)}}{\partial \phi}
	- 2 \frac{\sinh\xi}{\sin\phi_0} \tilde{u}_\xi^{(k)}
	- 2 \frac{\sin\phi}{\sin\phi_0} \tilde{u}_\phi^{(k)}
	+ i \alpha \tilde{u}_z = 0,
\end{equation}
\begin{subequations}
	\begin{align} \label{Eq: Stability_xi}
		&\begin{aligned}
			\lambda \tilde{u}_\xi^{(k)}
			&+ i \alpha U_z^{(k)} \tilde{u}_\xi^{(k)}
			= - \frac{\rho_1}{\rho_{12} \rho_k} 
			\frac{1}{H_\xi}
			\frac{\partial \tilde{p}^{(k)}}{\partial \xi}
			\\
			&+ \frac{1}{\Rey}\frac{\rho_1}{\rho_{12} \rho_k} \frac{\mu_{12} \mu_k}{\mu_1} 	\Biggl[
			\frac{1}{H_\xi H_\phi}
			\frac{\partial}{\partial \xi} 
			\biggl(\frac{H_\phi}{H_\xi}\frac{\partial \tilde{u}_\xi^{(k)}}{\partial \xi}\biggr)
			+ \frac{1}{H_\xi H_\phi}
			\frac{\partial}{\partial \phi}
			\biggl(\frac{H_\xi}{H_\phi} 
			\frac{\partial \tilde{u}_\xi^{(k)}}{\partial \phi}\biggr)
			- \alpha ^2 \tilde{u}_\xi^{(k)}
			\\
			&- 4 \frac{\sin\phi}{\sin\phi_0}
			\frac{1}{H_\xi}\frac{\partial \tilde{u}_\phi^{(k)}}{\partial \xi} 	
			+ 4 \frac{\sinh\xi}{\sin\phi_0}
			\frac{1}{H_\phi} \frac{\partial \tilde{u}_\phi^{(k)}}{\partial \phi}
			+ 4 \frac{\cos^2\phi - \cosh^2\xi}{\sin^2\phi_0} 
			\tilde{u}_\xi^{(k)}
			\Biggr],
		\end{aligned}
	\end{align}
	\begin{align} \label{Eq: Stability_phi}
		&\begin{aligned}
			\lambda \tilde{u}_\phi^{(k)}
			&+ i \alpha U_z^{(k)} \tilde{u}_\phi^{(k)}			
			=
			- \frac{\rho_1}{\rho_{12} \rho_k} 
			\frac{1}{H_\xi}
			\frac{\partial \tilde{p}^{(k)}}{\partial \phi}
			\\
			&+ \frac{1}{\Rey}\frac{\rho_1}{\rho_{12} \rho_k} \frac{\mu_{12} \mu_k}{\mu_1} 
			\Biggl[
			\frac{1}{H_\xi H_\phi}
			\frac{\partial}{\partial \xi}
			\biggl(\frac{H_\phi}{H_\xi}\frac{\partial \tilde{u}_\phi^{(k)}}{\partial \xi}\biggr)
			+ \frac{1}{H_\xi H_\phi}
			\frac{\partial}{\partial \phi}	\biggl(
			\frac{H_\xi}{H_\phi}
			\frac{\partial \tilde{u}_\phi^{(k)}}{\partial \phi}\biggr)	
			- \alpha ^2 \tilde{u}_\phi^{(k)}
			\\
			&+ 4 \frac{\sin\phi}{\sin\phi_0}
			\frac{1}{H_\xi} 
			\frac{\partial \tilde{u}_\xi^{(k)}}{\partial \xi}
			- 4 \frac{\sinh\xi}{\sin\phi_0}
			\frac{1}{H_\phi}
			\frac{\partial \tilde{u}_\xi^{(k)}}{\partial \phi}
			+ 4 \frac{\cos^2\phi - \cosh^2\xi}{\sin^2\phi_0}
			\tilde{u}_\phi^{(k)}
			\Biggr],
		\end{aligned}
	\end{align}
	\begin{align} \label{Eq: Stability_z}
		&\begin{aligned}
			\lambda \tilde{u}_z^{(k)}
			&+ \frac{\tilde{u}_\xi^{(k)}}{H_\xi}\frac{\partial U_z^{(k)}}{\partial \xi}
			+ \frac{\tilde{u}_\phi^{(k)}}{H_\phi}\frac{\partial U_z^{(k)}}{\partial \phi}			
			+ i \alpha U_z^{(k)} \tilde{u}_z^{(k)}
			= - i \alpha \frac{\rho_1}{\rho_{12} \rho_k} \tilde{p}^{(k)}
			\\
			&+ \frac{1}{\Rey} \frac{\rho_1}{\rho_{12} \rho_k} \frac{\mu_{12} \mu_k}{\mu_1} \Biggl[
			\frac{1}{H_\xi H_\phi}
			\frac{\partial}{\partial \xi}
			\left(\frac{H_\phi}{H_\xi} \frac{\partial \tilde{u}_z^{(k)}}{\partial \xi}\right)
			+ \frac{1}{H_\xi H_\phi} 
			\frac{\partial}{\partial \phi}
			\left(\frac{H_\xi}{H_\phi} \frac{\partial \tilde{u}_z^{(k)}}{\partial \phi}\right)
			\\
			&- \alpha ^2 \tilde{u}_z^{(k)}\Biggr].
		\end{aligned}
	\end{align}
\end{subequations}

\subsection{Linearized boundary conditions}

In this section, we list the full set of the linearized boundary conditions. The no-slip condition at the pipe wall is formulated for the perturbation amplitude (Eq.\ \ref{Eq: Perturbation}) as
\begin{equation} \label{Eq: BC_no-slip_lin}
	\vec{\tilde{u}}^{(1)}(\phi=\phi_0 + \pi) 
	= \vec{\tilde{u}}^{(2)}(\phi=\phi_0)
	= 0.
\end{equation}

The interfacial boundary conditions that were formulated in sec.\ \ref{Sec: BC} are  linearized using the Taylor expansion \ref{Eq: Taylor} around $\phi=\pi$. The kinematic boundary condition (Eq.\ \ref{Eq: BC_kinematic}) for the perturbed flow reads
\begin{equation} \label{Eq: BC_kinematic_lin}
	\lambda H_\phi \tilde{\eta}
	= \tilde{u}_\phi^{(1)}
	- i \alpha U_z^{(1)} (\phi=\pi) H_\phi \tilde{\eta}
	= \tilde{u}_\phi^{(2)}
	- i \alpha U_z^{(2)} (\phi=\pi) H_\phi \tilde{\eta}.
\end{equation}
The continuity of velocity components yields
\begin{subequations} \label{Eq: BC_continuity_velocity}
	\begin{align} 
		\tilde{u}_\xi^{(1)}(\phi=\pi) 
		&= \tilde{u}_\xi^{(2)}(\phi=\pi),
		\\
		\tilde{u}_\phi^{(1)}(\phi=\pi) 
		&= \tilde{u}_\phi^{(2)}(\phi=\pi),
		\\
		\tilde{u}_z^{(1)}(\phi=\pi)
		&= \tilde{u}_z^{(2)}(\phi=\pi) 
		+ \bigg(\frac{1}{H_\phi} \frac{\partial U_z^{(2)}}{\partial \phi} 
		\biggr|_{\phi=\pi}
		- \frac{1}{H_\phi} \frac{\partial U_z^{(1)}}{\partial \phi} 
		\biggr|_{\phi=\pi}\bigg) 
		H_\phi \tilde{\eta}.
	\end{align}
\end{subequations}
The interfacial boundary conditions require also continuity of the tangential stresses in planes $\bigl(\xi,\phi\bigr)$ and $\bigl(\phi,z\bigr)$ and a jump in the normal shear stress (Eqs. \ref{Eq: BC_shear_stress}a-c, respectively), and their linearized versions read
\begin{subequations} \label{Eq: BC_shear_stress_lin}
	\begin{align} 
		&\begin{aligned}
			&\mu_{1 2}\frac{1}{H_\phi}\frac{\partial \tilde{u}_\xi^{(1)}}{\partial \phi}\Biggr|_{\phi = \pi}
			- \frac{1}{H_\phi}\frac{\partial \tilde{u}_\xi^{(2)}}{\partial \phi}\Biggr|_{\phi = \pi}
			+ (\mu_{1 2}-1) \frac{1}{H_\xi}\frac{\partial \tilde{u}_\phi}{\partial \xi}\Biggr|_{\phi = \pi}
			\\
			&- i \alpha (\mu_{1 2}-1) \frac{1}{H_\xi} \frac{\partial U_z}{\partial \xi} H_\phi \tilde{\eta} \Biggr|_{\phi = \pi}
			+ 2 (\mu_{1 2}-1)  \frac{\sinh\xi}{\sin\phi_0} \tilde{u}_\phi \left(\phi = \pi\right)
			= 0,
		\end{aligned}
	\end{align}
	\begin{align}
		\begin{aligned}
			&i \alpha \left(\mu_{1 2}-1\right) \tilde{u}_\phi (\phi = \pi)
			- (\mu_{1 2}-1) \frac{1}{H_\xi}
			\frac{\partial U_z}{\partial \xi} \frac{1}{H_\xi} \frac{\partial \big(H_\phi \tilde{\eta}\big)}{\partial \xi} \Biggr|_{\phi = \pi}
			\\
			&+ \Biggl[\mu_{1 2} \frac{1}{H_\phi} \frac{\partial}{\partial \phi} \biggl(\frac{1}{H_\phi}\frac{\partial U_z^{(1)}}{\partial \phi}\biggr)
			- \mu_{1 2} \frac{1}{H_\xi} \frac{\partial U_z^{(1)}}{\partial \xi} \frac{2 \sinh\xi}{\sin\phi_0} 
			\\
			&- \frac{1}{H_\phi} \frac{\partial}{\partial \phi} \biggl(\frac{1}{H_\phi}\frac{\partial U_z^{(2)}}{\partial \phi}\biggr)
			+ \frac{1}{H_\xi}
			\frac{\partial U_z^{(2)}}{\partial \xi} \frac{2 \sinh\xi}{\sin\phi_0} \Biggr]_{\phi = \pi} H_\phi \tilde{\eta}
			\\
			&+ \biggl(\mu_{1 2} \frac{1}{H_\phi} \frac{\partial \tilde{u}_z^{(1)}}{\partial \phi} 
			- \frac{1}{H_\phi} \frac{\partial \tilde{u}_z^{(2)}}{\partial \phi}\biggr)\Biggr|_{\phi = \pi}
			= 0,
		\end{aligned}
	\end{align}
	\begin{align}
		&\begin{aligned}
			&\left(\tilde{p}^{(2)} - \tilde{p}^{(1)}\right)\biggr|_{\phi = \pi}
			+ \frac{1}{\Fr} \left(\rho_{1 2}-1\right)
			\left(-\cosh\xi + \frac{\sinh^2\xi}{\cosh\xi -\cos\phi}\right) 
			H_\phi \tilde{\eta} \Biggr|_{\phi = \pi}
			\\
			&+ \frac{2}{\Rey} \biggl(
			\frac{\mu_{1 2}}{H_\phi}\frac{\partial \tilde{u}_\phi^{(1)}}{\partial \phi}	
			- \frac{1}{H_\phi}\frac{\partial \tilde{u}_\phi^{(2)}}{\partial \phi}\biggr) \Biggr|_{\phi = \pi}
			- \frac{4}{\Rey} \left(\mu_{1 2}-1\right) \frac{\sinh\xi}{\sin\phi_0} \tilde{u}_\xi \Biggr|_{\phi = \pi}
			\\
			&= - \frac{1}{\We} \biggl[\frac{1}{H_\xi}
			\frac{\partial}{\partial \xi}\left(\frac{H_\phi}{H_\xi} \frac{\partial \tilde{\eta}}{\partial \xi}\right)
			- 2 \frac{H_\phi}{H_\xi} \frac{\partial \tilde{\eta}}{\partial \xi}
			\frac{\sinh\xi}{\sin\phi_0}
			- \alpha ^2 H_\phi \tilde{\eta}\biggr].
		\end{aligned}
	\end{align}
\end{subequations}

%%%%%%%%%%%%%%%%%%%%%%%%%%%%%%%%%%%%%%%%%%%%%%%%%%%%%%%%%%%%%%%%%%
\section{Numerical method} \label{Sec: Numerics}

\subsection{Computational grid} \label{Sec: Grid}
The stability problem formulated in the previous section is solved on a staggered grid  using the finite--volume method. The computational grid is represented by a pipe cross section divided into four-sided cells, whose faces coincide with the coordinate lines in the bipolar coordinates, so that each cell is rectangular as well as the whole domain. 

We use the staggered grid, so that the scalar value of pressure $p$, the base flow velocity $U_z^{(k)}$, and the axial velocity of the perturbation $u_z^{(k)}$ are calculated in the cell centers defined by integer indices $\big[\xi_i,\phi_j\big]$, where $i=0,1,...,N_\xi$ and $j=0,1,...,N_\phi$. For the sake of conciseness, we omit the tilde marks placed above the perturbation amplitude. The two other components of the perturbation velocity, $u_\xi^{(k)}$ and $u_\phi^{(k)}$, are calculated on the cell faces $\displaystyle\big[\xi_{i+1/2},\phi_j\big]$ and $\displaystyle\big[\xi_i,\phi_{j+1/2}\big]$, respectively, where $\displaystyle\xi_{i+1/2} = \big(\xi_i + \xi_{i+1}\big)/2$ and $\displaystyle\phi_{j+1/2} = \big(\phi_j + \phi_{j+1}\big)/2$. In addition, all the variables are calculated at the interface, which is aligned with the boundary between two rows of cells at $\phi=\pi$. Since the holdup is not known \textit{a priori} and $\phi_0$ depends on it (see sec.\ \ref{Sec: Geometry}), the mesh distribution in the $\phi$-direction is changing in the process of the base flow solution until the holdup, corresponding to the specified input flow rates of the phases, is found. The number of cells in the $\phi$-direction in each sublayer (phase) is set to be proportional to its relative height ($N_\text{lower}/N_\phi = (\pi-\phi_0)/\pi$).

For the purpose of the finite--volume discretization of the governing equations, described below, the arc lengths of each cell have to be introduced \citep[][]{Issa88}
\begin{subequations}
	\begin{align}
		&\begin{aligned} \label{Eq: Delta_XI}
			&\Delta\xi_{i,j}
			= \int_{\xi_{i-1/2}}^{\xi_{i+1/2}} H_\xi d\xi
			\\
			&= 
			\begin{cases}
				\dfrac{\sin\phi_0}{\sin\phi_j}
				\Bigg[\arctan \left(\sqrt{\dfrac{1 + \cos\phi_j}{1 - \cos\phi_j}} \tanh\left(\dfrac{\xi}{2}\right)\right)\Bigg]_{\xi_{i-1/2}}^{\xi_{i+1/2}}
				\qquad
				\text{if }\cos\phi_j \ne - 1
				\\
				\dfrac{1}{2}\sin\phi_0 \bigg[\tanh\left(\dfrac{\xi}{2}\right)\bigg]_{\xi_{i-1/2}}^{\xi_{i+1/2}}
				\qquad
				\text{if }\cos\phi_j = - 1
			\end{cases}
		\end{aligned}
		\\
		&\begin{aligned} \label{Eq: Delta_PHI}
			&\Delta\phi_{i,j}
			= \int_{\phi_{j-1/2}}^{\phi_{j+1/2}} H_\phi d\phi
			\\
			&= 
			\begin{cases}
				\dfrac{\sin\phi_0}{\sinh\xi}
				\Bigg[\arctan\left(\sqrt{\dfrac{\cosh\xi_i + 1}{\cosh\xi_i - 1}}\tan\left(\dfrac{\phi}{2}\right)\right)\Bigg]_{\phi_{j-1/2}}^{\phi_{j+1/2}}
				\qquad
				\text{if }\cosh\xi_i \ne 1
				\\
				- \dfrac{1}{2} \sin\phi_0 \bigg[\cot\left(\dfrac{\phi}{2}\right)\bigg]_{\phi_{j-1/2}}^{\phi_{j+1/2}}
				\qquad
				\text{if }\cosh\xi_i = 1
			\end{cases}
		\end{aligned}
	\end{align}
\end{subequations}

Note that some other arc lengths are also required for the discretization, and they are defined as, e.g., $\displaystyle\Delta\xi_{i+1/2,j} = \int_{\xi_{i}}^{\xi_{i+1}} H_\xi d\xi$ and  $\displaystyle\Delta\phi_{i,j+1/2} = \int_{\phi_{j}}^{\phi_{j+1}} H_\phi d\phi$, while $\displaystyle\Delta\xi_{i\_i-1/2,j} = \int_{\xi_{i-1/2}}^{\xi_{i}} H_\xi d\xi$ and $\displaystyle\Delta\phi_{i,j\_j-1/2} = \int_{\phi_{j-1/2}}^{\phi_{j}} H_\phi d\phi$. Their values are calculated in the same manner as $\displaystyle\Delta\xi_{i,j}$ and $\displaystyle\Delta\phi_{i,j}$ above.

\begin{figure}[h!]
	\centering
	\includegraphics[width=0.6\textwidth,clip]{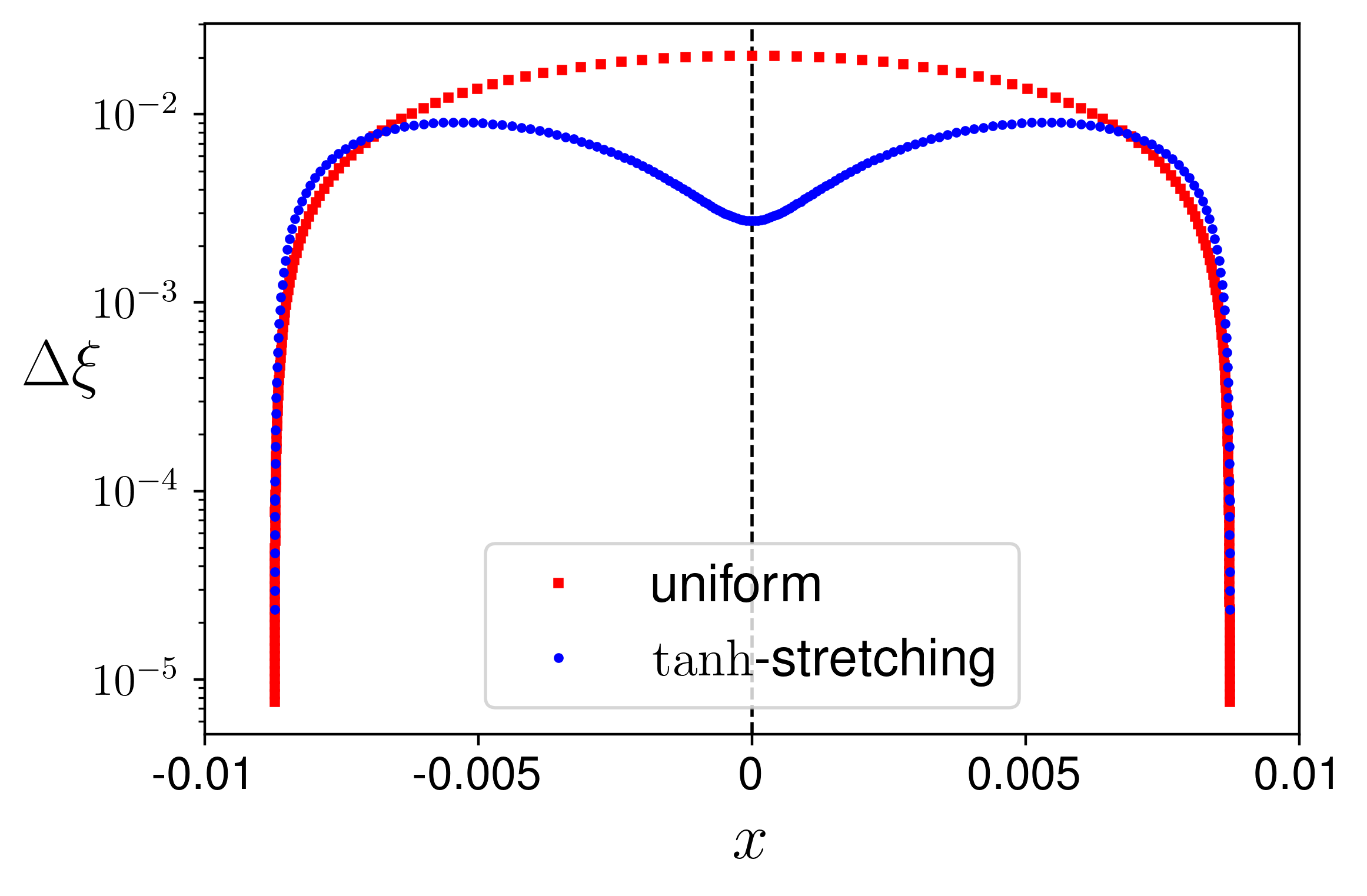}
	\caption{\label{Fig: different_grids}Arc lengths of computational cells in the $\xi$-direction at the interface for their uniform distribution (red squares) and for a distribution with the $\tanh$-stretching \ref{Eq: tanh_stretching} for $a_\xi=5$ (blue dots). The number of cells in $\xi$-direction is 200 for both cases.}	
\end{figure}

The uniform cell distribution in the $\xi$-direction leads to very strong clustering of the cells near the triple points in the physical space, since $\Delta\xi_{i,j} \to 0$ for $\xi\to \pm \infty$ (Eq.\ \ref{Eq: Delta_XI}). On the other hand, $\Delta\xi_{i,j}$ is relatively large near the pipe midplane resulting in relatively coarse grid. Therefore, we use hyperbolic tangent stretching 
\begin{equation} \label{Eq: tanh_stretching}
	\xi \to \xi_\text{max}\bigg(\frac{\tanh\big[a_\xi\big(\xi - 1\big)\big]}{\tanh(a_\xi)}
	+ \frac{\tanh(a_\xi \xi)}{\tanh(a_\xi)}\bigg),
\end{equation}
where $\xi$ is originally uniformly distributed grid cells and $a_\xi$($=5$) is the stretching parameter that determines the degree of clustering, so that the cells are redistributed to be more dense near the midplane (around $\xi=0$). This can be seen in Fig.\ \ref{Fig: different_grids} for the cells along the interface (cf. minimal and maximal values of $\Delta\xi$ for red squares and blue dots). In the $\phi$-direction, the grid is stretched near the pipe walls and at both sides of the interface using the $sin$ function
\begin{equation} \label{Eq: sin_stretching}
	\phi \to \phi - a_\phi \sin \big(2 \pi \phi\big).
\end{equation} 

In this work we use $a_\xi=5$ and $a_\phi=0.12$.

\subsection{Finite--volume base flow equations}
The finite--volume form of the base flow equation \ref{Eq: Base_flow_eqs}a is obtained by integration of the differential equation over a computational cell with $\displaystyle\xi \in \big[\xi_{i-1/2},\xi_{i+1/2}\big]$ and $\displaystyle\phi \in \bigl[\phi_{j-1/2},\phi_{j+1/2}\bigr]$. By multiplying the result of integration by $(\Delta\xi \Delta\phi)_{i,j}$ we get \citep[c.f. ][]{Issa88}
\begin{align} \label{Eq: FV_Base}
	&\begin{aligned}
		&\frac{1}{\Rey} 
		\Biggl\{
		\bigg[\bigg(\frac{\Delta\phi}{\Delta\xi}\bigg)_{i+1/2,j}
		\bigg(U_{i+1,j}^{(k)} - U_{i,j}^{(k)}\bigg)
		- \bigg(\frac{\Delta\phi}{\Delta\xi}\bigg)_{i-1/2,j}
		\bigg(U_{i,j}^{(k)} - U_{i-1,j}^{(k)}\bigg)\bigg]
		\\
		&+ \bigg[\bigg(\frac{\Delta\xi}{\Delta\phi}\bigg)_{i,j+1/2}
		\bigg(U_{i,j+1}^{(k)} - U_{i,j}^{(k)}\bigg)
		- \bigg(\frac{\Delta\xi}{\Delta\phi}\bigg)_{i,j-1/2}
		\bigg(U_{i,j}^{(k)} - U_{i,j-1}^{(k)}\bigg)\bigg]
		\Biggr\}
		\\
		&= \frac{\mu_1}{\mu_{12} \mu_k} (\Delta\xi \Delta\phi)_{i,j}
		\frac{\partial P^{(k)}_{i,j}}{\partial z}.
	\end{aligned}
\end{align}

\subsection{Finite--volume continuity equation}
The finite--volume form of the continuity equation \ref{Eq: Stability_continuity} is obtained by integration of the differential equation over a computational cell with $\displaystyle\xi \in \big[\xi_{i-1/2},\xi_{i+1/2}\big]$ and $\displaystyle\phi \in \bigl[\phi_{j-1/2},\phi_{j+1/2}\bigr]$

\begin{align} \label{Eq: FV_continuity}
	&\begin{aligned}
		& \frac{1}{\Delta\xi_{i,j}}
		\bigg(1
		- \frac{2 \sinh\xi_i}{\sin\phi_0} 
		\Delta\xi_{i\_i-1/2,j}\bigg)
		u_{\xi\_{i+1/2,j}}^{(k)}	
		- \frac{1}{\Delta\xi_{i,j}}
		\bigg(1
		+ \frac{2 \sinh\xi_i}{\sin\phi_0} 
		\Delta\xi_{i+1/2\_i,j}\bigg) 
		u_{\xi\_{i-1/2,j}}^{(k)}
		\\
		&+ \frac{1}{\Delta\phi_{i,j}}
		\bigg(1
		- \frac{2 \sin\phi_j}{\sin\phi_0} 
		\Delta\phi_{i,j\_j-1/2}\bigg)
		u_{\phi\_{i,j+1/2}}^{(k)}		
		\\
		&- \frac{1}{\Delta\phi_{i,j}}
		\bigg(1
		+ \frac{2 \sin\phi_j}{\sin\phi_0} 
		\Delta\phi_{i,j+1/2\_j}
		\bigg) 
		u_{\phi\_{i,j-1/2}}^{(k)}
		+ i \alpha u_{z\_i,j}^{(k)}
		= 0. 
	\end{aligned}
\end{align}

\subsection{Finite--volume momentum equations}

Integration of the momentum equation in $\xi$-direction \ref{Eq: Stability_xi} over  $\displaystyle\xi \in \big[\xi_{i},\xi_{i+1}\big]$ and $\displaystyle\phi \in \bigl[\phi_{j-1/2},\phi_{j+1/2}\bigr]$ yields
\begin{align} \label{Eq: FV_xi}
	&\begin{aligned}
		&\lambda \left(\Delta\xi \Delta\phi\right)_{i+1/2,j} u_{\xi\_{i+1/2,j}}^{(k)} 
		\\
		&= -i \alpha \Delta\phi_{i+1/2,j}
		\bigg(U_{i,j}^{(k)} \Delta\xi_{i+1\_i+1/2,j}
		+ U_{i+1,j}^{(k)} \Delta\xi_{i+1/2\_i,j}\bigg)
		u_{\xi\_i+1/2,j}^{(k)}
		\\
		&- \frac{\rho_1}{\rho_{12} \rho_k} 
		\Delta\phi_{i+1/2,j}
		\big(p_{i+1,j}^{(k)} - p_{i,j}^{(k)}\big)
		\\
		&+ \frac{1}{\Rey} \frac{\rho_1}{\rho_{12} \rho_k} \frac{\mu_{12} \mu_k}{\mu_1}
		\bigg[\left(\frac{\Delta\phi}{\Delta\xi}\right)_{i+1,j}
		\bigg(u_{\xi\_{i+3/2,j}}^{(k)} - u_{\xi\_{i+1/2,j}}^{(k)}\bigg)
		\\
		&- \left(\frac{\Delta\phi}{\Delta\xi}\right)_{i,j}
		\bigg(u_{\xi\_{i+1/2,j}}^{(k)} - u_{\xi\_{i-1/2,j}}^{(k)}\bigg)\bigg]
		\\
		&+ \frac{1}{\Rey} \frac{\rho_1}{\rho_{12} \rho_k} \frac{\mu_{12} \mu_k}{\mu_1}
		\bigg[\left(\frac{\Delta\xi}{\Delta\phi}\right)_{i+1/2,j+1/2}
		\left(u_{\xi\_{i+1/2,j+1}}^{(k)} - u_{\xi\_{i+1/2,j}}^{(k)}\right)
		\\
		&- \left(\frac{\Delta\xi}{\Delta\phi}\right)_{i+1/2,j-1/2}
		\left(u_{\xi\_{i+1/2,j}}^{(k)} - u_{\xi\_{i+1/2,j-1}}^{(k)}\right)\bigg]
		\\	
		&- \frac{1}{\Rey} \frac{\rho_1}{\rho_{12} \rho_k} \frac{\mu_{12} \mu_k}{\mu_1}
		\alpha ^2 \left(\Delta\xi \Delta\phi\right)_{i+1/2,j} u_{\xi\_{i+1/2,j}}^{(k)} 
		\\
		&- \frac{4}{\Rey} \frac{\rho_1}{\rho_{12} \rho_k} \frac{\mu_{12} \mu_k}{\mu_1}
		\frac{\sin\phi_{j} \Delta\phi_{i+1/2,j}}{\sin\phi_0}
		\\
		&\bigg(
		\frac{u_{\phi\_i+1,j-1/2}^{(k)} \Delta\phi_{i+1,j+1/2\_j}
			+ u_{\phi\_i+1,j+1/2}^{(k)} \Delta\phi_{i+1,j\_j-1/2}}{\Delta\phi_{i+1,j}}
		\\
		&- \frac{u_{\phi\_i  ,j-1/2}^{(k)} \Delta\phi_{i,j+1/2\_j} 
			+ u_{\phi\_i  ,j+1/2}^{(k)} 
			\Delta\phi_{i,j\_j-1/2}}{\Delta\phi_{i,j}}
		\bigg)
		\\
		&+ \frac{4}{\Rey} \frac{\rho_1}{\rho_{12} \rho_k} \frac{\mu_{12} \mu_k}{\mu_1}
		\frac{\Delta\xi_{i+1/2,j} \sinh\xi_{i+1/2}}{\sin\phi_0}
		\\
		&\bigg(\frac{u_{\phi\_i,j+1/2}^{(k)} 
			\Delta\xi_{i+1\_i+1/2,j+1/2}
			+ u_{\phi\_i+1,j+1/2}^{(k)} \Delta\xi_{i+1/2\_i,j+1/2}}{\Delta\xi_{i+1/2,j+1/2}}
		\\
		&- \frac{u_{\phi\_i,j-1/2}^{(k)} \Delta\xi_{i+1\_i+1/2,j-1/2}
			+ u_{\phi\_i+1,j-1/2}^{(k)} \Delta\xi_{i+1/2\_i,j-1/2}}{\Delta\xi_{i+1/2,j-1/2}}
		\bigg)
		\\
		&+ \frac{4}{\Rey} \frac{\rho_1}{\rho_{12} \rho_k} \frac{\mu_{12} \mu_k}{\mu_1}
		\frac{\cos^2\phi_j - \cosh^2\xi_{i+1/2}}{\sin^2\phi_0}
		\left(\Delta\xi\Delta\phi\right)_{i+1/2,j}  
		u_{\xi\_i+1/2,j}^{(k)}. 
	\end{aligned}
\end{align}
Integration of the momentum equation in the $\phi$-direction \ref{Eq: Stability_phi} over  $\displaystyle\xi \in \big[\xi_{i-1/2},\xi_{i+1/2}\big]$ and $\displaystyle\phi \in \bigl[\phi_{j},\phi_{j+1}\bigr]$ yields
\begin{align} \label{Eq: FV_phi}
	&\begin{aligned}
		&\lambda 
		(\Delta\xi \Delta\phi)_{i,j+1/2}
		u_{\phi\_i,j+1/2}^{(k)}
		\\
		&=- i \alpha \Delta\xi_{i,j+1/2}
		\biggl(U_{i,j}^{(k)} \Delta\phi_{i,j+1\_j+1/2}
		+ U_{i,j+1}^{(k)} \Delta\phi_{i,j+1/2\_j}\biggr)		
		u_{\phi\_i,j+1/2}^{(k)}
		\\
		&- \frac{\rho_1}{\rho_{12} \rho_k}
		\Delta\xi_{i,j+1/2} 
		\bigl(p_{i,j+1}^{(k)} - p_{i,j}^{(k)}\bigr)
		\\
		&+ \frac{1}{\Rey}\frac{\rho_1}{\rho_{12} \rho_k} \frac{\mu_{12} \mu_k}{\mu_1} 	
		\biggl[\left(\frac{\Delta\phi}{\Delta\xi}\right)_{i+1/2,j+1/2}
		\left(u_{\phi\_{i+1,j+1/2}}^{(k)} - u_{\phi\_{i,j+1/2}}^{(k)}\right)
		\\
		&- \left(\frac{\Delta\phi}{\Delta\xi}\right)_{i-1/2,j+1/2}
		\left(u_{\phi\_{i,j+1/2}}^{(k)} - u_{\phi\_{i-1,j+1/2}}^{(k)}\right)\biggr]
		\\
		&+ \frac{1}{\Rey}\frac{\rho_1}{\rho_{12} \rho_k} \frac{\mu_{12} \mu_k}{\mu_1} 
		\biggl[\left(\frac{\Delta\xi}{\Delta\phi}\right)_{i,j+1}
		\left(u_{\phi\_{i,j+3/2}}^{(k)} - u_{\phi\_{i,j+1/2}}^{(k)}\right)
		\\
		&- \left(\frac{\Delta\xi}{\Delta\phi}\right)_{i,j}
		\left(u_{\phi\_{i,j+1/2}}^{(k)} - u_{\phi\_{i,j-1/2}}^{(k)}\right)\biggr]
		\\
		&- \frac{1}{\Rey}\frac{\rho_1}{\rho_{12} \rho_k} \frac{\mu_{12} \mu_k}{\mu_1}
		\alpha ^2 
		(\Delta\xi \Delta\phi)_{i,j+1/2}
		u_{\phi\_i,j+1/2}^{(k)}
		\\
		&+ \frac{4}{\Rey}\frac{\rho_1}{\rho_{12} \rho_k} \frac{\mu_{12} \mu_k}{\mu_1}
		\frac{\sin\phi_{j+1/2}}{\sin\phi_0}	
		\\	
		&\Delta\phi_{i,j+1/2}
		\biggl(
		\frac{u_{\xi\_i+1/2,j}^{(k)} \Delta\phi_{i+1/2,j+1\_j+1/2}
			+ u_{\xi\_i+1/2,j+1}^{(k)} \Delta\phi_{i+1/2,j+1/2\_j}}{\Delta\phi_{i+1/2,j+1/2}}
		\\
		-& \frac{u_{\xi\_i-1/2,j}^{(k)} \Delta\phi_{i-1/2,j+1\_j+1/2} 
			+ u_{\xi\_i-1/2,j+1}^{(k)} 
			\Delta\phi_{i-1/2,j+1/2\_j}}{\Delta\phi_{i-1/2,j+1/2}}
		\biggr)	
		\\
		&- \frac{4}{\Rey}\frac{\rho_1}{\rho_{12} \rho_k} \frac{\mu_{12} \mu_k}{\mu_1}
		\frac{\sinh\xi_{i}}{\sin\phi_0}	
		\\	
		&\Delta\xi_{i,j+1/2}
		\biggl(\frac{u_{\xi\_i-1/2,j+1}^{(k)} 
			\Delta\xi_{i+1/2\_i,j+1}
			+ u_{\xi\_i+1/2,j+1}^{(k)} \Delta\xi_{i\_i-1/2,j+1}}{\Delta\xi_{i,j+1}}
		\\
		-& \frac{u_{\xi\_i-1/2,j}^{(k)} \Delta\xi_{i+1/2\_i,j}
			+ u_{\xi\_i+1/2,j}^{(k)} \Delta\xi_{i\_i-1/2,j}}{\Delta\xi_{i,j}}
		\biggr)
		\\
		&+ \frac{4}{\Rey}\frac{\rho_1}{\rho_{12} \rho_k} \frac{\mu_{12} \mu_k}{\mu_1}
		\frac{\cos^2\phi_{j+1/2} - \cosh^2\xi_i}{\sin^2\phi_0} 
		(\Delta\xi \Delta\phi)_{i,j+1/2}		
		u_{\phi\_i,j+1/2}^{(k)}.
	\end{aligned}
\end{align}
Integration of the momentum equation in the $z$-direction \ref{Eq: Stability_z} over  $\displaystyle\xi \in \big[\xi_{i-1/2},\xi_{i+1/2}\big]$ and $\displaystyle\phi \in \bigl[\phi_{j-1/2},\phi_{j+1/2}\bigr]$ yields
\begin{align} \label{Eq: FV_z}
	&\begin{aligned}
		&\lambda (\Delta\xi \Delta\phi)_{i,j} u_{z\_i,j}^{(k)} 
		\\		
		=&- \biggl[\frac{(\Delta\xi)_{i-1/2,j}}{(\Delta\xi)_{i+1/2,j}}\left(U_{i+1,j}^{(k)} - U_{i,j}^{(k)}\right) 
		+ \frac{(\Delta\xi)_{i+1/2,j}}{(\Delta\xi)_{i-1/2,j}}\left(U_{i,j}^{(k)} - U_{i-1,j}^{(k)}\right)\biggr] \Delta\phi_{i,j}
		\\
		&\frac{u_{\xi\_i-1/2,j}^{(k)} \Delta\xi_{i+1/2\_i,j}
			+ u_{\xi\_i+1/2,j}^{(k)} \Delta\xi_{i\_i-1/2,j}}{(\Delta\xi)_{i+1/2,j} + (\Delta\xi)_{i-1/2,j}}
		\\
		&- \biggl[ \frac{(\Delta\phi)_{i,j-1/2}}{(\Delta\phi)_{i,j+1/2}}\left(U_{i,j+1}^{(k)}
		- U_{i,j}^{(k)}\right)
		+   \frac{(\Delta\phi)_{i,j+1/2}}{(\Delta\phi)_{i,j-1/2}}\left(U_{i,j}^{(k)}
		- U_{i,j-1}^{(k)}\right) \biggr] \Delta\xi_{i,j}
		\\
		&\frac{u_{\phi\_i,j-1/2}^{(k)} \Delta\phi_{i,j+1/2\_j}
			+        u_{\phi\_i,j+1/2}^{(k)} \Delta\phi_{i,j\_j-1/2}}{(\Delta\phi)_{i,j+1/2} + (\Delta\phi)_{i,j-1/2}}
		\\
		&- i \alpha U_{z\_i,j}^{(k)} (\Delta\xi \Delta\phi)_{i,j} u_{z\_i,j}^{(k)} 
		\\
		&-i \alpha \frac{\rho_1}{\rho_{12} \rho_k} (\Delta\xi \Delta\phi)_{i,j} p^{(k)}_{i,j} 
		\\
		&+ \frac{1}{\Rey} \frac{\rho_1}{\rho_{12} \rho_k} \frac{\mu_{12} \mu_k}{\mu_1}
		\biggl[\left(\frac{\Delta\phi}{\Delta\xi}\right)_{i+1/2,j}
		\left(u_{z\_{i+1,j}}^{(k)} - u_{z\_{i,j}}^{(k)}\right)
		\\
		&- \left(\frac{\Delta\phi}{\Delta\xi}\right)_{i-1/2,j}
		\left(u_{z\_{i,j}}^{(k)} - u_{z\_{i-1,j}}^{(k)}\right)\biggr]
		\\
		&+ \frac{1}{\Rey} \frac{\rho_1}{\rho_{12} \rho_k} \frac{\mu_{12} \mu_k}{\mu_1}
		\biggl[\left(\frac{\Delta\xi}{\Delta\phi}\right)_{i,j+1/2}
		\left(u_{z\_{i,j+1}}^{(k)} - u_{z\_{i,j}}^{(k)}\right)
		\\
		&- \left(\frac{\Delta\xi}{\Delta\phi}\right)_{i,j-1/2}
		\left(u_{z\_{i,j}}^{(k)} - u_{z\_{i,j-1}}^{(k)}\right)\biggr]
		\\
		&- \frac{1}{\Rey} \frac{\rho_1}{\rho_{12} \rho_k} \frac{\mu_{12} \mu_k}{\mu_1}
		\alpha ^2 (\Delta\xi \Delta\phi)_{i,j} 
		u_{z\_i,j}^{(k)}. 
	\end{aligned}
\end{align}

\subsection{Finite--volume interfacial boundary conditions}

The finite--volume discretization of the non-trivial (linear) boundary conditions at the interface (Eqs.\ \ref{Eq: BC_kinematic_lin}, \ref{Eq: BC_continuity_velocity}c, \ref{Eq: BC_shear_stress_lin}a-c), where all the variables are calculated at the boundary between two rows of cells at $\phi_j=\pi$. An infinitesimal interface deformation in the $y$-direction (in phisycal space) is defined through the Lam\'e coefficient
\begin{equation}
	\Delta\eta_i
	= \int_{\phi_{j-1/2}}^{\phi_{j+1/2}} \eta H_\phi d\phi.
\end{equation}

Then the finite--volume form of the kinematic boundary condition (Eq.\ \ref{Eq: BC_kinematic_lin}) is
\begin{align}
	&\begin{aligned}
		\lambda \Delta\eta_i &=
		u_{\phi\_i,j(\phi=\pi)} 
		- i \alpha U_{i,j(\phi=\pi)} \Delta\eta_i.
	\end{aligned}
\end{align}
and the jump in the axial velocity of the perturbation at the interface (Eq.\ \ref{Eq: BC_kinematic_lin}c) can be written as
\begin{equation}
	u_{z\_i,j(\phi = \pi)}^{(1)}
	- u_{z\_i,j(\phi = \pi)}^{(2)}
	+ \Biggl(\frac{U_{i,j+1}^{(1)} - U_{i,j(\phi=\pi)}}{\Delta\phi_{i,j+1/2}} 
	- \frac{U_{i,j(\phi=\pi)} - U_{i,j-1}^{(2)}}{\Delta\phi_{i,j-1/2}}\Biggr)
	\Delta\eta_i
	= 0.
\end{equation}

The finite--volume discretization of the linearized boundary conditions on the tangential stresses in planes $\bigl(\xi,\phi\bigr)$ and $\bigl(\phi,z\bigr)$ (Eqs.\ \ref{Eq: BC_shear_stress_lin}a-c)) yields
\begin{subequations}
	\begin{align}  \label{Eq: FV_tan_xi_phi_BC}
		&\begin{aligned}
			&\mu_{1 2}
			\frac{u_{\xi\_i+1/2,j+1}^{(1)} 
				- u_{\xi\_i+1/2,j}}{\Delta\phi_{i+1/2,j+1/2}}\Bigr|_{\phi = \pi}
			- \frac{u_{\xi\_i+1/2,j} 
				- u_{\xi\_i+1/2,j-1}^{(2)}}{\Delta\phi_{i+1/2,j-1/2}}\Bigr|_{\phi = \pi}
			\\
			&+ (\mu_{1 2}-1) \frac{u_{\phi\_i+1,j} 
			- u_{\phi\_i,j}}{\Delta\xi_{i+1/2,j}}\Bigr|_{\phi = \pi}
			\\
			&- i \alpha 
			\frac{U_{z\_i+1,j} 
				- U_{z\_i,j}}{\Delta\xi_{i+1/2,j}} 
			\frac{\Delta\xi_{i+1\_i+1/2,j} \Delta\eta_i
				+ \Delta\xi_{i+1/2\_i,j} \Delta\eta_{i+1}}{\Delta\xi_{i+1/2,j}} \Bigr|_{\phi = \pi}
			\\
			&+ 2 (\mu_{1 2}-1) \frac{\sinh\xi_{i+1/2}}{\sin\phi_0}
			\frac{u_{\phi\_i,j} \Delta\xi_{i+1\_i+1/2,j}
				+ u_{\phi\_i+1,j} \Delta\xi_{i+1/2\_i,j}}{\Delta\xi_{i+1/2,j}} \Bigr|_{\phi = \pi}
			= 0.
		\end{aligned}
	\end{align}
	\begin{align}  \label{Eq: FV_tan_phi_z_BC}
		&\begin{aligned}
			&i \alpha \left(\mu_{1 2}-1\right)u_{\phi\_i,j(\phi = \pi)}
			\\
			&- \frac{\mu_{1 2} - 1}{\Delta\xi_{i-1/2,j} + \Delta\xi_{i+1/2,j}}
			\Biggl[\frac{\Delta\xi_{i-1/2,j}}{\Delta\xi_{i+1/2,j}} \left(\Delta\eta_{i+1} - \Delta\eta_i\right) 
			+ \frac{\Delta\xi_{i+1/2,j}}{\Delta\xi_{i-1/2,j}} \left(\Delta\eta_i - \Delta\eta_{i-1}\right) 
			\Biggr]_{\phi = \pi}			
			\\
			&+ \frac{\mu_{1 2} \Delta\eta_{i}}{\Delta\xi_{i,j}\Delta\phi_{i,j+1/2}}
			\left[\biggl(\frac{\Delta\xi}{\Delta\phi}\biggr)_{i,j+1} \left(U_{i,j+2}^{(1)} - U_{i,j+1}^{(1)}\right) 
			- \biggl(\frac{\Delta\xi}{\Delta\phi}\biggr)_{i,j} 
			\left(U_{i,j+1}^{(1)} - U_{i,j}\right) \right]_{\phi = \pi}
			\\
			&- \frac{\Delta\eta_{i}}{\Delta\xi_{i,j}\Delta\phi_{i,j-1/2}}
			\left[\biggl(\frac{\Delta\xi}{\Delta\phi}\biggr)_{i,j}\left(U_{i,j} - U_{i,j-1}^{(2)}\right) 
			- \biggl(\frac{\Delta\xi}{\Delta\phi}\biggr)_{i,j-1}\left(U_{i,j-1}^{(2)} - U_{i,j-2}^{(2)}\right) \right]_{\phi = \pi}
			\\
			&- \frac{\big(\mu_{1 2} - 1\big) \Delta\eta_{i}}{\Delta\xi_{i-1/2,j} + \Delta\xi_{i+1/2,j}}
			\Biggl[\frac{\Delta\xi_{i-1/2,j}}{\Delta\xi_{i+1/2,j}} \left(U_{i+1,j} - U_{i,j}\right)
			+ \frac{\Delta\xi_{i+1/2,j}}{\Delta\xi_{i-1/2,j}} \left(U_{i,j} - U_{i-1,j}\right)
			\Biggr]_{\phi = \pi}			
			\\
			&+ \mu_{1 2} \frac{u_{z\_i,j+1}^{(1)} 
				- u_{z\_i,j}^{(1)}}{\Delta\phi_{i,j+1/2}}\Bigr|_{\phi = \pi}
			- \frac{u_{z\_i,j}^{(2)} 
				- u_{z\_i,j-1}^{(2)}}{\Delta\phi_{i,j-1/2}}\Bigr|_{\phi = \pi}
			= 0.
		\end{aligned}
	\end{align}
\end{subequations}

Due to the surface tension, the pressure field is discontinuous across the interface. Therefore, at the interface the pressure in each phase has to be extrapolated from the neighboring cells
\begin{subequations}
	\begin{align}
	&\begin{aligned}
	p_{i,j(\phi=\pi)}^{(1)} 
	&= p_{i,j+1/2}^{(1)}\biggr|_{\phi = \pi} 
	- \frac{\Delta\phi_{i,j+1/2\_j}}{\Delta\phi_{i,j+1}} 
	\left(p_{i,j+3/2}^{(1)} - p_{i,j+1/2}^{(1)}\right)\biggr|_{\phi = \pi},
	\end{aligned}
	\end{align}
	\begin{align}
	&\begin{aligned}
	p_{i,j(\phi=\pi)}^{(2)} 
	&= p_{i,j-1/2}^{(2)}\biggr|_{\phi = \pi} 
	- \frac{\Delta\phi_{i,j\_j-1/2}}{\Delta\phi_{i,j-1}} 
	\left(p_{i,j-3/2}^{(2)} - p_{i,j-1/2}^{(2)}\right)\biggr|_{\phi = \pi}, 
	\end{aligned}
	\end{align}
\end{subequations}
and the finite--volume discretization of \ref{Eq: BC_shear_stress_lin}c reads
\begin{align} \label{Eq: FV_normal_BC}
	&\begin{aligned}
		&\frac{p_{i,j-1/2}^{(2)} \left(\Delta \phi_{i,j-1/2} + \Delta \phi_{i,j-1}\right)
			- p_{i,j-3/2}^{(2)} \Delta \phi_{i,j-1/2}}{\Delta\phi_{i,j-1}}\Biggr|_{\phi = \pi}
		\\
		&- \frac{p_{i,j+1/2}^{(1)} \left(\Delta \phi_{i,j+1/2} + \Delta \phi_{i,j+1}\right)
			- p_{i,j+3/2}^{(1)} \Delta \phi_{i,j+1/2}}{\Delta\phi_{i,j+1}}\Biggr|_{\phi = \pi}
		\\
		&+ \frac{1}{\Fr} \left(\rho_{1 2}-1\right)
		\left(-\cosh\xi_i + \frac{\sinh^2\xi_i}{\cosh\xi_i + 1}\right) 
		\Delta\eta_i
		\\
		&+ \frac{2}{\Rey} \mu_{1 2} 
		\frac{u_{\phi\_i,j+1}^{(1)} - u_{\phi\_i,j}}{\Delta\phi_{i,j+1/2}} \biggr|_{\phi = \pi}
		- \frac{2}{\Rey}
		\frac{u_{\phi\_i,j} - u_{\phi\_i,j-1}^{(2)}}{\Delta\phi_{i,j-1/2}} \biggr|_{\phi = \pi}
		\\
		&-\frac{4}{\Rey} \left(\mu_{12}-1\right)
		\frac{\sinh\xi_i}{\sin\phi_0}
		\frac{u_{\xi\_i-1/2,j} \Delta\xi_{i+1/2\_i,j} 
			+ u_{\xi\_i+1/2,j} \Delta\xi_{i\_i-1/2,j}}{\Delta\xi_{i,j}} \biggr|_{\phi = \pi}
		\\
		&+ \frac{1}{\We} 
		\bigg[\frac{2}{\Delta\xi_{i+1/2,j} \big(\Delta\xi_{i-1/2,j} + \Delta\xi_{i+1/2,j}\big)} \Biggr|_{\phi=\pi} \Delta\eta_{i+1}
		\\	
		&- \bigg[\frac{2}{\Delta\xi_{i+1/2,j} \big(\Delta\xi_{i-1/2,j} + \Delta\xi_{i+1/2,j}\big)}
		+ \frac{2}{\big(\Delta\xi_{i-1/2,j} + \Delta\xi_{i+1/2,j}\big) \Delta\xi_{i-1/2,j}}\bigg]_{\phi=\pi} \Delta\eta_{i}
		\\
		&+ \frac{2}{\big(\Delta\xi_{i-1/2,j} + \Delta\xi_{i+1/2,j}\big) \Delta\xi_{i-1/2,j}} \Biggr|_{\phi=\pi} \Delta\eta_{i-1}\bigg]
		\\
		&- \frac{2}{\We}\frac{1}{\sin\phi_0}
		\biggl[\biggl(\sinh\xi_{i+1/2} \frac{\Delta\xi_{i+1\_i+1/2} \Delta\eta_i + \Delta\xi_{i+1/2\_i} \Delta\eta_{i+1}}{\Delta\xi_{i+1/2,j}}
		\\
		&- \sinh\xi_{i-1/2} \frac{\Delta\xi_{i\_i-1/2} \Delta\eta_{i-1} 
			+ \Delta\xi_{i-1/2\_i-1} \Delta\eta_i}{\Delta\xi_{i-1/2,j}}
		\biggr)\biggr|_{\phi = \pi}	
		\\
		&- \biggl(\sinh\xi_{i+1/2}
		- \sinh\xi_{i-1/2}\biggr)\biggr|_{\phi = \pi}
		\Delta\eta_i	\biggr]
		- \frac{\alpha ^2}{\We} \Delta\eta_i
		= 0.
	\end{aligned}
\end{align}

\subsection{Linear stability analysis }

The discretization of the linearized governing equations (Eqs.\ \ref{Eq: FV_continuity} - \ref{Eq: FV_z}) and boundary conditions(Eqs.\ \ref{Eq: BC_no-slip_lin} - \ref{Eq: BC_shear_stress_lin}c) reduces the linear stability problem to a generalized eigenvalue problem
\begin{equation} \label{Eq: Eigenvalue_problem}
	\lambda \cdot \vec{B} 
	\cdot \vec{v}
	= \vec{J} 
	\cdot \vec{v}
\end{equation}
with the complex time increment $\lambda = \lambda_{R} + i \lambda_{I}$ being the eigenvalue and the perturbation $\displaystyle\vec{v}= \big(u_\xi,u_\phi,u_z, p,\Delta\eta\big)^T$ being the corresponding eigenvector. $\vec{B}$ is a diagonal matrix with the elements corresponding to time derivatives of $\tilde{u}$ and $\eta$ equal to one. On the other hand, the diagonal elements corresponding to $p$ and the boundary conditions without time derivatives (all except for the kinematic one) are zeros, so that $\det \vec{B} = 0$. The real part of the eigenvalue, $\lambda_{R}$ determines the growth rate of the perturbation, so that the flow is considered unstable when there exists an eigenvalue with a positive real part. Considering all possible wavenumbers, the eigenvalue with the largest $\lambda_{R}$ and the corresponding eigenvector are referred as leading or the most unstable one.

The generalized eigenvalue problem \ref{Eq: Eigenvalue_problem} is solved by the Arnoldi iteration in the shift-and-inverse mode
\begin{equation}
	(\vec{J} - \lambda_0 \cdot \vec{B})^{-1} 
	\cdot \vec{B} 
	\cdot \vec{v}
	= \vartheta 
	\cdot \vec{v},
\end{equation}
where $\displaystyle\vartheta = 1 /\bigl(\lambda - \lambda_0\bigr)$ and $\lambda_0$ is a complex shift. Following the approach of \citep{Gelfgat07a} and in similar manner as in \citep{Gelfgat20a}, we use the ARPACK package of \citep{Lechouq98} in FORTRAN for the Arnoldi iteration, computing the LU decomposition of the complex matrix $\displaystyle(\vec{J} - \lambda_0 \cdot \vec{B})^{-1}$, so that the calculation of the next Krylov vector for the Arnoldi
method is reduced to one backward and one forward substitution.

%%%%%%%%%%%%%%%%%%%%%%%%%%%%%%%%%%%%%%%%%%%%%%%%%%%%%%%%%%%%%%%%%%
\section{Immersed boundary method}

An alternative numerical approach to solve the stability problem for two-phase stratified flow in a circular pipe is to use Cartesian formulation same as in \cite{Gelfgat20a} and to treat the pipe wall by the immersed boundary method (IBM). In this approach, the computational domain is chosen as a square with the length of the side equals to the pipe diameter.  Momentum forcing and mass source are applied outside of the flow domain and the velocity is interpolated across the pipe wall to satisfy no-slip boundary conditions. For this purpose, we apply interpolations of \cite{Kim01} for the conservation of mass in the continuity equation and those of \cite{Wu19} for velocities in the momentum equations. In the following section, the results obtained by IBM are cross verified with those in the bipolar coordinates.

%%%%%%%%%%%%%%%%%%%%%%%%%%%%%%%%%%%%%%%%%%%%%%%%%%%%%%%%%%%%%%%%%%
\section{Results and discussion}

To the best of our knowledge, a rigorous linear stability analysis of two-phase stratified flow in a circular pipe has never been performed before. Since there are no independent results for comparison, we use two independent numerical approaches to verify our results. In addition, we can verify the numerical solution for the base flow by comparison with the exact analytical solution reported in \cite{Goldstein15}. Since the considered problem involves too many governing parameters, we focus on two representative examples of two-phase flows to demonstrate the proposed numerical approach. The physical properties and pipe diameters of the considered flows are summarized in Table\ \ref{Tab: Two-phase_systems}.  

{\renewcommand{\arraystretch}{1.8}
	\begin{table}[h!]
		\caption{\label{Tab: Two-phase_systems}Physical properties of the two-phase pipe flows considered.}				
		\centering		
		\begin{tabular}{|c|c|c|c|c|c|c|c|c|}
			\hline\hline
			& $\mu_1$ & $\mu_2$ & $\mu_{1 2}$ & $\rho_1$ & $\rho_2$ & $\rho_{1 2}$ & $\sigma$ & $D$ 
			\\
			& $\bigg[\dfrac{kg}{m \cdot s}\bigg]$ & $\bigg[\dfrac{kg}{m \cdot s}\bigg]$ & & $\bigg[\dfrac{kg}{m^3}\bigg]$ & $\bigg[\dfrac{kg}{m^3}\bigg]$ &  & $\bigg[\dfrac{N}{m}\bigg]$ & $[m]$ \\
			\hline
			oil-water & 0.001 & 0.0097 & 0.103 & 1000 & 835 & 1.197 &
			0.03 & 0.02\\
			\hline
			air-water & 0.001 & 1.818$\times 10^{-5}$ & 55.556 & 1000 & 1 & 1000 &
			0.072 & 0.014 \\
			\hline\hline
		\end{tabular}
	\end{table}
}

%%%%%%%%%%%%%%%%%%%%%%%%%%%%%%%%%%%%%%%%%%%%%%%%%%%%%%%%%%%%%%%%%%
\subsection{Liquid--liquid pipe flow}

As a benchmark liquid-liquid system, an oil-water flow in a circular pipe of diameter $D=0.02$m is chosen (see the first line in table\ \ref{Tab: Two-phase_systems}). The light upper phase is oil with density and viscosity slightly different from water, $\rho_{1 2}=1.198$ and $\mu_{1 2}=0.103$. Once the physical properties and pipe diameter are defined, the holdup and base flow velocity profile are uniquely determined by the flow rate ratio \citep{Goldstein15}. In this work, we evaluate the base flow by solving numerically equation\ \ref{Eq: FV_Base} in the same manner as in \cite{Gelfgat20a}.  In table\ \ref{Tab: L-L_BF_covergence}, we demonstrate convergence of $U_{1S}/U_{2S}$ for two different holdups $h$ using the proposed numerical methods in the bipolar (denoted as bipolar FVM in the table) and Cartesian (denoted as IBM) coordinates. It can be seen that $U_{1S}/U_{2S}$ converges to the third decimal place (0.379) for the computational grids with 200 and more cells in each direction, for both numerical methods ($N_\xi (N_x) = 200, N_\phi (N_y) = 200$). The analytical solution of \cite{Goldstein15} is also reported in table\ \ref{Tab: L-L_BF_covergence}, and it can be seen that deviation of the numerical solution from the analytical one is much less than one percent. 
 
\begin{figure}[h!]
	\centering
	\subfloat[]{\includegraphics[width=0.34\textwidth,clip]{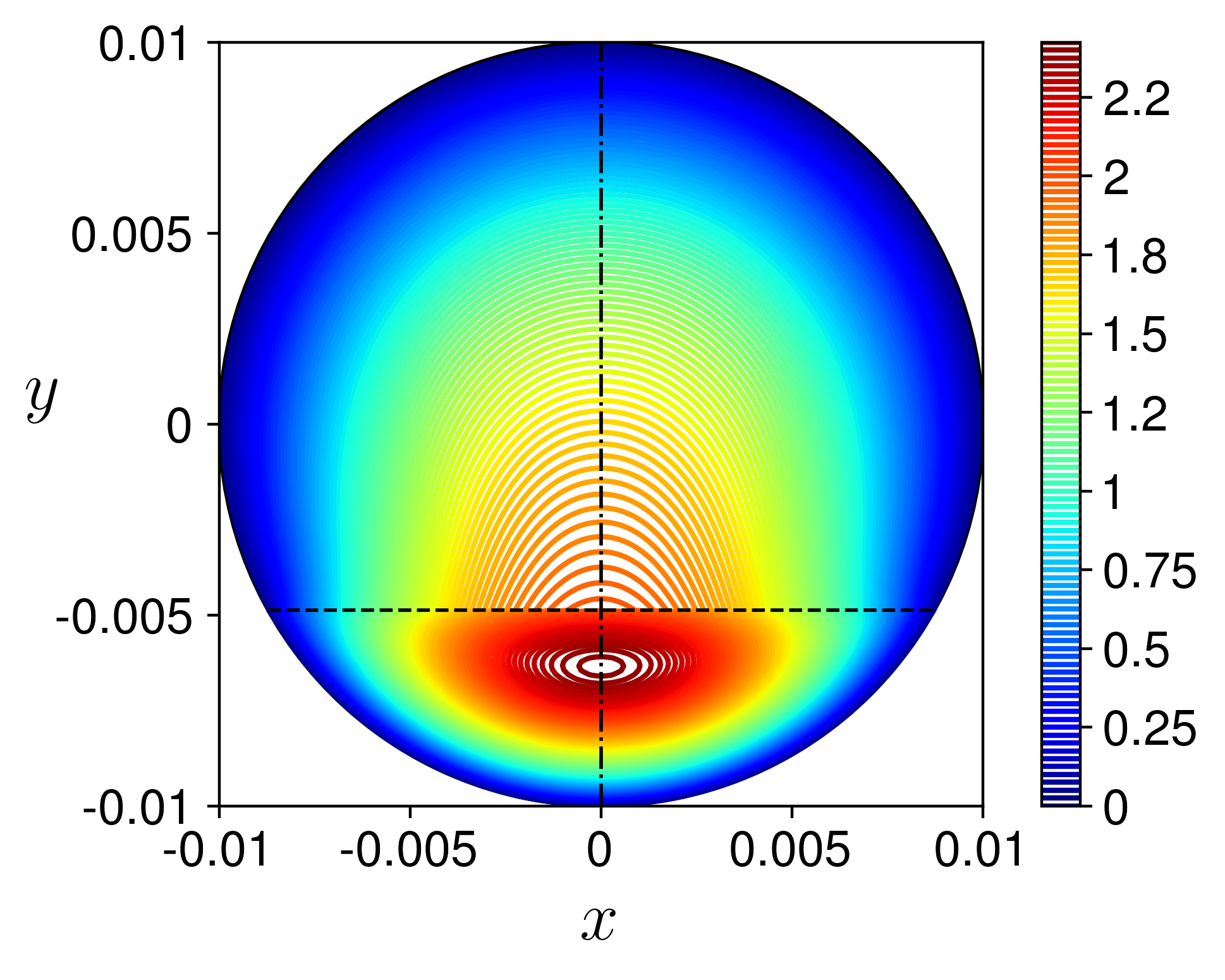}}
	\subfloat[]{\includegraphics[width=0.34\textwidth,clip]{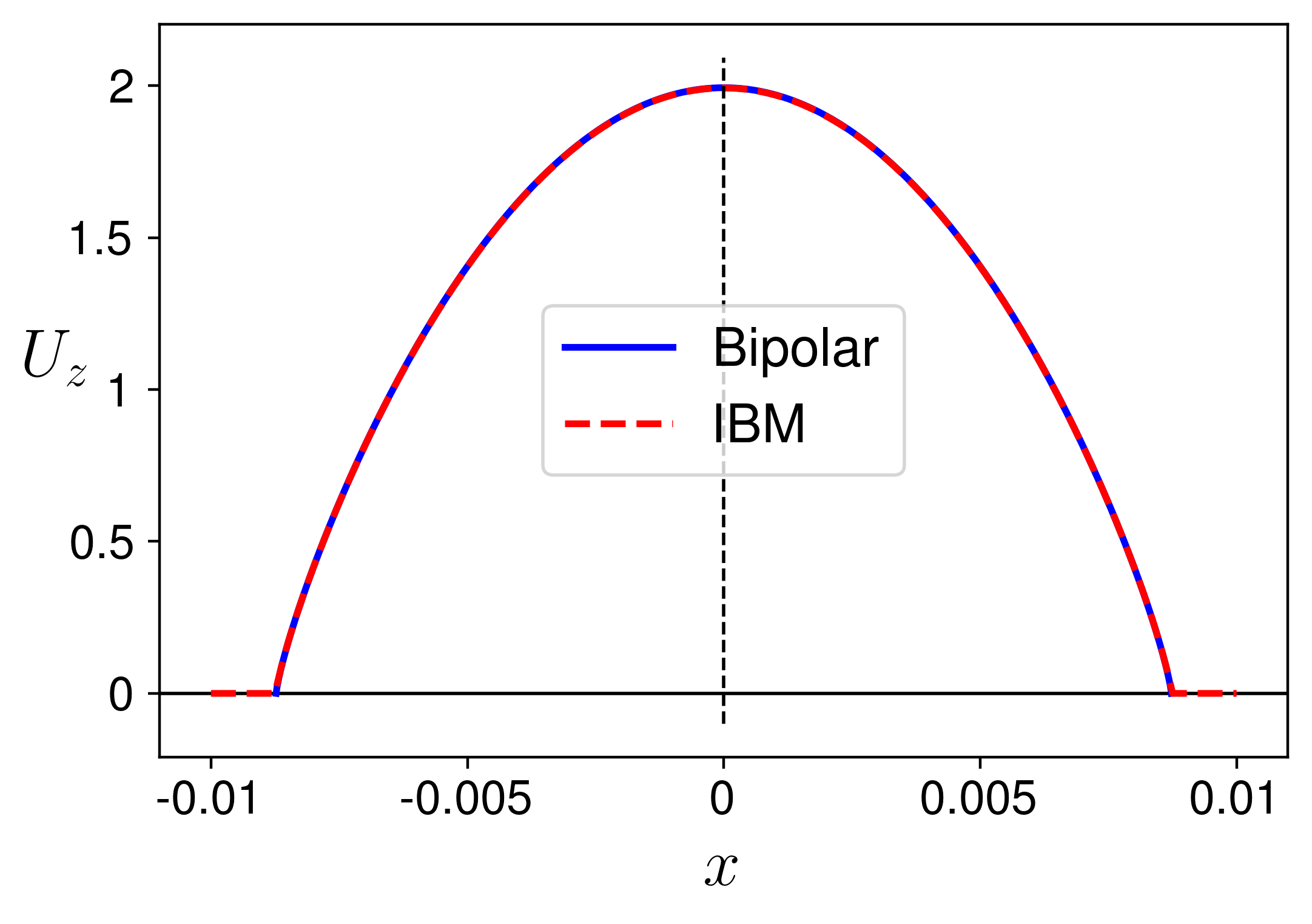}}
	\subfloat[]{\includegraphics[width=0.34\textwidth,clip]{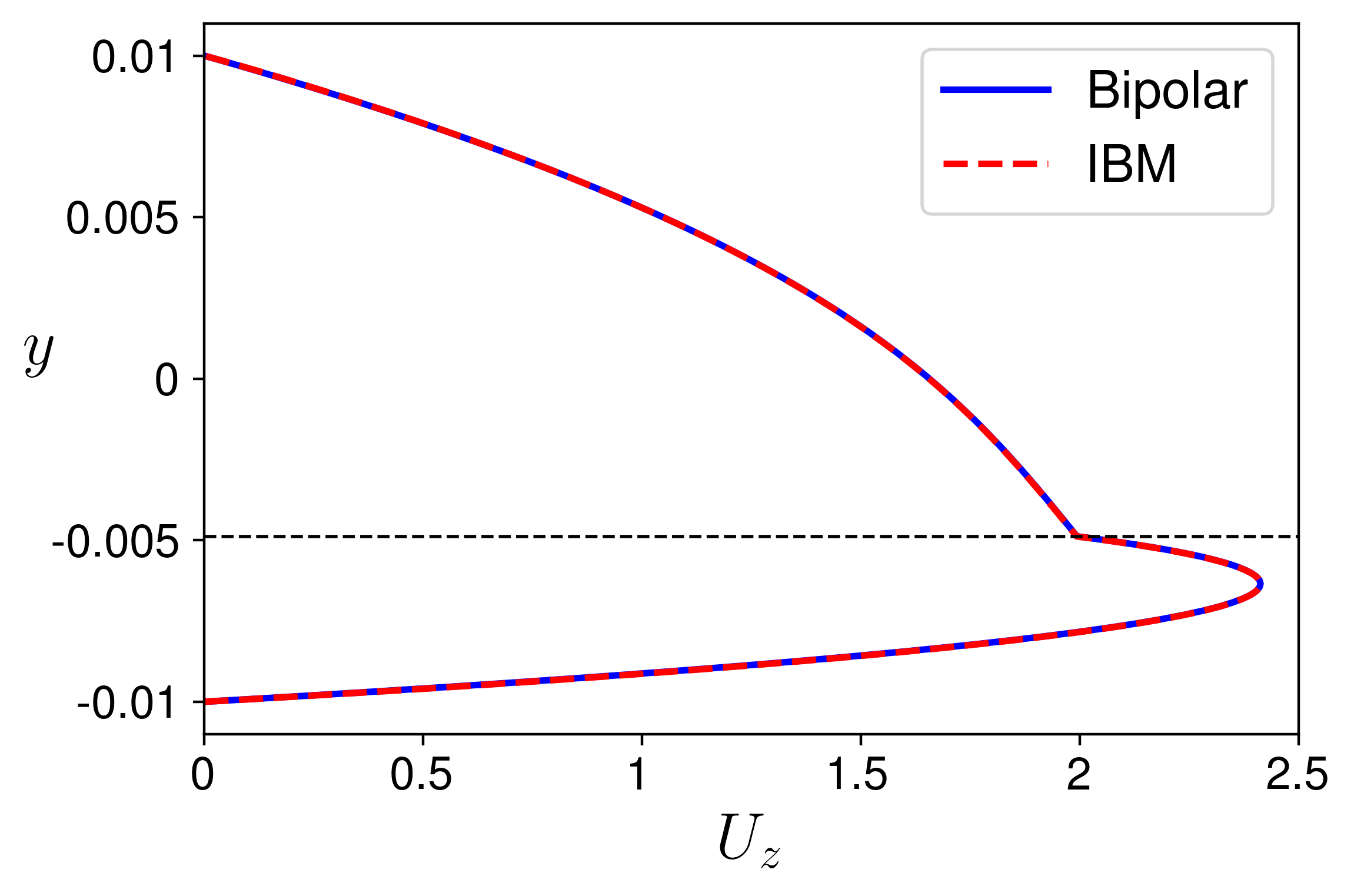}}
	\caption{\label{Fig: Base_flow}Base flow velocity, $U_z$. Liquid-liquid ($m=0.103$) flow in a circular pipe of diameter $D=0.02$m, $h=0.202$. (a) cross-sectional contours; (b) velocity at the interface; (c) velocity at the midplane $x=0$. The interface is denoted by horizontal dashed black line. The cross-section centerline is denoted by a vertical dash-dot black line.}
\end{figure}

The base flow velocity $U_z$ profile in the pipe cross-section is shown in Fig.\ \ref{Fig: Base_flow}. It is important to note that for $h=0.202$, although the lower (water) phase occupies only about a fifth of the pipe cross-sectional area as defined by its holdup, the height of the interface equals to about a fourth of the pipe diameter. Symmetry of the problem with respect to the midplane, $x=0$, prescribes symmetry of the base flow velocity that has only axial non-zero component. The base flow velocity at the liquid-liquid interface (horizontal dashed black line in Fig.\ \ref{Fig: Base_flow}a) is a function of the horizontal coordinate $x$ and reaches the maximum at the midplane, $x=0$, from which it decays to zero at the triple points (Fig. 3b). The overall maximum of the base flow velocity is located at the midplane below the interface in the bulk of water phase (Fig.\ \ref{Fig: Base_flow}a,c). These graphs also show that for the chosen grids there is no visible difference between the base flow velocity profiles for the two numerical methods. 

{\renewcommand{\arraystretch}{1.2}
	\begin{table}[h!] 
		\caption{\label{Tab: L-L_BF_covergence}Base flow convergence. Liquid-liquid ($m=0.103$) flow in a circular pipe of diameter $D=0.02$m.}	
		\centering		
		\begin{tabular}{|c|c|c|c|c|c|} 
			\hline\hline
			& & \multicolumn{4}{c|}{$U_{1S}/U_{2S}$}  
			\\
			\cline{3-6}
			$N_\xi$ & $N_\phi$ & \multicolumn{2}{c|}{$h=0.202$} & 
			\multicolumn{2}{c|}{$h=0.601$}			
			\\ 
			\cline{3-6}
			$(N_x)$ & $(N_y)$ & \multicolumn{1}{c|}{Bipolar FVM} & \multicolumn{1}{c|}{IBM} &  \multicolumn{1}{c|}{Bipolar FVM} & \multicolumn{1}{c|}{IBM} 
			\\
			\hline
			50  &  50 & 0.3813 & 0.3828
			& 6.2936 & 6.3054 \\
			100 & 100 & 0.3801 & 0.3808
			& 6.2821 & 6.2815 \\
			200 & 200 & 0.3797 & 0.3800
			& 6.2788 & 6.2787 \\
			300 & 300 & 0.3796 & 0.3798
			& 6.2782 & 6.2782 \\
			400 & 400 & 0.3796 & 0.3797
			& 6.2780 & 6.2780 \\
			\hline
			\multicolumn{2}{|c|}{Analytical} & \multicolumn{2}{c|}{0.3801}
			& \multicolumn{2}{c|}{6.2864} \\
			\hline\hline
		\end{tabular}
	\end{table}
}

{\renewcommand{\arraystretch}{1.2}
	\begin{table}[h!]
		\caption{\label{Tab: Convergence_L-L}Critical flow rates and the critical eigenvalue ($\alpha_{cr}=5.6$) for $h=0.202$ for liquid-liquid ($m=0.103$, $r=1.198$) flow in a circular pipe of diameter $D=0.02$m.}
		\centering
		\begin{tabular}{|c|c|c|c|c|c|c|c|}
			\hline\hline
			$N_\xi$ & $N_\phi$  & \multicolumn{3}{c|}{Bipolar FVM} & \multicolumn{3}{c|}{IBM} 
			\\ 
			\cline{3-8}
			$(N_x)$ & $(N_y)$ & $U_{1S}$ & $U_{2S}$ & $\lambda_{I}$ & $U_{1S}$ & $U_{2S}$ & $\lambda_{I}$
			\\
			\hline
			50  &  50  & 0.0653 & 0.1720 & -10.9131 
			& 0.0759 & 0.2000 & -10.5352  \\
			100 & 100  & 0.0664 & 0.1751 & -10.8338
			& 0.0699 & 0.1841 & -10.7157  \\
			200 & 200  & 0.0666 & 0.1755 & -10.8297
			& 0.0678 & 0.1786 & -10.7841  \\
			300 & 300  & 0.0667 & 0.1756 & -10.8293
			& 0.0673 & 0.1773 & -10.8021  \\
			400 & 400  & 0.0667 & 0.1756 & -10.8293
			& 0.0672 & 0.1769 & -10.8085  \\
			\hline\hline
		\end{tabular}
	\end{table}
}

Once the base flow solution is obtained, the critical operational conditions are found by changing the superficial velocities of both phases, while keeping their ratio constant, so that the holdup remains constant. The critical superficial velocities are defined as those for which the perturbations of all wavenumbers are damped ($\displaystyle\lambda_{R} < 0$), except for a critical one $\displaystyle \alpha_{cr}$, for which the perturbation is neutrally stable, i.e., $\displaystyle Re[\lambda(\alpha_{cr})]=0$. For a holdup $h=0.202$, the critical wavenumber is found to be $\alpha_{cr} = 5.6$. In table\ \ref{Tab: Convergence_L-L}, we report the critical superficial velocities and their grid convergence. The critical conditions are also characterized by the imaginary part of the critical eigenvalue (see its convergence in table\ \ref{Tab: Convergence_L-L}), which provides the oscillation frequency, $\displaystyle|\lambda_{I}|$, and the propagation speed of the critical perturbation, $c_{cr} = -\lambda_{I}/\alpha_{cr}$.   

\begin{figure}[h!]
	\centering
	\subfloat[]{\includegraphics[width=0.48\textwidth,clip]{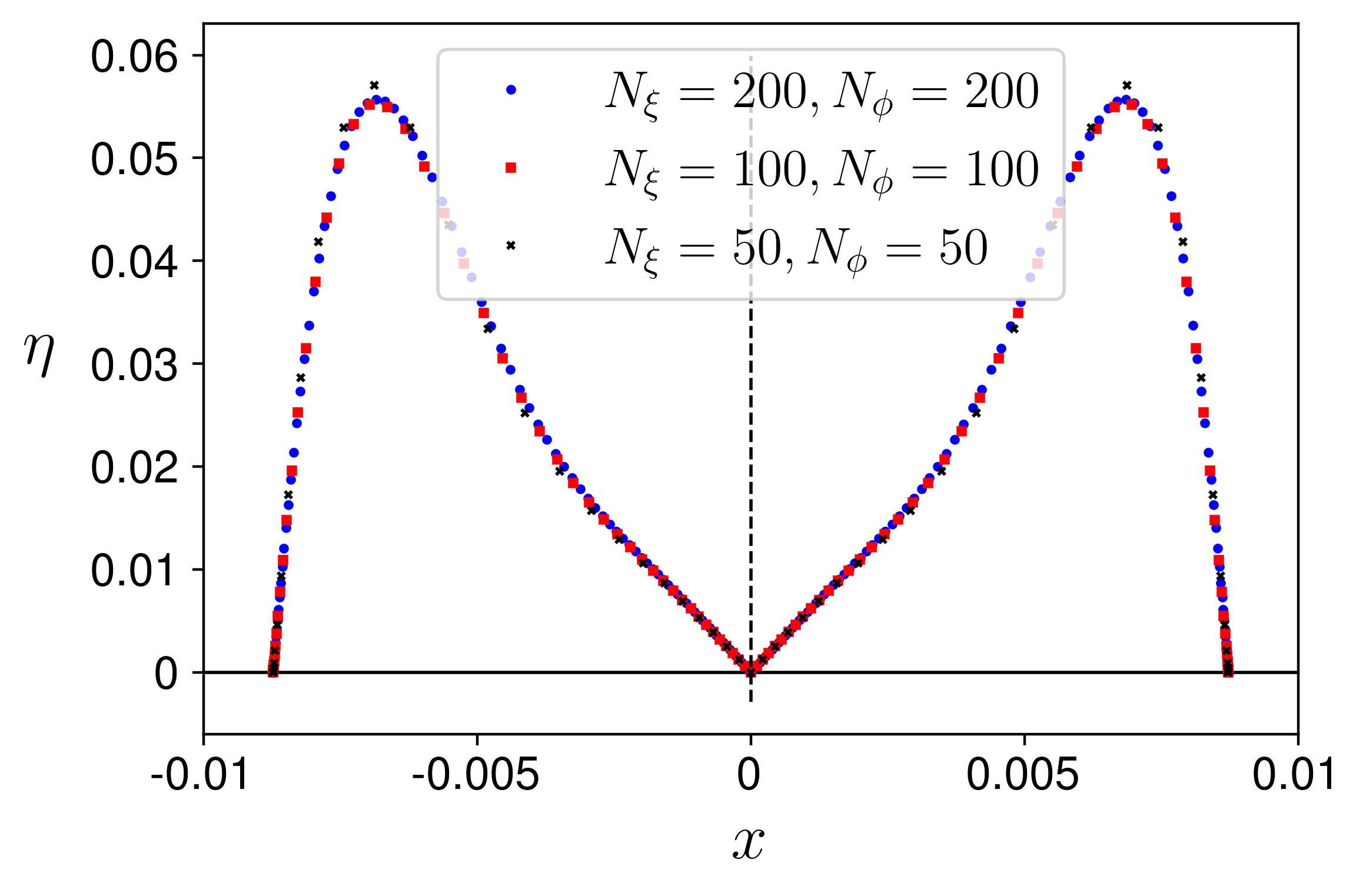}}
	\subfloat[]{\includegraphics[width=0.48\textwidth,clip]{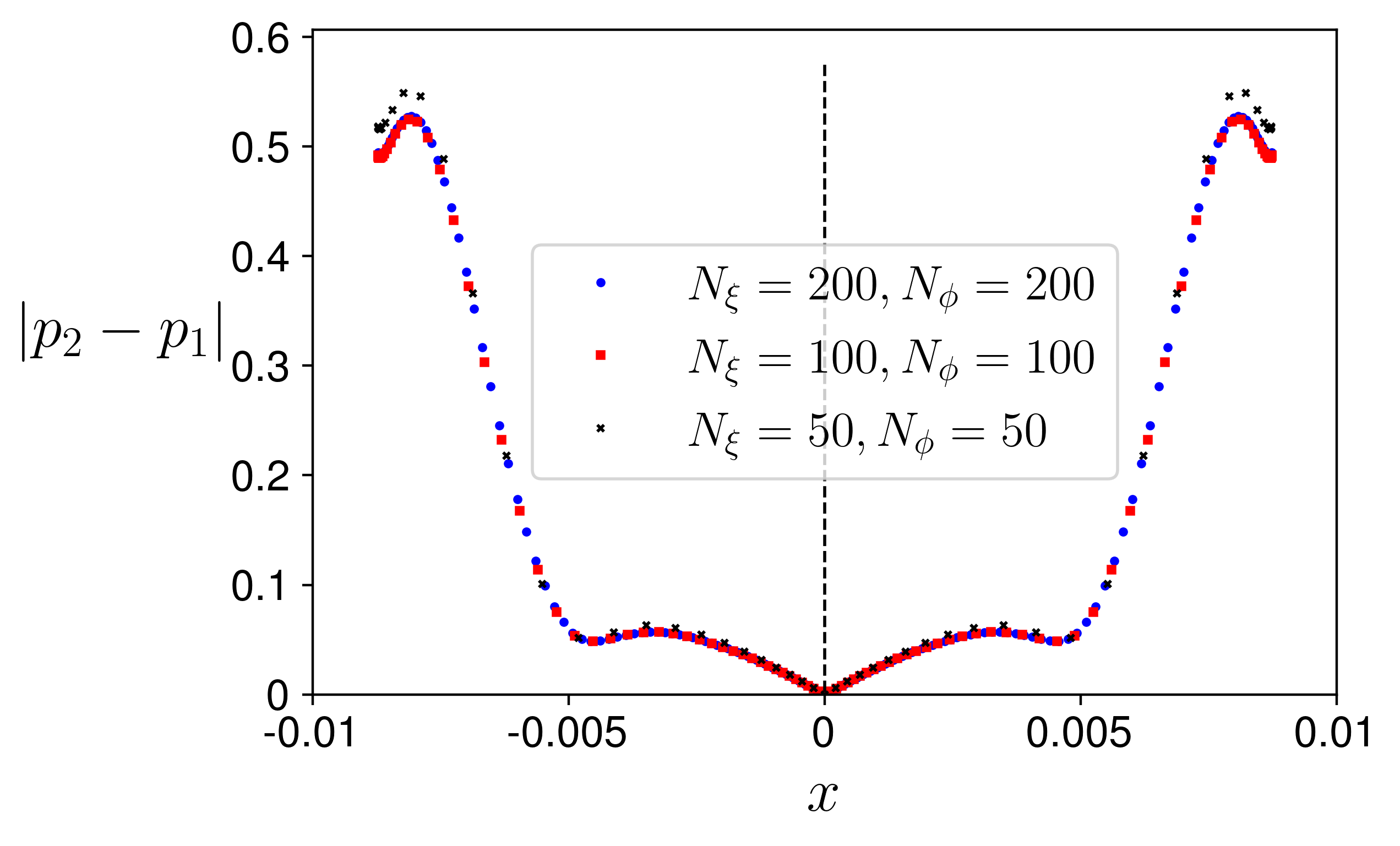}}
	\caption{\label{Fig: crit_perturbation_convergence}Grid convergence of the critical perturbation amplitude. Liquid-liquid ($m=0.103$, $r=1.198$) flow in a circular pipe of diameter $D=0.02$m, $h=0.202$ ($U_{1S} = 0.067$, $U_{2S} = 0.176$), $\alpha = 5.6$. (a) Interface displacement; (b) pressure jump across the interface.}	
\end{figure}

\begin{figure}[h!]
	\centering
	\subfloat[]{\includegraphics[width=0.48\textwidth,clip]{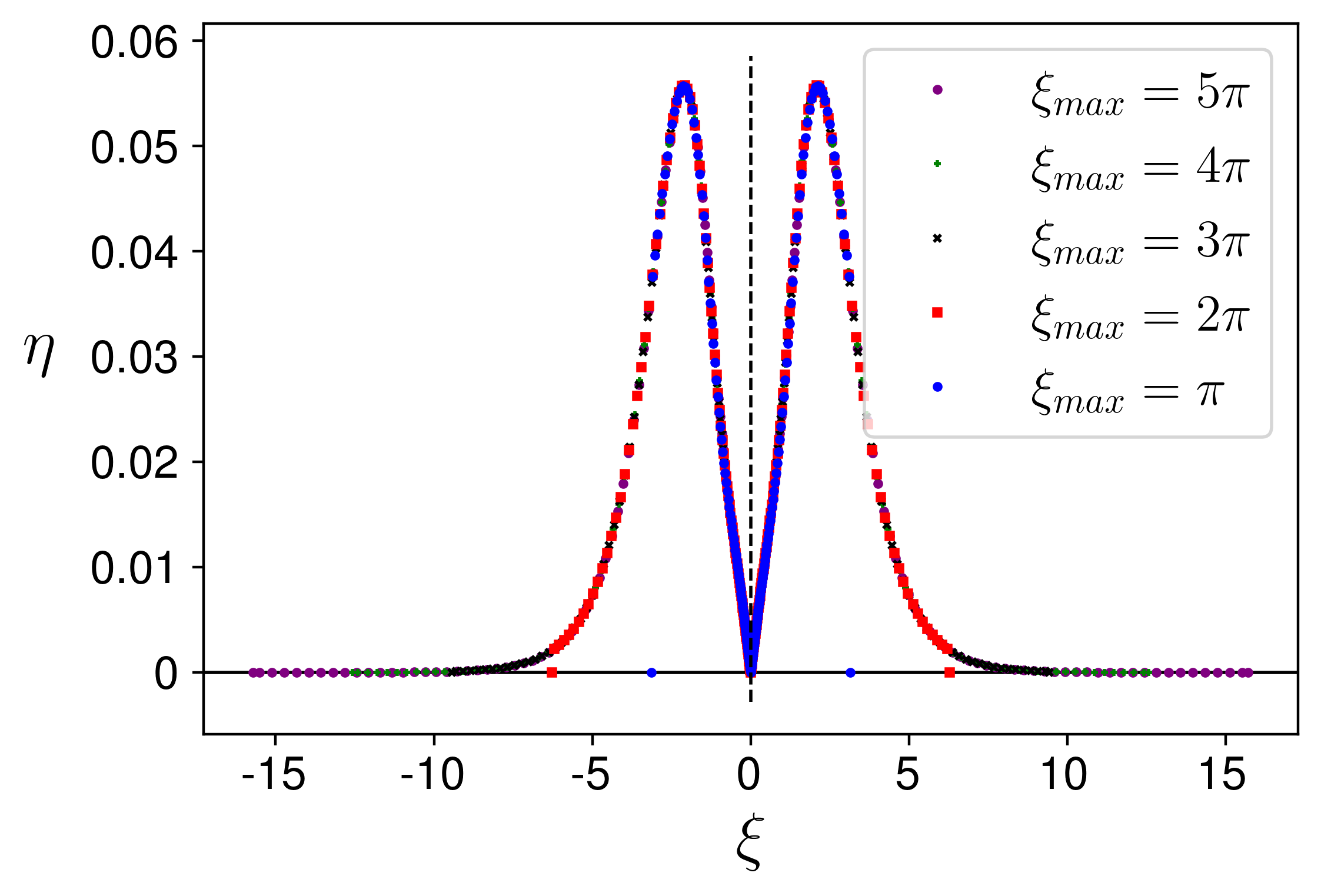}}
	\subfloat[]{\includegraphics[width=0.5\textwidth,clip]{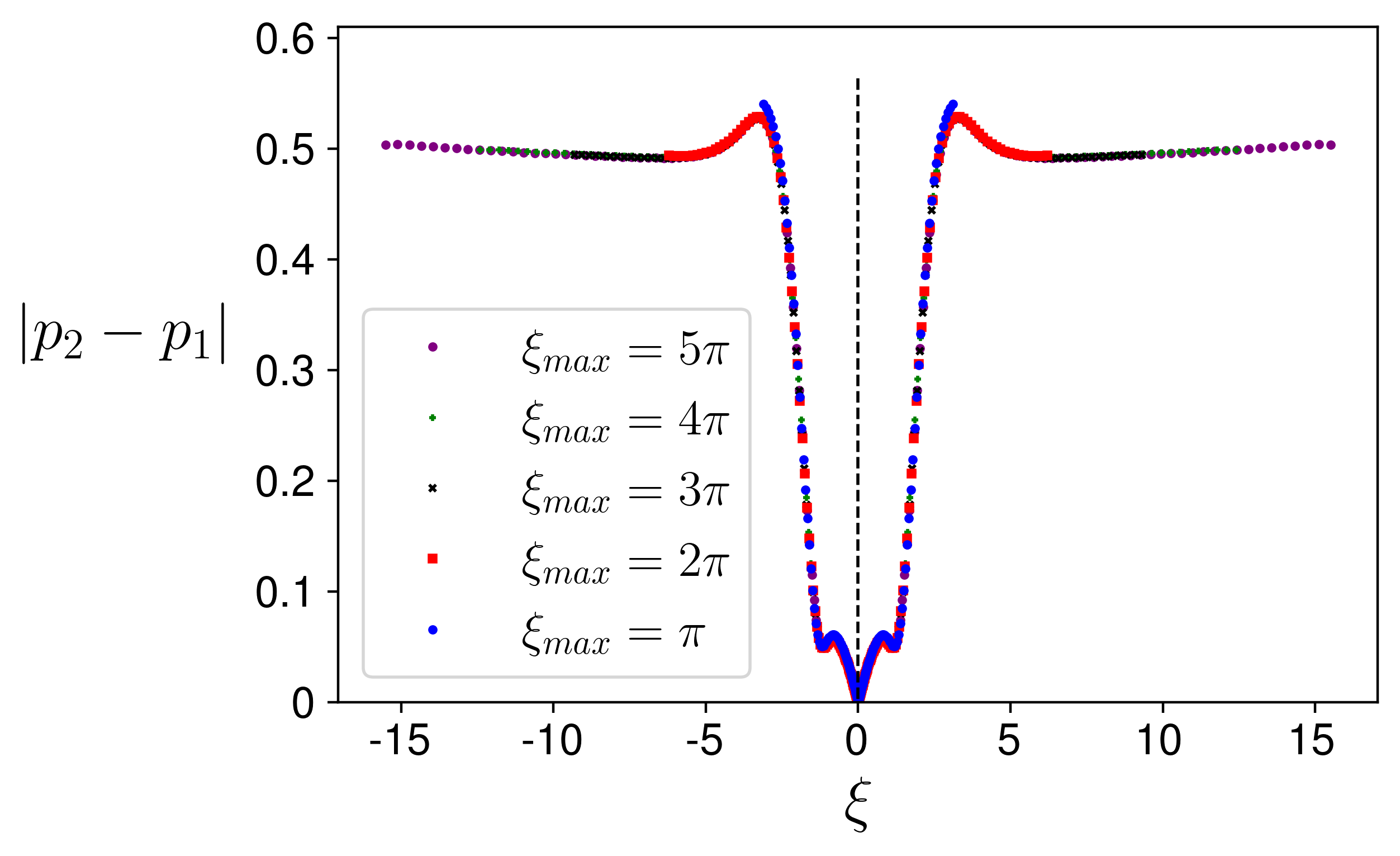}}
	\caption{\label{Fig: crit_pert_bipolar_convergence}Amplitude of the critical perturbation for different $\xi_\text{max}$. Liquid-liquid ($m=0.103$, $r=1.198$) flow in a circular pipe of diameter $D=0.02$m, $h=0.202$ ($U_{1S} = 0.067$, $U_{2S} = 0.176$), $\alpha = 5.6$. (a) Interface displacement; (c) pressure jump across the interface.}	
\end{figure}

The results of the stability analysis are obtained on the computational grid with 200 cells in each direction, i.e., $\displaystyle N_\xi=N_\phi=200$, using the stretching mentioned in sec.\ \ref{Sec: Grid}. Such grid ensures convergence within three decimal places (table \ \ref{Tab: Convergence_L-L}). Additionally, Fig.\ \ref{Fig: crit_perturbation_convergence} illustrates convergence of the most unstable perturbation of the interface displacement (Fig. \ref{Fig: crit_perturbation_convergence}a) and of a jump in the pressure field across the interface due to the surface tension (Fig. \ref{Fig: crit_perturbation_convergence}b). An additional advantage of the bipolar coordinates compared to the Cartesian ones is a possibility to calculate flow fields very close to the triple points and to prescribe exactly the boundary conditions there, i.e., at $\xi\to\pm\infty$. The cells adjacent to the triple points have their center at $\pm\xi_\text{max}$. The base flow and critical eigenvalues become independent of $\xi_\text{max}$, when it is greater than $2\pi$ as demonstrated in table\ \ref{Tab: Aspect_ratio_l-l}. Same convergence behavior in bipolar coordinates is found for the critical perturbation profiles, as shown for the interface displacement (a) and for the pressure jump at the interface (b) in Fig.\ \ref{Fig: crit_pert_bipolar_convergence}. Note that the black dots ($\xi_\text{max}=3\pi$) in the bipolar coordinates in Fig.\ \ref{Fig: crit_pert_bipolar_convergence}  correspond to the blue dots in Cartesian coordinates in Fig.\ \ref{Fig: crit_perturbation_convergence}. Contrarily to the base flow velocity reaching the maximum at the midplane, $x=0$,  the interface displacement reaches its maximal value at $\xi \approx \pm 2$ that corresponds to $x \approx \pm 0.7$, while the maximum jump in pressure is attained closer to the triple point, at  $\xi \pm 3.2$ ($x \approx \pm 0.8$). In this work, we use $\xi_\text{max}=3\pi$ (distance to the triple point $e^{-3\pi}\approx8\times10^{-5}$), so that the smallest grid size is $\approx2\times10^{-5}$. For comparison, the smallest grid size near the wall is approximately $0.005$ in the Cartesian grid (IBM) with $N_x=N_y=200$ and $sin$-stretching next to the walls (similar to Eq.\ \ref{Eq: sin_stretching}). The finite distance from the triple point ($\approx e^{-\xi_\text{max}}$) necessarily introduces an error that is reflected in the values of $\lambda_{R}$ (table\ \ref{Tab: Aspect_ratio_l-l}). At the instability threshold, these values are expected to be numerical zeroes (as for $\xi_\text{max}=3\pi$), so that the other values imply numerical errors. Far from the instability threshold, $\lambda_{R}$ converges similarly to the holdup and $\lambda_{I}$. 

{\renewcommand{\arraystretch}{1.2}
	\begin{table}[h!]
		\caption{\label{Tab: Aspect_ratio_l-l}Holdup and critical eigenvalue for different $\xi_\text{max}$. Liquid-liquid ($m=0.103$, $r=1.198$) flow in a circular pipe of diameter $D=0.02$m, $U_{1S} = 0.0666$, $U_{2S} = 0.1755$, $\alpha=5.6$, $N_\xi=200$, $N_\phi=200$.}				
		\centering		
		\begin{tabular}{|c|c|c|c|}
			\hline\hline
			$\xi_\text{max}$ &  $h$ & $\lambda_{R}$ & $\lambda_{I}$ 
			\\
			\hline
			$\pi$   & 0.2017 & 0.0201 				     & -10.7683 \\
			$2\pi$  & 0.2021 & -0.7070 $\times 10^{-4}$  & -10.8294 \\
			$3\pi$  & 0.2021 & 0.5632 $\times 10^{-10}$  & -10.8297 \\
			$4\pi$  & 0.2021 & 0.8255  $\times 10^{-4}$  & -10.8295 \\
			$5\pi$  & 0.2021 & 0.000144					 & -10.8293 \\
			\hline\hline
		\end{tabular}
	\end{table}
}

\begin{figure}[h]
	\centering
	\subfloat[$|u_x|/\max(|u_z|)$]{\includegraphics[width=0.34\textwidth,clip]{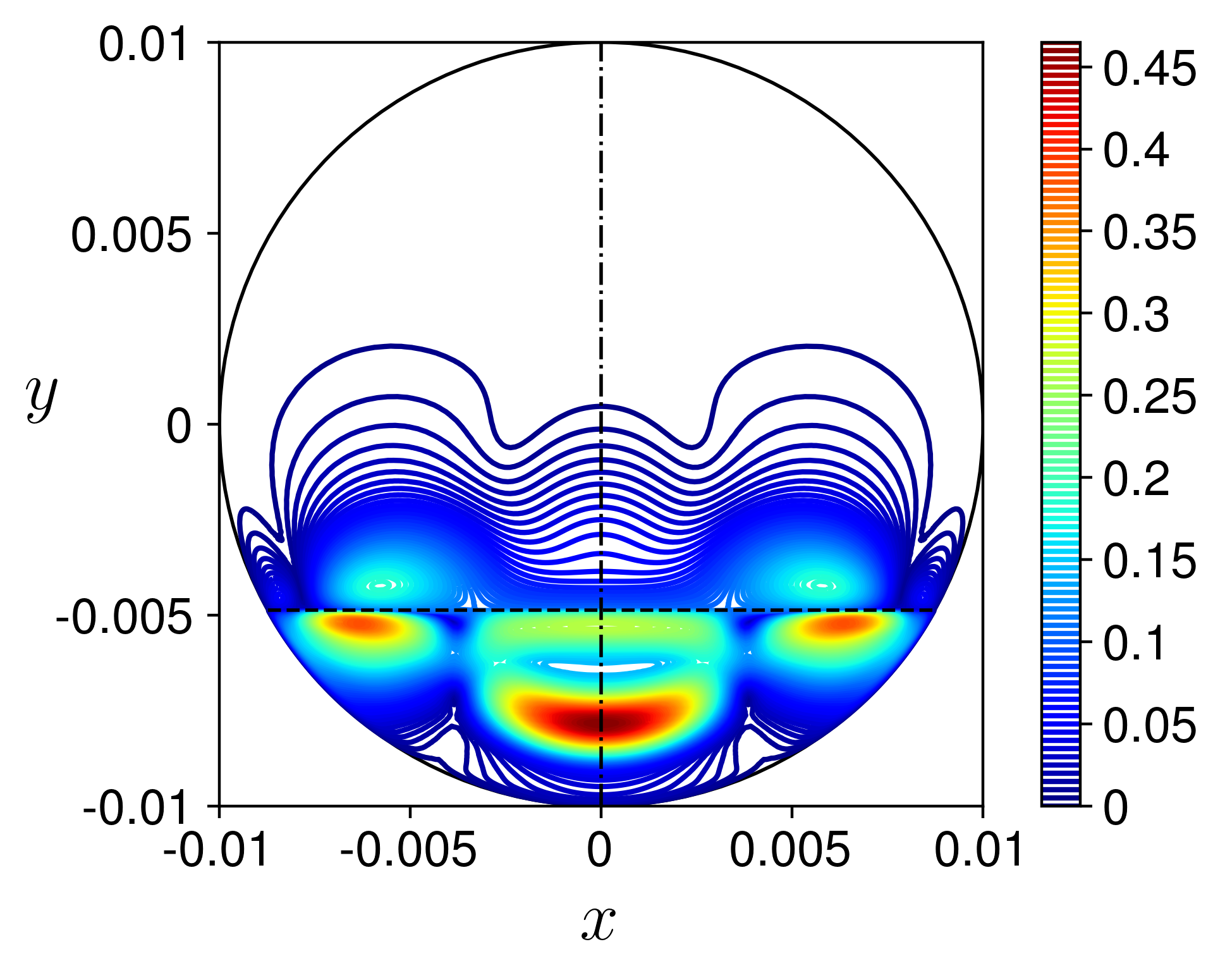}}
	\subfloat[$|u_y|/\max(|u_z|)$]{\includegraphics[width=0.34\textwidth,clip]{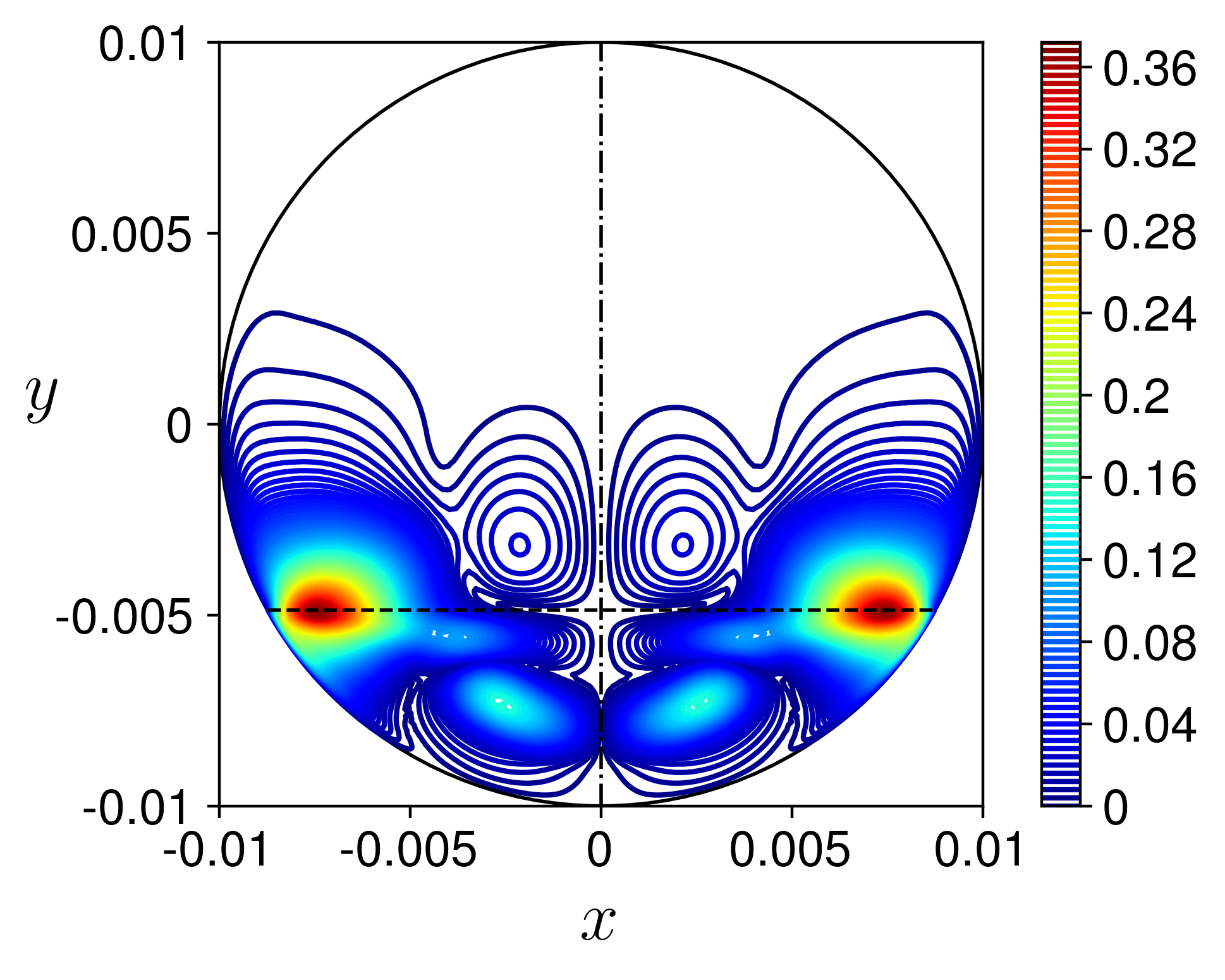}}
	\subfloat[$|u_z|/\max(|u_z|)$]{\includegraphics[width=0.34\textwidth,clip]{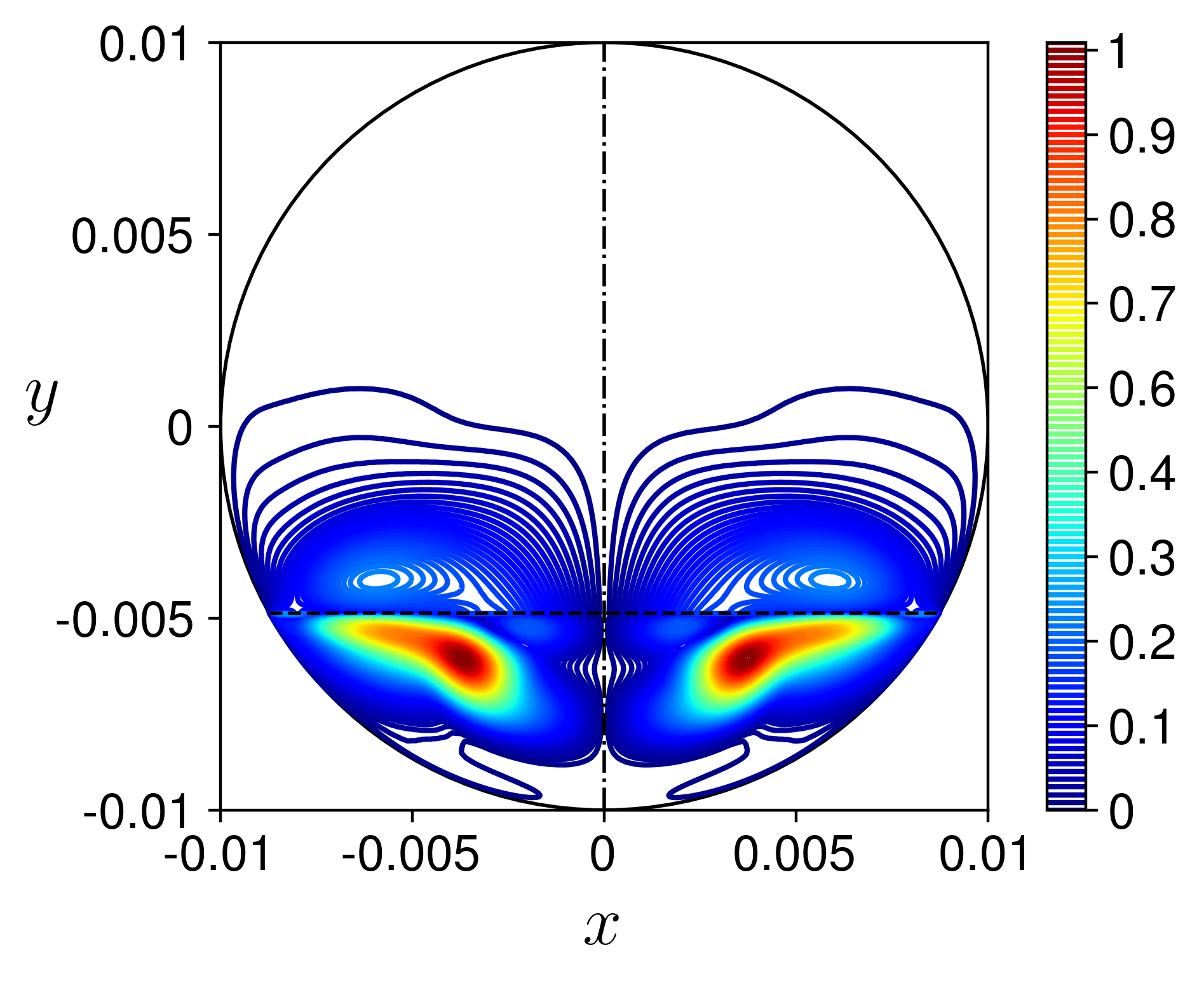}}
	\caption{\label{Fig: Critical_perturbation}Amplitude contours of the critical perturbation. Liquid-liquid ($m=0.103$, $r=1.198$) flow in a circular pipe of diameter $D=0.02$m, $h=0.202$ ($U_{1S} = 0.067$, $U_{2S} = 0.176$), $\alpha = 5.6$. (a) horizontal component of velocity; (b) vertical component of velocity; (c) axial velocity. The unperturbed interface is denoted by a horizontal dashed black line, while the cross-section centerline - by a vertical dash-dot black line.}
\end{figure}

The velocity perturbation is three-dimensional, while its amplitude depends only on the two cross-sectional coordinates. Its amplitude is defined up to multiplication by a complex constant (see Eq.\ \ref{Eq: Perturbation}). Therefore, for the purpose of presentation of the critical perturbation, we scale the amplitudes by the maximal absolute value of the axial perturbation velocity $max\big(|u_z|\big)$, as is shown in Fig.\ \ref{Fig: Critical_perturbation}. The amplitude of the axial velocity component is the largest (note that the limits of color bars, $\approx0.45$ and $\approx0.36$ in Fig.\ \ref{Fig: Critical_perturbation}a,b, respectively, are less than one). Its two maxima are in the heavy phase between the midplane (vertical dash-dot black line) and the pipe wall Fig.\ \ref{Fig: Critical_perturbation}a. The horizontal component, $u_x$, reaches the maximum at the midplane, while two secondary maxima (orange contours) develop slightly below the interface at $x \approx \pm 0.007$. The maxima of the vertical component of velocity, $u_y$, which has the smallest amplitude among the velocity components, is located at the interface at $x \approx \pm 0.008$. 

\begin{figure}[h]
	\centering
	\subfloat[]{\includegraphics[width=0.32\textwidth,clip]{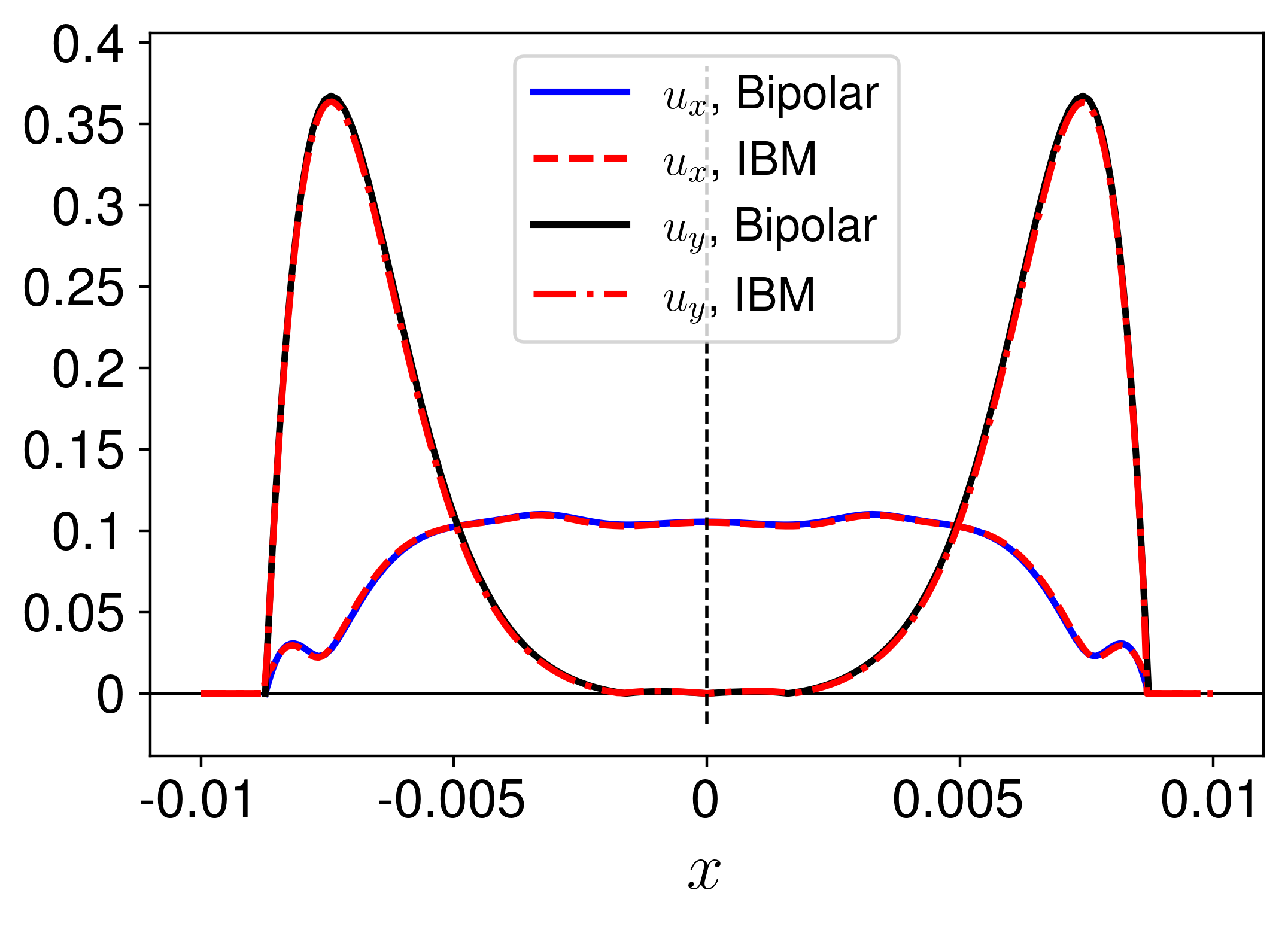}}
	\subfloat[]{\includegraphics[width=0.32\textwidth,clip]{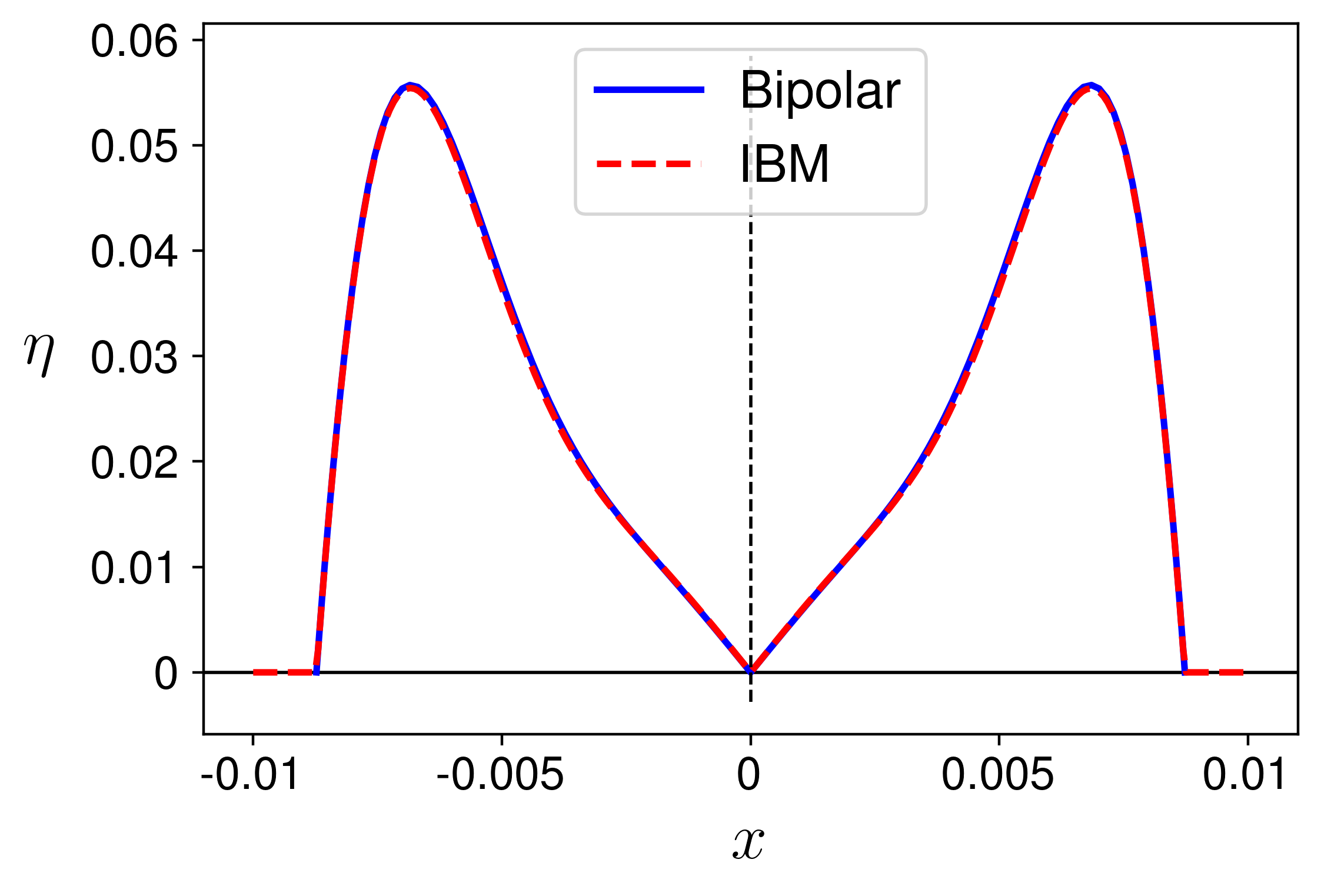}}
	\subfloat[]{\includegraphics[width=0.35\textwidth,clip]{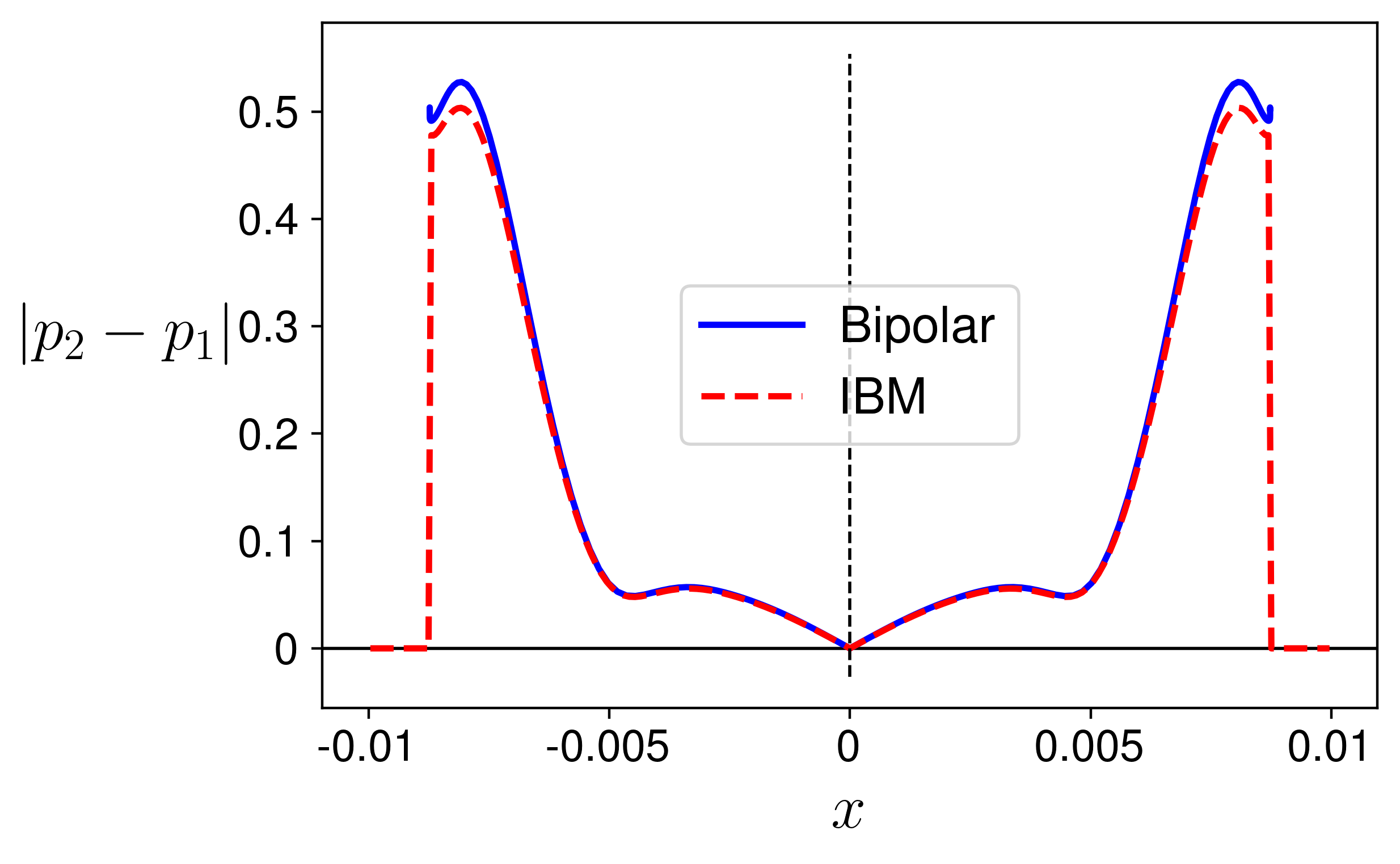}}
	\caption{\label{Fig: crit_perturbation_bipolar_v_IBM}Amplitude of the critical perturbation. Liquid-liquid ($m=0.103$, $r=1.198$) flow in a circular pipe of diameter $D=0.02$m, $h=0.202$ ($U_{1S} = 0.067$, $U_{2S} = 0.176$), $\alpha = 5.6$. (a) interfacial velocity components; (b) interface displacement; (c) pressure jump at the interface.}	
\end{figure}

Isolines of the most unstable perturbation obtained by the IBM with the grid of two hundred cells in each direction are indistinguishable from those shown in Fig.\ \ref{Fig: Critical_perturbation}. The similarity of the perturbation profiles at the interface is shown in Fig.\ \ref{Fig: crit_perturbation_bipolar_v_IBM} for the horizontal and vertical components of velocity (a), the interface displacement (b), and the pressure difference across the interface (c). The discretization of the interface boundary conditions by the two numerical methods is noticeably different, however it does not affect the solutions behavior inside the pipe, except for the pressure jump in the vicinity of the triple points (Fig.\ \ref{Fig: crit_perturbation_bipolar_v_IBM}c).The latter is nevertheless more accurately represented by the solution in the bipolar coordinates. 

%%%%%%%%%%%%%%%%%%%%%%%%%%%%%%%%%%%%%%%%%%%%%%%%%%%%%%%%%%%%%%%%%%
\subsection{Gas--liquid pipe flow}

\begin{figure}[h!]
	\centering
	\subfloat[]{\includegraphics[width=0.34\textwidth,clip]{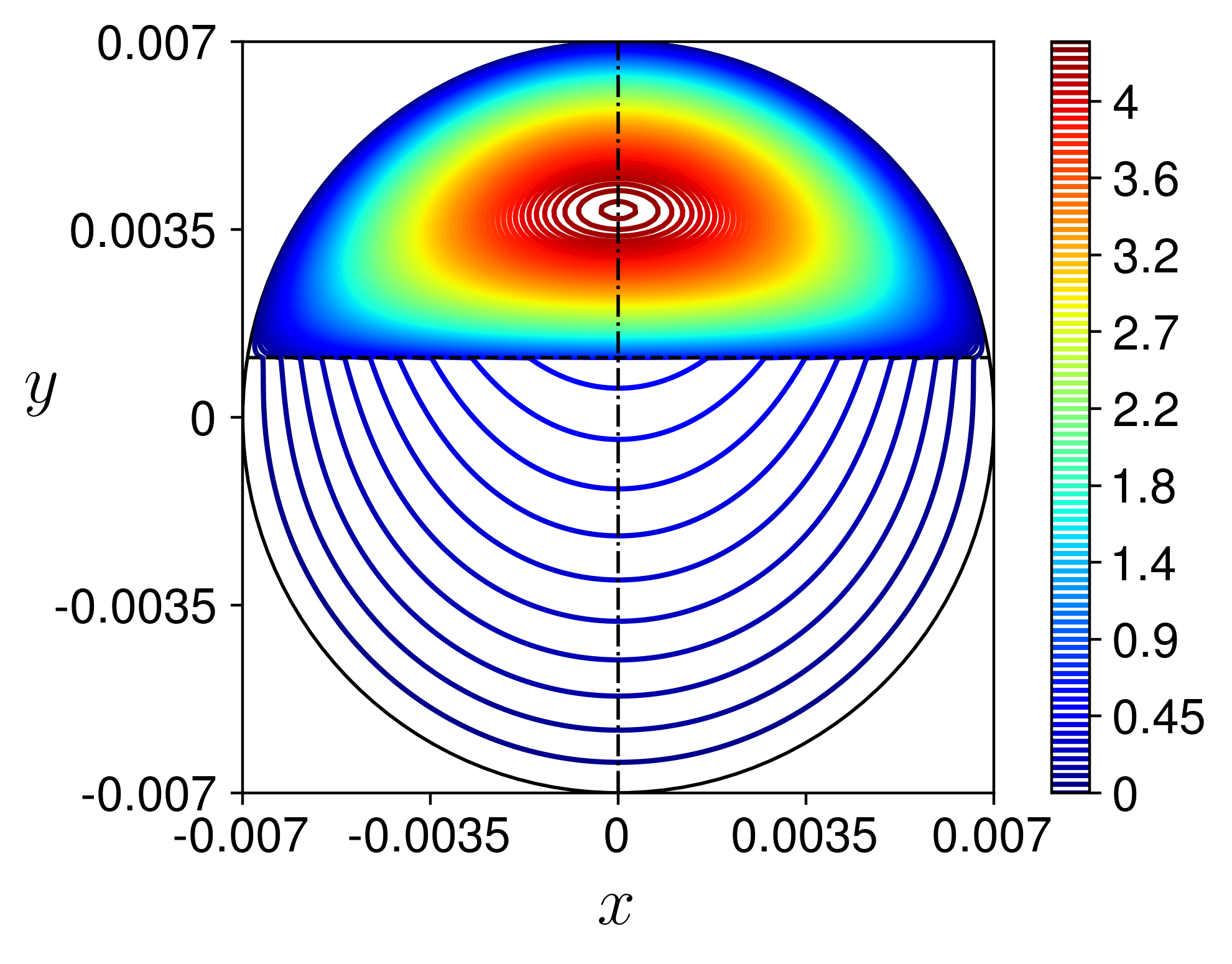}}
	\subfloat[]{\includegraphics[width=0.34\textwidth,clip]{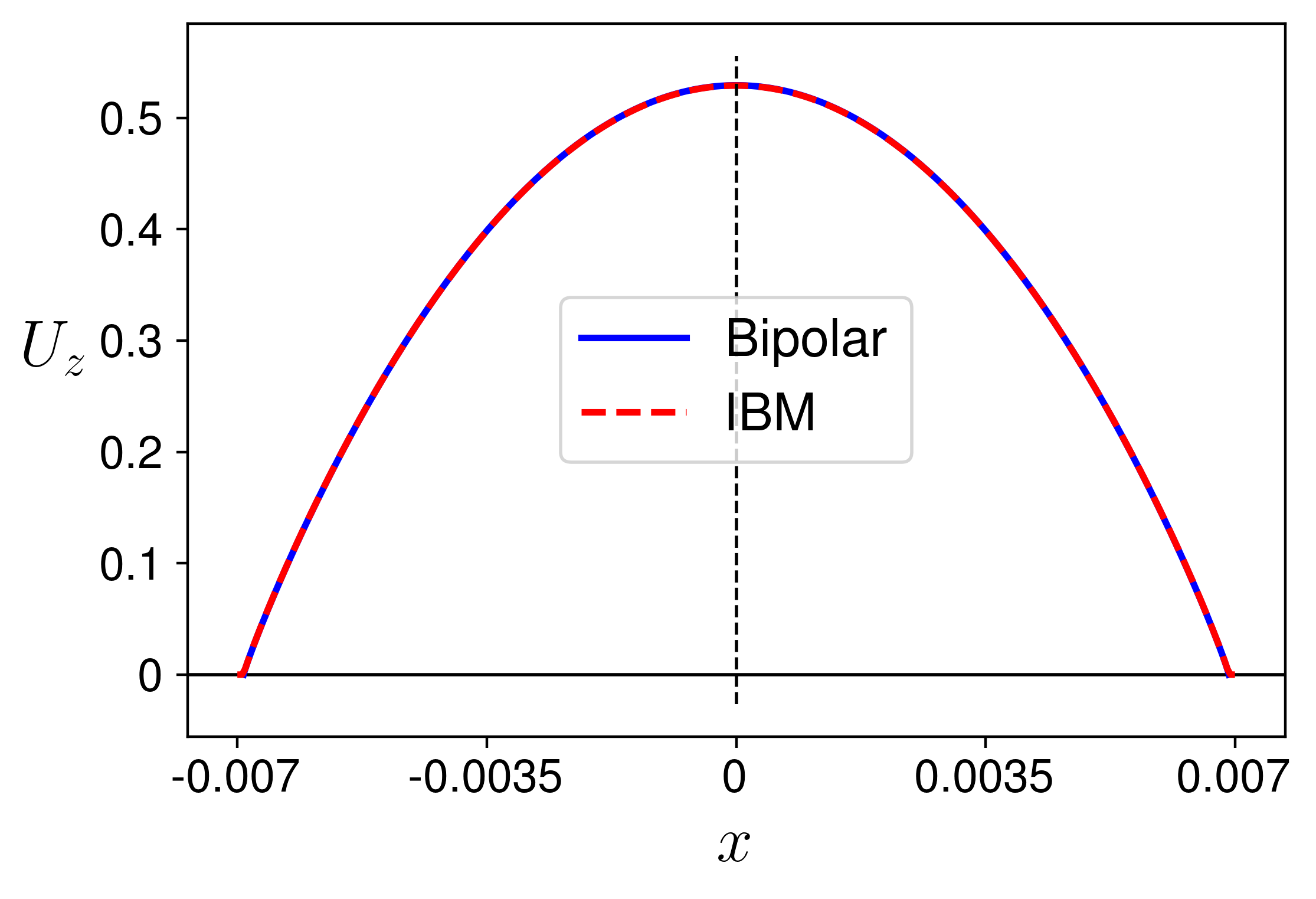}}
	\subfloat[]{\includegraphics[width=0.34\textwidth,clip]{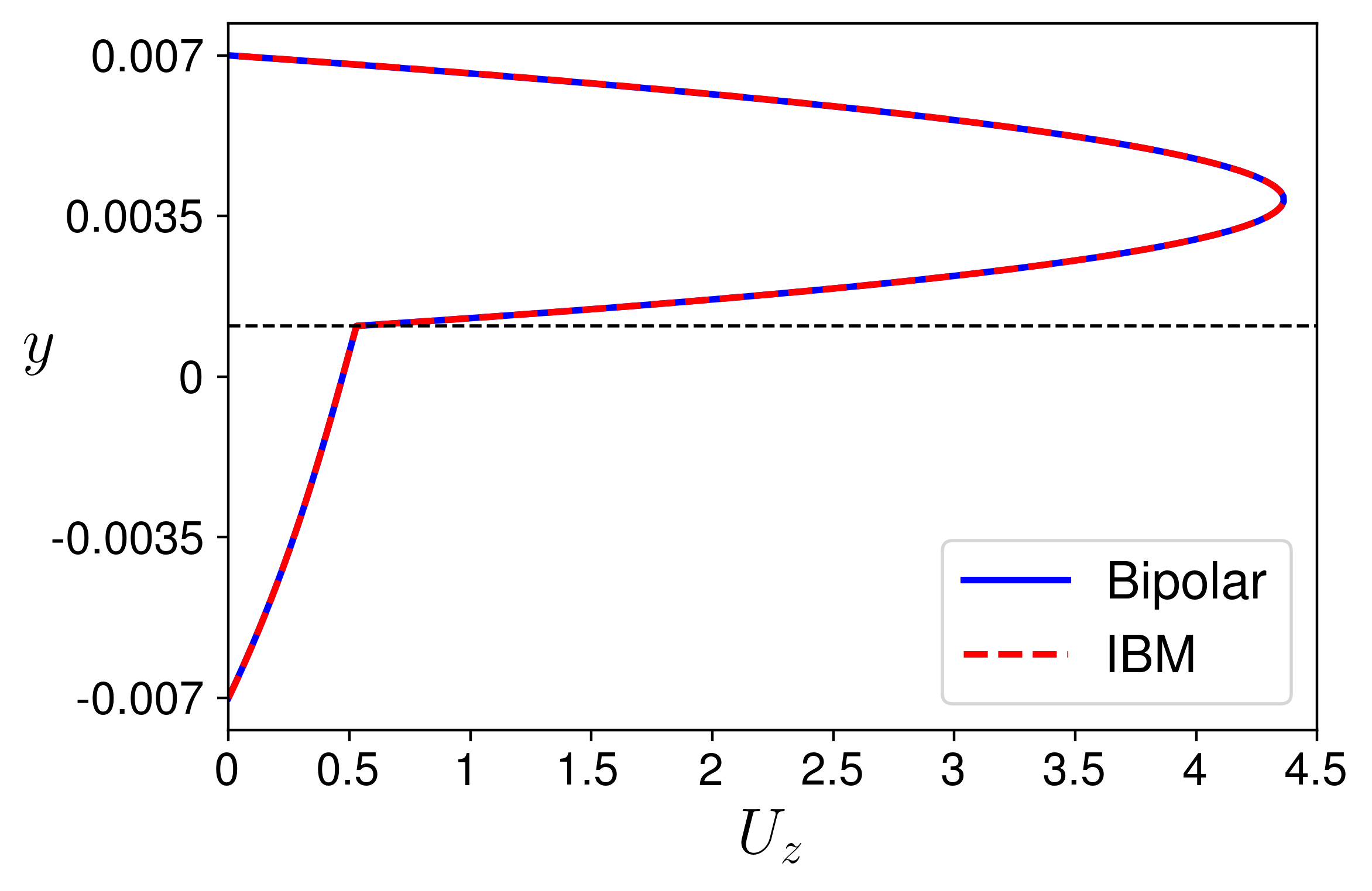}}
	\caption{\label{Fig: Base_flow_a-w}Base flow velocity, $U_z$. Air-water ($m=55$) flow in a circular pipe of diameter $D=0.014$m, $h=0.6$. (a) cross-sectional contours; (b) velocity at the interface; (c) velocity at the midplane $x=0$. The interface is denoted by a horizontal dashed black line. The cross-section centerline is denoted by a vertical dash-dot black line.}
\end{figure}

In this section we consider an air-water flow in a circular pipe of diameter $D=0.014$m (line 2 of table\ \ref{Tab: Two-phase_systems}). The air-water flow is characterized by large density and viscosity ratios, $\rho_{1 2}=1000$ and $\mu_{1 2}=55$, respectively. Compared to the oil-water flow considered above, a much lower water-to-air flow rate ratio is required for the same holdup of the heavy phase. For example, for a holdup $0.6$, the ratio of the superficial velocities of water to air $U_{1S}/U_{2S}=0.15$. The dimensionless base flow velocity in the pipe cross-section is shown in Fig.\ \ref{Fig: Base_flow_a-w}. As before, the velocity at the air-water interface (horizontal dashed black line in Fig.\ \ref{Fig: Base_flow_a-w}a) depends on the horizontal coordinate $x$ and is symmetrical around the midplane ($x=0$, Fig.\ \ref{Fig: Base_flow_a-w}b), where it reaches its maximum in the bulk of the light (air) phase (Fig.\ \ref{Fig: Base_flow_a-w}a,c). Due to the large water-to-air viscosity ratio, the water layer is perceived almost as a solid wall by the upper phase (air), except for the nonzero interfacial velocity of the water. There, the water reaches its maximal velocity, which is much lower than the maximum air velocity (compare $U_z(y=0)\approx0.5$ with $\max(U_z)\approx4.3$). The air velocity profile is similar to that of the single-phase air flowing through the part of the pipe cross section it occupies. In horizontal air-water flows, a similar picture can be observed for the base flow for other water holdups (not shown). 

{\renewcommand{\arraystretch}{1.2}
	\begin{table}[h!]
		\caption{\label{Tab: Long-wave_A-W} Long-wave critical superficial velocities and critical eigenvalue for $h=0.6$. Air-water ($m=55$, $r=1000$) flow in a circular pipe of diameter $D=0.014$m ($N_\xi=N_\phi=200$).}
		\centering
		\begin{tabular}{|c|c|c|c|c|}
			\hline\hline
			$\alpha$ & $U_{1S}$ & $U_{2S}$ & $\lambda_{I}$ & $c$
			\\
			\hline
			0.01    & 0.0488 & 0.3247 & -0.010863 
			& 1.0862 \\
			0.001   & 0.0481 & 0.3197 & -0.001098 
			& 1.0975 \\
			0.0001  & 0.0481 & 0.3197 & -0.000110 
			& 1.0976 \\
			0.00001 & 0.0481 & 0.3197 & -0.1097$\times 10^{-4}$ 
			& 1.0976 \\
			\hline\hline
		\end{tabular}
	\end{table}
}

{\renewcommand{\arraystretch}{1.2}
	\begin{table}[h!]
		\caption{\label{Tab: Convergence_A-W}Critical superficial velocities and the critical eigenvalue ($\alpha_{cr}=0.001$) for $h=0.6$. Air-water ($m=55$, $r=1000$) flow in a circular pipe of diameter $D=0.014$m.}
		\centering
		\begin{tabular}{|c|c|c|c|c|c|c|c|}
			\hline\hline
			$N_\xi$ & $N_\phi$  & \multicolumn{3}{c|}{Bipolar FVM} & \multicolumn{3}{c|}{IBM} 
			\\ 
			\cline{3-8}
			$(N_x)$ & $(N_y)$ & $U_{1S}$ & $U_{2S}$ & $\lambda_{I}$ & $U_{1S}$ & $U_{2S}$ & $\lambda_{I}$
			\\
			\hline
			50  & 50  & 0.0476 & 0.3168 & -0.001085 
			& 0.0504 & 0.3351 & -0.001062  \\
			100 & 100 & 0.0480 & 0.3191 & -0.001098
			& 0.0503 & 0.3348 & -0.001064  \\
			200 & 200 & 0.0481 & 0.3197 & -0.001098
			& 0.0504 & 0.3352 & -0.001063  \\
			300 & 300 & 0.0481 & 0.3198 & -0.001097
			& 0.0504 & 0.3352 & -0.001062  \\
			400 & 400 & 0.0481 & 0.3199 & -0.001098
			& 0.0504 & 0.3353 & -0.001062  \\
			\hline\hline
		\end{tabular}
	\end{table}
}

\begin{figure}[h!]
	\centering
	\subfloat[]{\includegraphics[width=0.33\textwidth,clip]{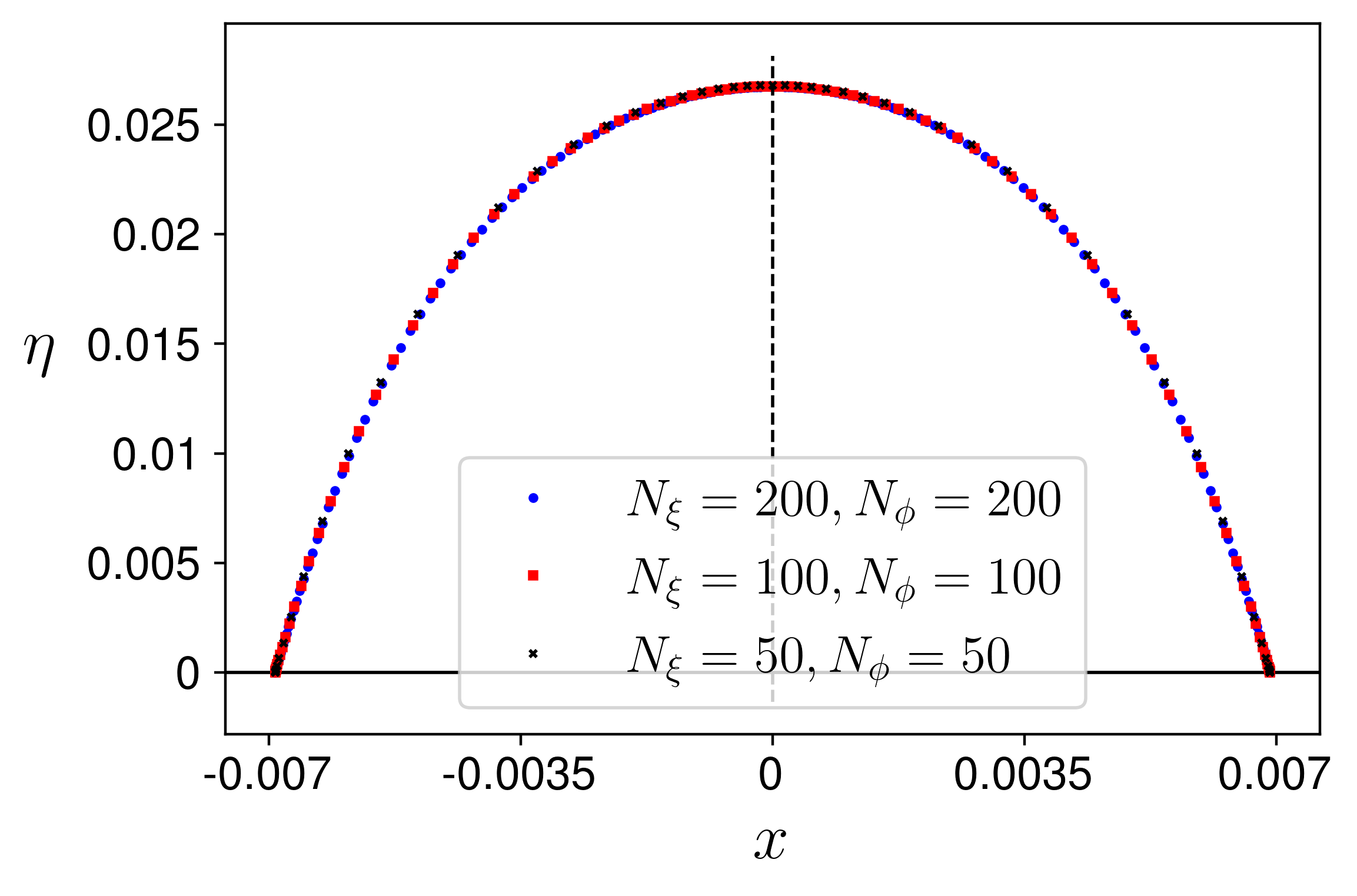}}
	\subfloat[]{\includegraphics[width=0.35\textwidth,clip]{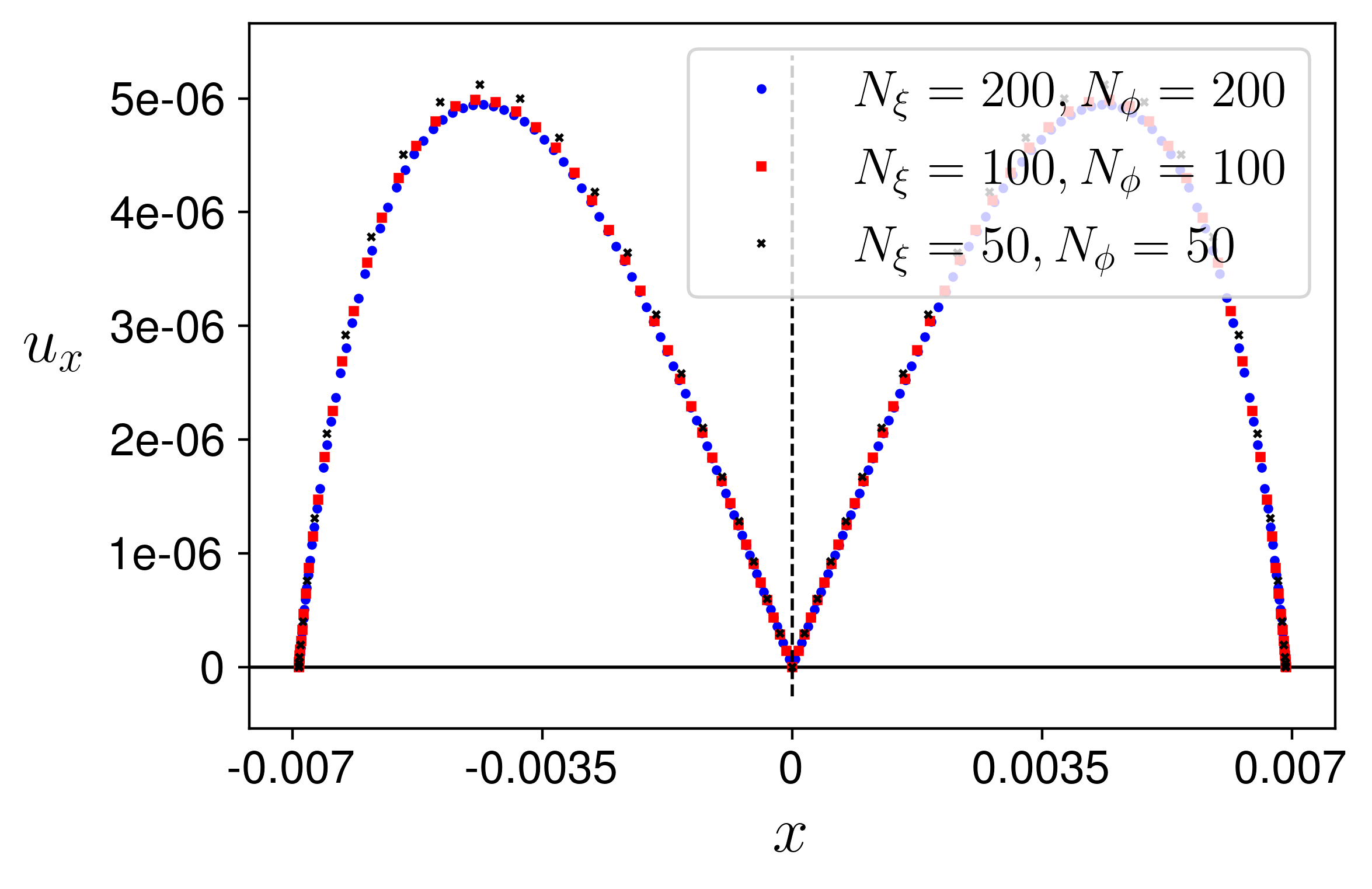}}
	\subfloat[]{\includegraphics[width=0.35\textwidth,clip]{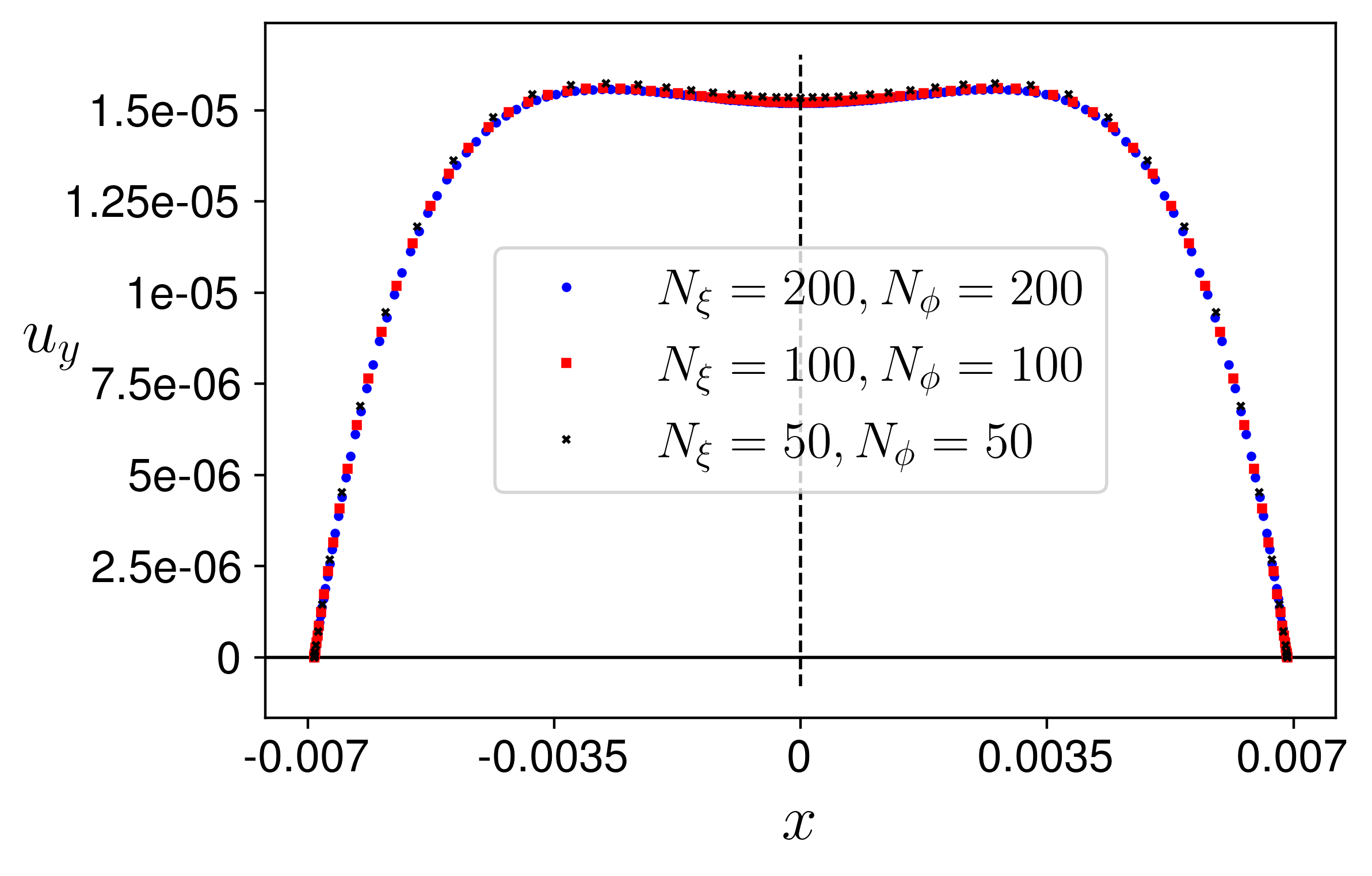}}
	\caption{\label{Fig: crit_perturbation_convergence_a-w}Grid convergence of the critical perturbation amplitude. Air-water ($m=55$, $r=1000$) flow in a circular pipe of diameter $D=0.014$m, $h=0.6$ ($U_{1S} = 0.048$, $U_{2S} = 0.319$), $\alpha=0.001$. (a) interface displacement; (b) horizontal velocity at the interface; (c) vertical velocity at the interface.}	
\end{figure}

While the short-wave perturbation ($\alpha_{cr}=5.6$) was found to be critical for the oil-water flow discussed in the previous section, for the considered air-water pipe flow, the long-wave perturbation ($\alpha_{cr}\to0$) is critical for a holdup of $0.6$. To find small, but nonzero, value of the wavenumber $\alpha$ (i.e., large but finite wavelength), we compute the critical superficial velocities and the propagation speed $c=-Im(\lambda)/\alpha$ of the critical perturbation for several small $\alpha$ as summarized in table\ \ref{Tab: Long-wave_A-W}. In the following, the long-wave instability can be obtained starting from $\alpha = 0.001$, from which the convergence up to the third decimal place for decreasing $\alpha$ is reached in the zero-wavenumber limit. The convergence of results for computational grids with different number of cells for FVM in the bipolar coordinates and with the IBM in the Cartesian coordinates are summarized in table\ \ref{Tab: Convergence_A-W}. The grid convergence of the air-water flow is similar to that of the oil-water flow, and $N_\xi=N_\phi=200$ is sufficient for convergence of the critical superficial velocities and perturbation frequency. The convergence of perturbation profiles is shown in Fig.\ \ref{Fig: crit_perturbation_convergence_a-w} for the interface displacement (a) and the horizontal (b) and vertical (c) velocities at the interface. Even though the components of velocity in the cross section have small absolute values compared to the axial velocity (as discussed below), the grid convergence can be clearly seen in the graphs.

\begin{figure}[h!]
	\centering
	\subfloat[$|u_x|/\max(|u_z|)$]{\includegraphics[width=0.34\textwidth,clip]{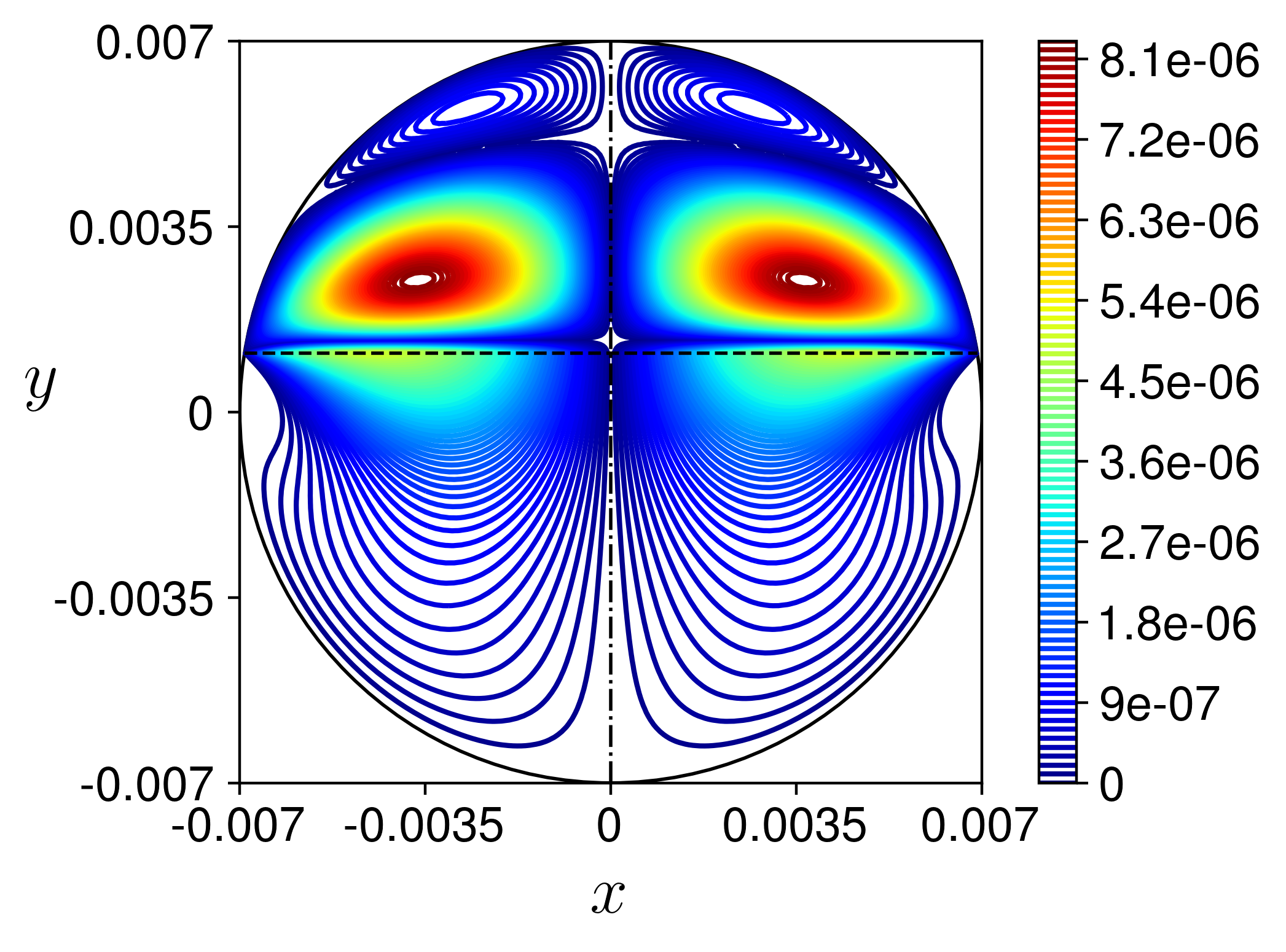}}
	\subfloat[$|u_y|/\max(|u_z|)$]{\includegraphics[width=0.34\textwidth,clip]{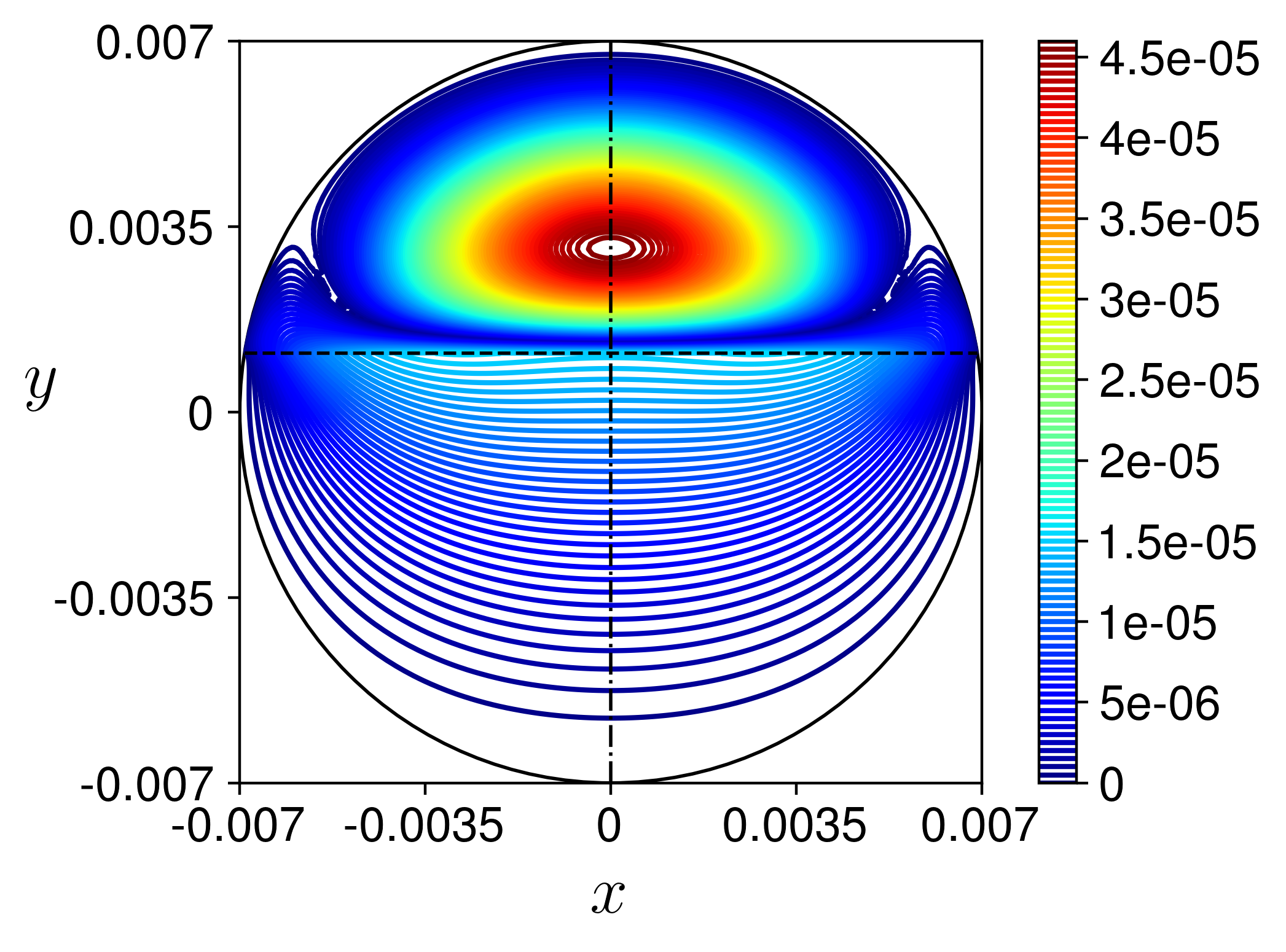}}
	\subfloat[$|u_z|/\max(|u_z|)$]{\includegraphics[width=0.34\textwidth,clip]{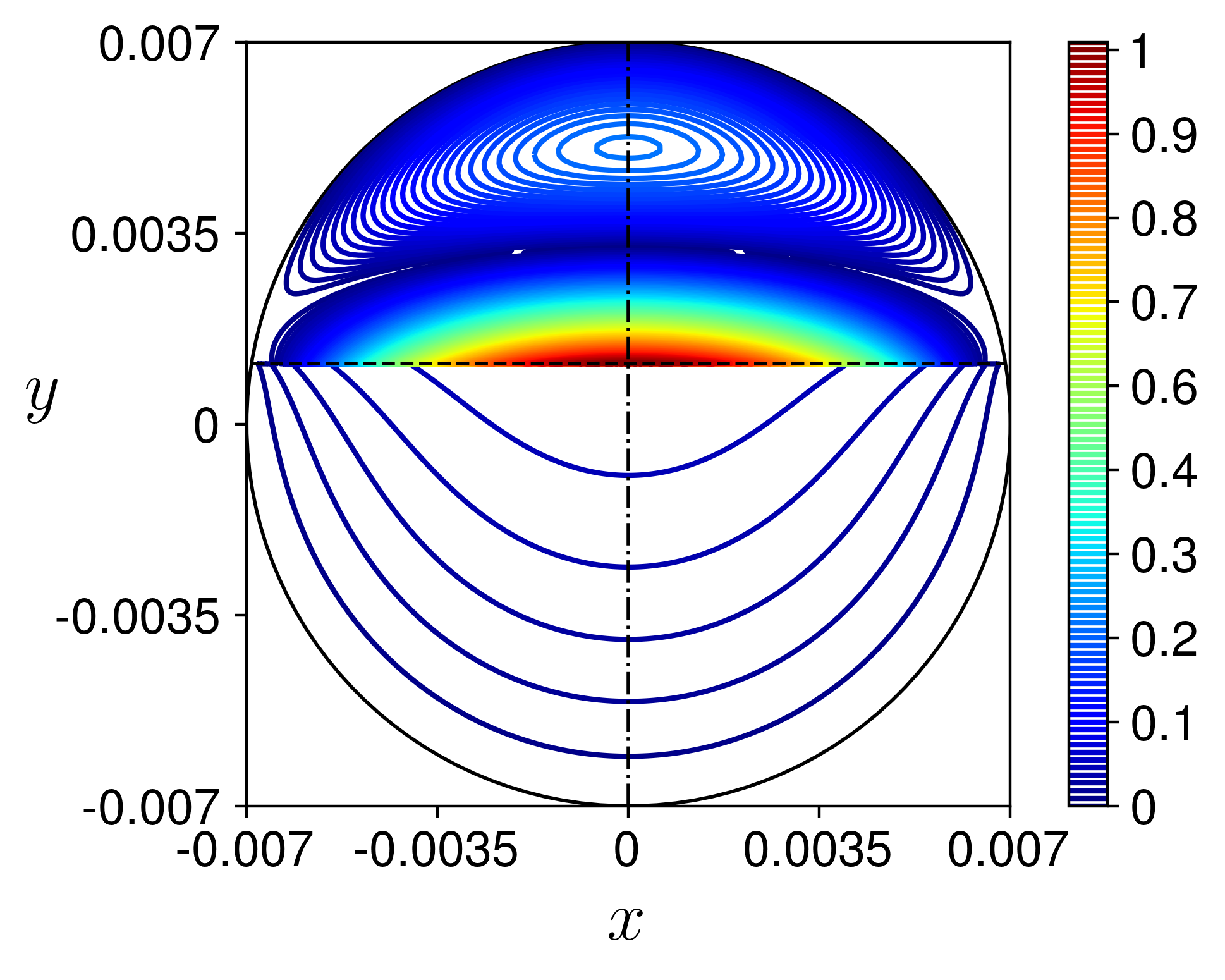}}
	\caption{\label{Fig: Critical_perturbation_a-w}Amplitude contours of the critical perturbation. Air-water ($m=55$, $r=1000$) flow in a circular pipe of diameter $D=0.014$m, $h=0.6$ ($U_{1S} = 0.048$, $U_{2S} = 0.319$), $\alpha = 0.001$. (a) horizontal component of velocity; (b) vertical component of velocity; (c) axial velocity. The unperturbed interface is denoted by a horizontal dashed black line, while the cross-section centerline - by a vertical dash-dot black line.}
\end{figure}
\begin{figure}[h!]
	\centering
	\subfloat[]{\includegraphics[width=0.45\textwidth,clip]{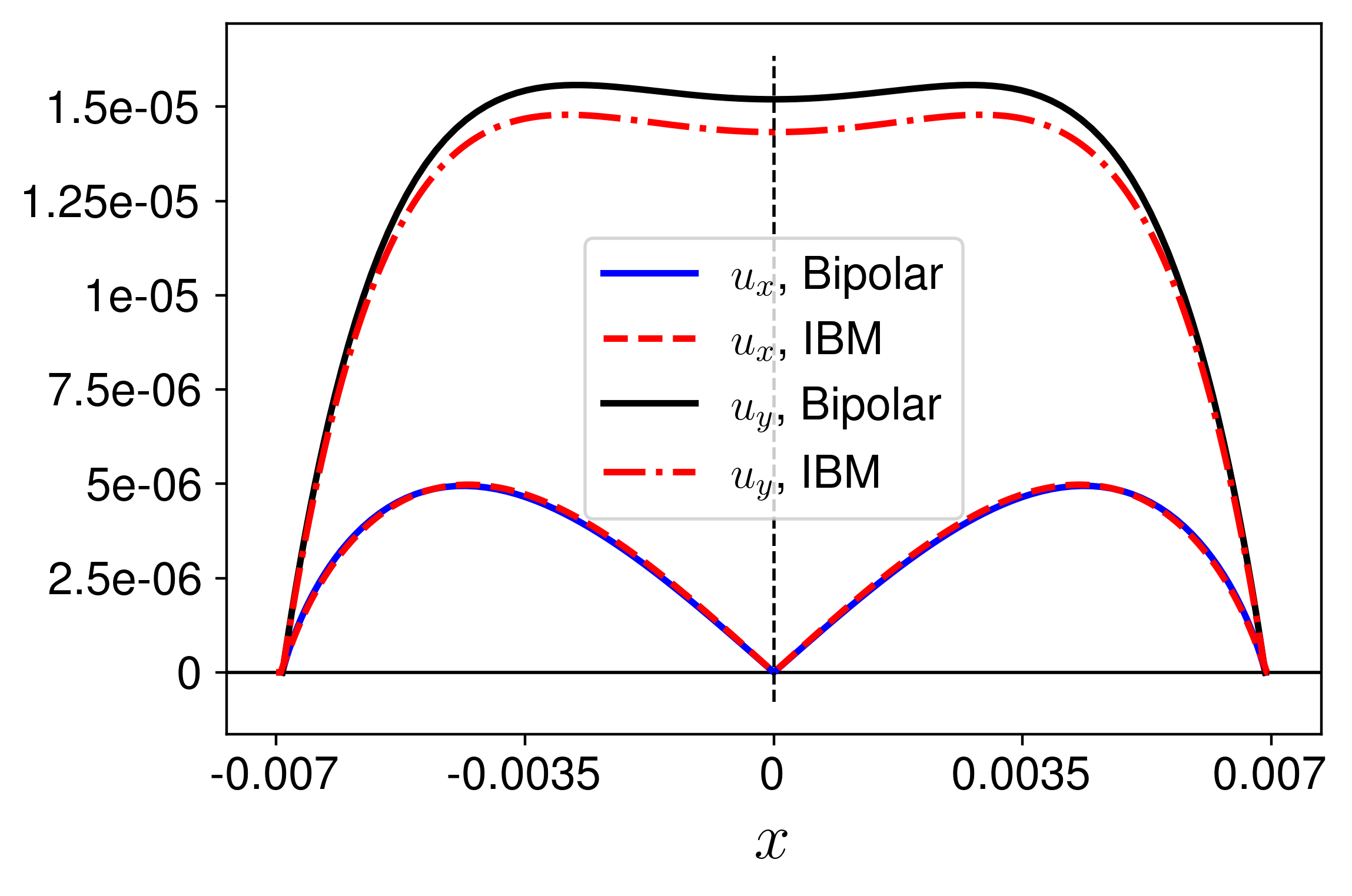}}
	\subfloat[]{\includegraphics[width=0.45\textwidth,clip]{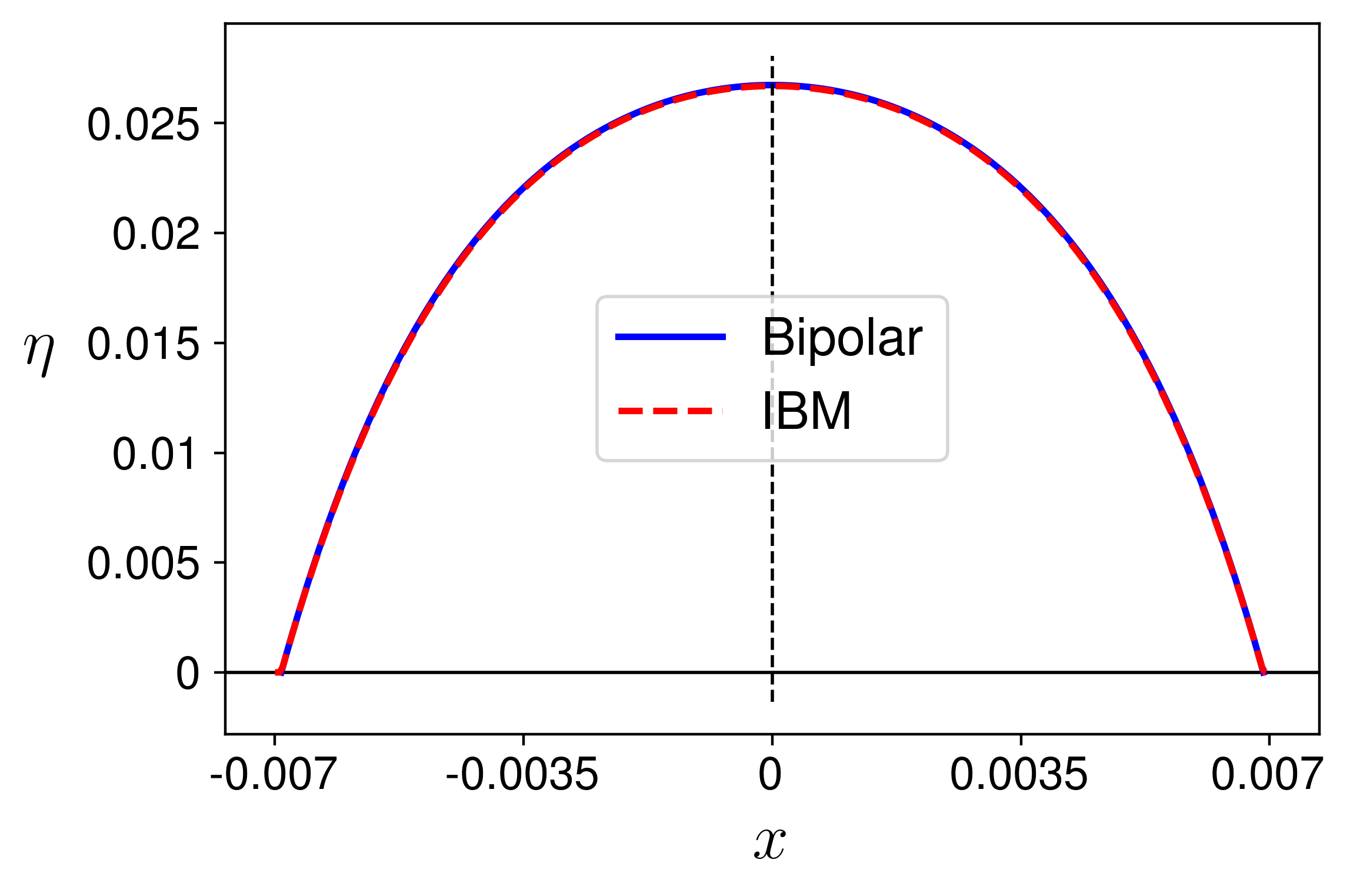}}
	\caption{\label{Fig: crit_perturbation_bipolar_v_IBM_a-w}Amplitude of the critical perturbation. Air-water ($m=55$, $r=1000$) flow in a circular pipe of diameter $D=0.014$m, $h=0.6$ ($U_{1S} = 0.048$, $U_{2S} = 0.319$), $\alpha = 0.001$. (a) interfacial velocity components; (b) interface displacement.}	
\end{figure}

One of the features of long-wave perturbation is that its axial component is dominating and is larger by several orders of magnitude than the two other components (Fig.\ \ref{Fig: Critical_perturbation_a-w}, cf. the color bar limits). Moreover, for the long-wave perturbations, the amplitudes of the horizontal and vertical components scale down with $\alpha$ for long waves, so that $|u_x|/\alpha$ and $|u_y|/\alpha$ are independent on $\alpha\ll1$. The maximum of the axial velocity (red contours) is at the center of the interface, and the flow on the air side is perturbed stronger, as depicted in Fig.\ \ref{Fig: Critical_perturbation_a-w}c. Another local maximum in the air phase is located at the midplane above the base flow maximum (compare Fig.\ \ref{Fig: Critical_perturbation_a-w}c with Fig.\ \ref{Fig: Base_flow_a-w}). Since the maximal value of the velocity field of the critical perturbation is at the interface, one can expect the onset of instability triggered by growth of the interface deformation. This instability can be classified as interfacial instability as opposed to the shear instability that was examined above for the oil-water flow. Further nonlinear analysis is required to find out whether such instability leads to transition to stratified wavy or other flow pattern. More details on the perturbation flow field at the interface can be seen in Fig.\ \ref{Fig: crit_perturbation_bipolar_v_IBM_a-w}a (horizontal and vertical velocities), while the interface displacement is shown in Fig.\ \ref{Fig: crit_perturbation_bipolar_v_IBM_a-w}b. These graphs show that the flow fields obtained by the two numerical methods are found to be similar.
%\clearpage

%%%%%%%%%%%%%%%%%%%%%%%%%%%%%%%%%%%%%%%%%%%%%%%%%%%%%%%%%%%%%%%%%%
\section{Conclusions}

Linear stability analysis of two-phase stratified flow in a horizontal circular pipe was studied numerically for the first time. The analysis took into account all infinitesimal three-dimensional disturbances, including disturbances of the interface shape. To avoid difficulties connected with cylindrical axis and definition of the interface position in commonly used Cartesian or cylindrical coordinates, the problem was treated in bipolar cylindrical coordinates. In these coordinates, both the pipe wall and the plane interface coincide with the coordinate surfaces. This allows us to apply the no-slip boundary conditions explicitly and impose the interfacial boundary conditions relatively easy. Additionally, it enables investigation of the local behavior of the flow fields and shear stresses in the vicinity of the triple point. 

The results obtained in the bipolar coordinates were verified by an independent numerical solution based on the problem formulation in the Cartesian coordinates and the immersed boundary method. In this approach, the pipe wall is approximated by the immersed boundary technique, which slightly reduces the accuracy, while the plane unperturbed fluid-fluid interface remains a coordinate surface. The results computed by the two approaches are in good agreement, which verifies their correctness. 

Two representative examples of gas-liquid and liquid-liquid flows with particular holdups were considered to demonstrate applicability of the proposed numerical technique. These cases were chosen such that we could verify the numerical solution for the base flow with the analytical solution derived in \cite{Goldstein15}, which was obtained in the same bipolar coordinate system. 

In this work, we report results on the critical operational conditions and perturbations and show that our numerical approach accounting for all-wavelength perturbations, ranging from short- to very long-waves, allows computation of the shear (as in case of the air-water flow considered) and interfacial instabilities, as for the oil-water flow. Further comprehensive parametric studies, including computation of the stability boundaries for the entire range of flow holdups, are to be addressed in the future. An additional issue that is out of the scope of the present work is the case of nonplane unperturbed interface \citep[e.g., constant-curvature interface in capillary-dominated flows, see ][]{Gorelik99}. Since the constant-curvature interface aligns with coordinate line of bipolar coordinate system, the proposed numerical method can be easily adjusted to such flows. In light of a recent study of \cite{Goldstein21a} on the local flow behavior near the triple point, future works should also address the effect of boundary conditions at the triple points on stability of two-phase stratified flow. 

%%%%%%%%%%%%%%%%%%%%%%%%%%%%%%%%%%%%%%%%%%%%%%%%%%%%%%%%%%%%%%%%%%
\section*{Acknowledgments}

This research was supported by Israel Science Foundation (ISF) grant No 415/18.

%%%%%%%%%%%%%%%%%%%%%%%%%%%%%%%%%%%%%%%%%%%%%%%%%%%%%%%%%%%%%%%%%%
\appendix 

\begin{appendices}

	\section{Governing equations in orthogonal coordinates} 
	\label{sec: Appendix_gov}
	\numberwithin{equation}{section}
	\setcounter{equation}{0}
	
	An orthogonal coordinate system $(\xi,\phi,z)$ is defined such that it is curvilinear in the cross-sectional coordinates $\xi$ and $\phi$, while $z$ is an axis coordinate identical to that in Cartesian coordinate system. In these coordinates, the velocity vector can be written through its coordinate components as $\displaystyle\vec{u}^{(k)} = u_\xi^{(k)} \vec{e_\xi} + u_\phi^{(k)} \vec{e_\phi} + u_z^{(k)} \vec{e_z}$, where $\displaystyle \left\{\vec{e_\xi},\vec{e_\phi},\vec{e_z}\right\}$ form the orthonormal basis. The scale factors, also known as Lam\'e coefficients, are:
	
	\begin{subequations}
		\begin{align} \label{Eq: Lame}
			H_\xi 
			&= \sqrt{\left(\frac{\partial x}{\partial \xi}\right)^2 
				+ \left(\frac{\partial y}{\partial \xi}\right)^2},
			\\
			H_\phi 
			&= \sqrt{\left(\frac{\partial x}{\partial \phi}\right)^2 
				+ \left(\frac{\partial y}{\partial \phi}\right)^2},
		\end{align}
	\end{subequations}
	and $H_z=1$. The vector equations\ \ref{Eq: Continuity_vector} and \ref{Eq: N-S_vector_form} can be rewritten using the expressions of vector differential operators in orthogonal coordinates \citep{Kochin64}:
	\begin{equation} \label{Eq: Continuity_orthogonal}
		\frac{\partial (u_\xi^{(k)} H_\phi) }{\partial \xi}
		+ \frac{\partial (u_\phi^{(k)} H_\xi) }{\partial \phi}
		+ H_\xi H_\phi \frac{\partial u_z^{(k)}}{\partial z} = 0,
	\end{equation}
	
	\begin{subequations} \label{Eq: N-S_orthogonal}	
		\begin{align}
			&\begin{aligned}
				&\frac{\partial u_\xi^{(k)}}{\partial t}
				+ \frac{u_\xi^{(k)}}{H_\xi}\frac{\partial u_\xi^{(k)}}{\partial \xi}
				+ \frac{u_\phi^{(k)}}{H_\phi}\frac{\partial u_\xi^{(k)}}{\partial \phi}
				+ u_z^{(k)} \frac{\partial u_\xi^{(k)}}{\partial z}
				+ \frac{u_\phi^{(k)} u_\xi^{(k)}}{H_\phi H_\xi}\frac{\partial H_\xi}{\partial \phi}
				- \frac{\big(u_\phi^{(k)}\big)^2}{H_\phi H_\xi} \frac{\partial H_\phi}{\partial \xi}
				\\
				&= - \frac{\rho_1}{\rho_{12} \rho_k} \frac{1}{H_\xi} \frac{\partial p^{(k)}}{\partial \xi}
				+ \frac{1}{\Rey} \frac{\rho_1}{\rho_{12} \rho_k} \frac{\mu_{12} \mu_k}{\mu_1} 
				\Biggl\{
				\frac{1}{H_\phi H_\xi} 
				\frac{\partial}{\partial \xi} 
				\biggl[
				\frac{H_\phi}{H_\xi} \frac{\partial u_\xi^{(k)}}{\partial \xi}
				\biggr]
				\\
				&+\frac{1}{H_\phi} \frac{\partial}{\partial \phi} 
				\biggl[\frac{1}{H_\phi H_\xi}
				\frac{\partial (H_\xi u_\xi^{(k)})}{\partial \phi}\biggr]	
				+ \frac{\partial^2 u_\xi^{(k)}}{\partial z^2}
				+ \frac{2}{H_\phi H_\xi^2} \frac{\partial H_\xi}{\partial \phi} 
				\frac{\partial u_\phi^{(k)}}{\partial \xi}			
				- \frac{2}{H_\phi^2 H_\xi} \frac{\partial H_\phi}{\partial \xi} 
				\frac{\partial u_\phi^{(k)}}{\partial \phi}
				\\
				&+ \frac{1}{H_\phi H_\xi^2}
				\biggl[
				- \frac{1}{H_\xi} \frac{\partial H_\xi}{\partial \xi} \frac{\partial H_\phi}{\partial \xi}
				+ \frac{\partial^2 H_\phi}{\partial \xi^2}
				- \frac{1}{H_\phi} \left(\frac{\partial H_\phi}{\partial \xi}\right)^2
				\biggr]
				u_\xi^{(k)}
				\\			
				&+ \biggl[- \frac{1}{H_\phi^2 H_\xi} \frac{\partial^2 H_\phi}{\partial \xi \partial \phi}
				+ \frac{1}{H_\phi^3 H_\xi} \frac{\partial H_\phi}{\partial \phi} 
				\frac{\partial H_\phi}{\partial \xi}		
				+ \frac{1}{H_\phi H_\xi^2} \frac{\partial^2 H_\xi}{\partial \phi \partial \xi}
				- \frac{1}{H_\phi H_\xi^3}\frac{\partial H_\xi}{\partial \xi}
				\frac{\partial H_\xi}{\partial \phi}\biggr] u_\phi^{(k)}
				\Biggr\}
				\\			 
				&+ \frac{1}{\Fr} \frac{\vec{g}\cdot\vec{e_\xi}}{|\vec{g}|},
			\end{aligned}
		\end{align}
		\begin{align}
			&\begin{aligned}
				&\frac{\partial u_\phi^{(k)}}{\partial t}
				+ \frac{u_\xi^{(k)}}{H_\xi}\frac{\partial u_\phi^{(k)}}{\partial \xi}
				+\frac{u_\phi^{(k)}}{H_\phi}\frac{\partial u_\phi^{(k)}}{\partial \phi}
				+ u_z^{(k)} \frac{\partial u_\phi^{(k)}}{\partial z}
				+ \frac{u_\xi^{(k)} u_\phi^{(k)}}{H_\xi H_\phi}\frac{\partial H_\phi}{\partial \xi}
				- \frac{\big(u_\xi^{(k)}\big)^2}{H_\xi H_\phi} \frac{\partial H_\xi}{\partial \phi}
				\\
				=& - \frac{\rho_1}{\rho_{12} \rho_k} \frac{1}{H_\phi} \frac{\partial p^{(k)}}{\partial \phi}
				+ \frac{1}{\Rey} \frac{\rho_1}{\rho_{12} \rho_k} \frac{\mu_{12} \mu_k}{\mu_1} 
				\Biggl\{
				\frac{1}{H_\xi} \frac{\partial}{\partial \xi}
				\biggl[\frac{1}{H_\phi H_\xi} 
				\frac{\partial (H_\phi u_\phi^{(k)})}{\partial \xi}\biggr]
				\\
				&+ \frac{1}{H_\phi H_\xi}
				\frac{\partial}{\partial \phi}
				\biggl[\frac{H_\xi}{H_\phi}
				\frac{\partial u_\phi^{(k)}}{\partial \phi}\biggr]
				+ \frac{\partial^2 u_\phi^{(k)}}{\partial z^2}		
				+ \frac{2}{H_\phi^2 H_\xi} \frac{\partial H_\phi}{\partial \xi}
				\frac{\partial u_\xi^{(k)}}{\partial \phi}
				- \frac{2}{H_\phi H_\xi^2} 
				\frac{\partial H_\xi}{\partial \phi} 
				\frac{\partial u_\xi^{(k)}}{\partial \xi}
				\\
				&+ \biggl[-\frac{1}{H_\phi^3 H_\xi}\frac{\partial H_\phi}{\partial \phi}
				\frac{\partial H_\xi}{\partial \phi}
				+ \frac{1}{H_\phi^2 H_\xi} \frac{\partial^2 H_\xi}{\partial \phi^2}
				- \frac{1}{H_\phi^2 H_\xi^2} 
				\left(\frac{\partial H_\xi}{\partial \phi}\right)^2\biggr]  u_\phi^{(k)}				
				\\
				&+ \biggl[
				\frac{1}{H_\phi^2 H_\xi} \frac{\partial^2 H_\phi}{\partial \xi \partial \phi}
				-\frac{1}{H_\phi^3 H_\xi} \frac{\partial H_\phi}{\partial \phi}
				\frac{\partial H_\phi}{\partial \xi}
				- \frac{1}{H_\phi H_\xi^2} \frac{\partial^2 H_\xi}{\partial \phi \partial \xi}
				+ \frac{1}{H_\phi H_\xi^3} \frac{\partial H_\xi}{\partial \xi}
				\frac{\partial H_\xi}{\partial \phi}
				\biggr] u_\xi^{(k)}
				\Biggr\}
				\\
				&+ \frac{1}{\Fr} \frac{\vec{g}\cdot\vec{e_\phi}}{|\vec{g}|},
			\end{aligned}
		\end{align}
		\begin{align}
			&\begin{aligned}
				&\frac{\partial u_z^{(k)}}{\partial t}
				+ \frac{u_\xi^{(k)}}{H_\xi}\frac{\partial u_z^{(k)}}{\partial \xi} 
				+ \frac{u_\phi^{(k)}}{H_\phi}\frac{\partial u_z^{(k)}}{\partial \phi}				
				+ u_z^{(k)} \frac{\partial u_z^{(k)}}{\partial z} 
				= -\frac{\rho_1}{\rho_{12} \rho_k} \frac{\partial p^{(k)}}{\partial z}
				\\
				&+ \frac{1}{\Rey} \frac{\rho_1}{\rho_{12} \rho_k} \frac{\mu_{12} \mu_k}{\mu_1}
				\left\{\frac{1}{H_\phi H_\xi}
				\frac{\partial}{\partial \xi}
				\left(\frac{H_\phi}{H_\xi} \frac{\partial u_z^{(k)}}{\partial \xi}\right)
				+ \frac{1}{H_\phi H_\xi} 
				\frac{\partial}{\partial \phi}
				\left(\frac{H_\xi}{H_\phi} \frac{\partial u_z^{(k)}}{\partial \phi}\right)
				+ \frac{\partial^2 u_z^{(k)}}{\partial z^2}\right\} 
				\\
				&+ \frac{1}{\Fr} \frac{\vec{g}\cdot\vec{e_z}}{|\vec{g}|}.
			\end{aligned}
		\end{align}
	\end{subequations}
	
	%%%%%%%%%%%%%%%%%%%%%%%%%%%%%%%%%%%%%%%%%%%%%%%%%%%%%%%%%%%%%%%%%%
	\section{Interfacial boundary conditions}
	\label{sec: Appendix_BC}
	
	The fluid--fluid interface is defined as a surface
	\begin{equation}
		S(\xi,\phi,z) = \phi_\eta - \phi^* - \eta\left(\xi,z\right) = 0.
	\end{equation}
	In the present work, we considered a case of the plane unperturbed interface, i.e., $\displaystyle \phi^* = \pi$. The interface between the fluids is defined so that there is no mass transfer (i.e., fluid flow) across it. Therefore, the kinematic boundary condition at the interface, $\displaystyle \phi=\phi_\eta$, requires the material derivative of the surface with respect to time to be zero
	\begin{subequations}
		\begin{align}
			\frac{\DD S}{\DD t} 
			&= \frac{\partial S}{\partial t} 
			+ \vec{u}^{(k)} \cdot \nabla S 
			= 0,
			\\
			\frac{\DD S}{\DD t} &= \frac{\partial S}{\partial t} 
			+ \frac{u_\xi^{(k)}}{H_\xi} \frac{\partial S}{\partial \xi}
			+ \frac{u_\phi^{(k)}}{H_\phi}
			+ u_z^{(k)} \frac{\partial S}{\partial z}  = 0,
			\\
			&- \frac{\partial \eta}{\partial t} 
			- \frac{u_\xi^{(k)}}{H_\xi} \frac{\partial \eta}{\partial \xi}
			+ \frac{u_\phi^{(k)}}{H_\phi} 
			- u_z^{(k)} \frac{\partial \eta}{\partial z}  = 0,
			\\
			u_\phi^{(k)}  &= H_\phi \frac{\partial \eta}{\partial t} 
			+ u_\xi^{(k)} \frac{H_\phi}{H_\xi} \frac{\partial \eta}{\partial \xi}
			+ u_z^{(k)} H_\phi \frac{\partial \eta}{\partial z}.
		\end{align}
	\end{subequations}
	
	Continuity of the velocity at the interface reads
	\begin{equation}
		\vec{u}^{(1)}(\phi=\phi_\eta) 
		= \vec{u}^{(2)}(\phi=\phi_\eta).
	\end{equation}
	
	A unit vector normal to the surface is defined as
	\begin{align} \label{Eq: Normal_vector_bipolar}
		&\begin{aligned}
			\vec{n} 
			&= \frac{\nabla S\left(\xi, \phi,z\right)}{\left|\nabla S\left(\xi, \phi,z\right)\right|}
			= \frac{\left(\dfrac{1}{H_\xi}\dfrac{\partial S}{\partial \xi},
				\dfrac{1}{H_\phi}\dfrac{\partial S}{\partial \phi}, \dfrac{1}{H_z}\dfrac{\partial S}{\partial z}\right)}{\sqrt{\left(\dfrac{1}{H_\xi}\dfrac{\partial S}{\partial \xi}\right)^2
					+ \left(\dfrac{1}{H_\phi}\dfrac{\partial S}{\partial \phi}\right)^2
					+ \left(\dfrac{1}{H_z}\dfrac{\partial S}{\partial z}\right)^2}}
			\\		
			&= \dfrac{1}{\sqrt{\left(\dfrac{1}{H_\xi}\dfrac{\partial \eta}{\partial \xi}\right)^2 
					+ \left(\dfrac{1}{H_\phi}\right)^2
					+ \left(\dfrac{\partial \eta}{\partial z}\right)^2}} 
			\left(-\dfrac{1}{H_\xi}\dfrac{\partial \eta}{\partial \xi}, \dfrac{1}{H_\phi},-\dfrac{\partial \eta}{\partial z}\right),
		\end{aligned}
	\end{align}
	so the mean curvature of the interface is 
	\begin{align}
		&\begin{aligned}
			&\nabla \cdot \vec{n}			
			= \frac{1}{H_\xi H_\phi} \biggl[
			\frac{\partial}{\partial \xi}\left(n_\xi H_\phi\right)
			+ \frac{\partial}{\partial \phi}\left(n_\phi H_\xi\right)
			+ \frac{\partial}{\partial z}\left(n_z H_\xi H_\phi\right)
			\biggr]
			\\
			&= - \frac{1}{H_\xi H_\phi} 
			\bigg[\bigg(\dfrac{H_\phi}{H_\xi}\dfrac{\partial \eta}{\partial \xi}\bigg)^2 
			+ 1
			+ \bigg(H_\phi \frac{\partial \eta}{\partial z}\bigg)^2\bigg]^{-\frac{1}{2}}
			\bigg[\frac{\partial}{\partial \xi}
			\bigg(\frac{H_\phi^2}{H_\xi}
			\frac{\partial \eta}{\partial \xi}\bigg)
			- 2 H_\xi^2 \frac{\sin\phi}{\sin\phi_0}
			+ H_\xi H_\phi^2 \frac{\partial^2 \eta}{\partial z^2}\bigg]
			\\
			&- \frac{1}{2 H_\xi H_\phi}
			\bigg[\bigg(\dfrac{H_\phi}{H_\xi}\dfrac{\partial \eta}{\partial \xi}\bigg)^2 
			+ 1
			+ \bigg(H_\phi \frac{\partial \eta}{\partial z}\bigg)^2\bigg]^{-\frac{3}{2}}
			\Bigg[
			- \frac{H_\phi^2}{H_\xi}
			\frac{\partial \eta}{\partial \xi} 
			\biggl\{2 \frac{H_\phi}{H_\xi}\frac{\partial \eta}{\partial \xi} 
			\frac{\partial}{\partial \xi} \bigg(\frac{H_\phi}{H_\xi}\frac{\partial \eta}{\partial \xi}\bigg)
			\\
			&+ 2 H_\phi \frac{\partial \eta}{\partial z} \frac{\partial}{\partial \xi} 
			\bigg(H_\phi \frac{\partial \eta}{\partial z}\bigg)\biggr\}
			+ \frac{H_\xi}{H_\phi} 
			\biggl\{2 \frac{H_\phi}{H_\xi}\frac{\partial \eta}{\partial \xi} 
			\frac{\partial}{\partial \phi} 
			\bigg(\frac{1}{H_\xi}\frac{\partial \eta}{\partial \xi}\bigg)
			+ 2 H_\phi \frac{\partial \eta}{\partial z} \frac{\partial}{\partial \phi} 
			\bigg(H_\phi \frac{\partial \eta}{\partial z}\bigg)
			\biggr\}
			\\
			&- H_\xi H_\phi \frac{\partial \eta}{\partial z}
			\biggl\{2 \frac{H_\phi}{H_\xi}
			\frac{\partial \eta}{\partial \xi} 
			\frac{H_\phi}{H_\xi}
			\frac{\partial^2 \eta}{\partial \xi \partial z}
			+ 2 H_\phi^2 \frac{\partial \eta}{\partial z} 
			\frac{\partial^2 \eta}{\partial z^2}
			\biggr\}
			\Bigg].
		\end{aligned}
	\end{align}
	
	Linearization of the unit normal vector (Eq.\ \ref{Eq: Normal_vector_bipolar}) yields
	\begin{subequations}
		\begin{align}
			&\begin{aligned}
				\vec{n} 		
				&= \dfrac{1}{\sqrt{\cancel{\left(\dfrac{1}{H_\xi}\dfrac{\partial \eta}{\partial \xi}\right)^2} 
						+ \left(\dfrac{1}{H_\phi}\right)^2
						+ \cancel{\bigg(\dfrac{\partial \eta}{\partial z}\bigg)^2}}} 
				\bigg(-\dfrac{1}{H_\xi}\dfrac{\partial \eta}{\partial \xi}, \dfrac{1}{H_\phi},-\dfrac{\partial \eta}{\partial z}\bigg)
				\\
				&\approx \bigg(-\dfrac{H_\phi}{H_\xi}\dfrac{\partial \eta}{\partial \xi}, 1,- H_\phi \dfrac{\partial \eta}{\partial z}\bigg),
			\end{aligned}
		\end{align}
		\begin{equation}	
			\vec{n} = \bigg(n_\xi,n_\phi,n_z\bigg)
			\approx \bigg(-\dfrac{H_\phi}{H_\xi}\dfrac{\partial \eta}{\partial \xi}, 1,- H_\phi \dfrac{\partial \eta}{\partial z}\bigg),
		\end{equation}
	\end{subequations}
	the linearized version of the mean curvature of the interface then reads
	\begin{align}
		&\begin{aligned}
			\nabla \cdot \vec{n} \biggr|_{\phi=\phi_\eta}	
			&= \Delta S\left(\xi,\phi=\phi_\eta,z\right) 			
			\\
			&= - \frac{1}{H_\xi H_\phi}
			\biggl[\frac{\partial}{\partial \xi}\left(\frac{H_\phi^2}{H_\xi} 
			\frac{\partial \eta}{\partial \xi}\right)
			+ 2 H_\xi^2 \frac{\sin\phi_\eta}{\sin\phi_0}
			+ H_\xi H_\phi^2 \frac{\partial^2 \eta}{\partial z^2}
			\biggr].
		\end{aligned}
	\end{align}
	
	The linearized unit vectors tangential to the surface and located in the planes $(\xi,\phi)$ and $(\phi,z)$ are, respectively 
	\begin{align} 
		&\begin{aligned}
			\vec{t_{\xi \phi}} &= \dfrac{1}{\sqrt{\left(\dfrac{1}{H_\phi}\right)^2
					+ \cancel{\bigg(\dfrac{1}{H_\xi}\dfrac{\partial \eta}{\partial \xi}\bigg)^2}}} 
			\bigg(\dfrac{1}{H_\phi}, \dfrac{1}{H_\xi} \dfrac{\partial \eta}{\partial \xi},0\bigg)
			\approx
			\bigg(1,\frac{H_\phi}{H_\xi}\frac{\partial \eta}{\partial \xi},0\bigg),
		\end{aligned}
	\end{align}
	
	\begin{align}
		&\begin{aligned}
			\vec{t_{\phi z}} &= \dfrac{1}{\sqrt{\cancel{\left(\dfrac{\partial \eta}{\partial z}\right)^2}
					+ \bigg(\dfrac{1}{H_\phi}\bigg)^2}} 
			\bigg(0, \dfrac{\partial \eta}{\partial z}, \dfrac{1}{H_\phi}\bigg)
			\approx \bigg(0,H_\phi \dfrac{\partial \eta}{\partial z},1\bigg).
		\end{aligned}
	\end{align}
	
	The interfacial boundary conditions require continuity of the tangential components of the viscous stress tensor, $\displaystyle\vec{\mathcal{T}}=\frac{\mu_k \mu_{12}}{\mu_1} \left(\nabla \vec{u} + \left(\nabla \vec{u}\right)^T\right)$, while the discontinuity in the normal component is balanced by the surface tension
	\begin{subequations}
		\begin{align}
			\tau_{\xi \phi}^{(1)}(\phi=\phi_\eta) 
			&= \tau_{\xi \phi}^{(2)}(\phi=\phi_\eta),
			\\
			\tau_{\phi z}^{(1)}(\phi=\phi_\eta) 
			&= \tau_{\phi z}^{(2)}(\phi=\phi_\eta),
			\\		
			\biggl\llbracket -p + \sigma_n \biggr\rrbracket_{\phi=\phi_\eta} 
			&= - \frac{1}{\We} \nabla \cdot \vec{n}.	
		\end{align}
	\end{subequations}
	The tangential and normal viscous stresses are
	\begin{subequations} \label{Eq: Viscous_stresses}
		\begin{align}
			&\begin{aligned}
				\tau_{\xi \phi}
				&= \vec{t_{\xi \phi}} \cdot \vec{\mathcal{T}} \cdot \vec{n}
				= \mathcal{T}_{\xi  \xi} n_\xi  t_{\xi \phi\_\xi}
				+ \mathcal{T}_{\xi  \phi} \left(n_\phi t_{\xi \phi\_\xi}
				+ n_\xi t_{\xi \phi\_\phi}\right)
				+ \mathcal{T}_{\xi   z}   n_z    t_{\xi \phi\_\xi}
				\\
				&
				+ \mathcal{T}_{\phi \phi} n_\phi t_{\xi \phi\_\phi}
				+ \mathcal{T}_{\phi z}    n_z    t_{\xi \phi\_\phi},
			\end{aligned}
			\\
			&\begin{aligned}
				\tau_{\phi z}
				&= \vec{t_{\phi z}} \cdot \vec{\mathcal{T}} \cdot \vec{n}
				= \mathcal{T}_{\xi  \phi} n_\xi  t_{\phi z\_\phi}
				+  \mathcal{T}_{\xi   z}   n_\xi  t_{\phi z\_z}
				+  \mathcal{T}_{\phi \phi} n_\phi t_{\phi z\_\phi}			
				\\
				&+  \mathcal{T}_{\phi z} \left(n_z t_{\phi z\_\phi}
				+ n_\phi t_{\phi z\_z}\right)
				+  \mathcal{T}_{z    z}    n_z    t_{\phi z\_z},
			\end{aligned}
			\\
			\sigma_n 
			&= \vec{n} \cdot \vec{\mathcal{T}} \cdot \vec{n}
			= \mathcal{T}_{\xi\xi} n_\xi^2 + 2 \mathcal{T}_{\xi\phi} n_\xi n_\phi + 2 \mathcal{T}_{\xi z} n_\xi n_z + \mathcal{T}_{\phi\phi} n_\phi^2 + 2 \mathcal{T}_{\phi z} n_\phi n_z + \mathcal{T}_{zz} n_z^2.
			\end{align}
	\end{subequations}
	
	The components of the viscous stress tensor in bipolar coordinates are
	\begin{subequations} \label{Eq: Stress_tensor_components}
		\begin{align}
			\mathcal{T}_{\xi \xi} 
			=& 2 \frac{\mu_k \mu_{12}}{\mu_1} \biggl\{
			\frac{1}{H_\xi}\frac{\partial u_\xi}{\partial \xi} 
			+ \dfrac{u_\phi}{H_\xi H_\phi} \dfrac{\partial H_\xi}{\partial \phi}
			\biggr\}, 
			\\
			\mathcal{T}_{\xi \phi} 
			=& \mathcal{T}_{\phi \xi} 
			= \frac{\mu_k \mu_{12}}{\mu_1} \biggl\{\left(
			\frac{1}{H_\phi}\frac{\partial u_\xi}{\partial \phi} 			
			- \dfrac{u_\phi}{H_\xi H_\phi} \dfrac{\partial H_\phi}{\partial \xi}\right) 
			+ \left(
			\frac{1}{H_\xi}\frac{\partial u_\phi}{\partial \xi} 
			- \frac{u_\xi}{H_\xi H_\phi} \dfrac{\partial H_\xi}{\partial \phi} 
			\right)
			\biggr\},
			\\
			\mathcal{T}_{\xi z} 
			=& \mathcal{T}_{z \xi} 
			= \frac{\mu_k \mu_{12}}{\mu_1} \biggl\{
			\frac{\partial u_\xi}{\partial z} 
			+ \frac{1}{H_\xi}
			\frac{\partial u_z}{\partial \xi}
			\biggr\},
			\\
			\mathcal{T}_{\phi \phi} 
			=& 
			2 \frac{\mu_k \mu_{12}}{\mu_1} \biggl\{
			\frac{1}{H_\phi}\frac{\partial u_\phi}{\partial \phi}		
			+ \frac{u_\xi}{H_\xi H_\phi} \dfrac{\partial H_\phi}{\partial \xi} 
			\biggr\},
			\\
			\mathcal{T}_{\phi z} 
			=& \mathcal{T}_{z \phi}
			= \frac{\mu_k \mu_{12}}{\mu_1} \biggl\{
			\frac{\partial u_\phi}{\partial z} 
			+ \frac{1}{H_\phi}
			\frac{\partial u_z}{\partial \phi}
			\biggr\},
			\\
			\mathcal{T}_{z z} 
			=& 2 \frac{\mu_k \mu_{12}}{\mu_1} \frac{\partial u_z}{\partial z}.
		\end{align}
	\end{subequations}
	
	To derive boundary conditions for the viscous stresses, we substitute the components of the stress tensor ,\ref{Eq: Stress_tensor_components} and the normal and tangential unit vectors into Eq.\ \ref{Eq: Viscous_stresses} to obtain the final expressions:
	
	\begin{subequations} \label{Eq: BC_shear_stress}
		\begin{align}
			\begin{aligned}
				&\biggl \llbracket
				\tau_{\xi \phi}
				\biggr\rrbracket_{\phi = \phi_\eta} 
				=
				\Biggl \llbracket
				2 \frac{\mu \mu_{12}}{\mu_1} \biggl\{
				\frac{1}{H_\xi}\frac{\partial u_\xi}{\partial \xi} 
				+ \dfrac{u_\phi}{H_\xi H_\phi} \dfrac{\partial H_\xi}{\partial \phi}
				\biggr\} 
				\bigg(-\dfrac{1}{H_\xi}\dfrac{\partial \eta}{\partial \xi}\bigg)
				\\
				&+
				\frac{\mu \mu_{12}}{\mu_1} \biggl\{\left(
				\frac{1}{H_\phi}\frac{\partial u_\xi}{\partial \phi} 			
				- \dfrac{u_\phi}{H_\xi H_\phi} \dfrac{\partial H_\phi}{\partial \xi}\right) 
				+ \left(
				\frac{1}{H_\xi}\frac{\partial u_\phi}{\partial \xi} 
				- \frac{u_\xi}{H_\xi H_\phi} \dfrac{\partial H_\xi}{\partial \phi} 
				\right)
				\biggr\}
				\\
				&\bigg(\frac{1}{H_\phi}
				-\frac{1}{H_\xi} \frac{\partial \eta}{\partial \xi}
				\frac{1}{H_\xi}  \frac{\partial \eta}{\partial \xi}
				\bigg)
				+ \frac{\mu \mu_{12}}{\mu_1} \biggl\{
				\frac{\partial u_\xi}{\partial z} 
				+ \frac{1}{H_\xi}
				\frac{\partial u_z}{\partial \xi}
				\biggr\}
				\bigg(-\dfrac{\partial \eta}{\partial z}\bigg)
				\frac{1}{H_\phi}
				\\
				&+ 2 \frac{\mu \mu_{12}}{\mu_1} \biggl\{
				\frac{1}{H_\phi}\frac{\partial u_\phi}{\partial \phi}		
				+ \frac{u_\xi}{H_\xi H_\phi} \dfrac{\partial H_\phi}{\partial \xi} 
				\biggr\}
				\frac{1}{H_\phi}
				\frac{1}{H_\xi} \frac{\partial \eta}{\partial \xi}
				\\
				&+ \frac{\mu \mu_{12}}{\mu_1} \biggl\{
				\frac{\partial u_\phi}{\partial z} 
				+ \frac{1}{H_\phi}
				\frac{\partial u_z}{\partial \phi}
				\biggr\}
				\bigg(-\dfrac{\partial \eta}{\partial z}\bigg)
				\frac{1}{H_\xi} \frac{\partial \eta}{\partial \xi}
				\Biggr\rrbracket_{\phi = \phi_\eta} 
				= 0,
			\end{aligned}
		\end{align}
		\begin{align}
			&\begin{aligned}
				&\Biggl \llbracket
				\tau_{\phi z}
				\Biggr\rrbracket_{\phi = \phi_\eta} 
				= \Biggl \llbracket
				\frac{\mu \mu_{12}}{\mu_1} \biggl\{\left(
				\frac{1}{H_\phi}\frac{\partial u_\xi}{\partial \phi} 			
				- \dfrac{u_\phi}{H_\xi H_\phi} \dfrac{\partial H_\phi}{\partial \xi}\right) 
				+ \left(
				\frac{1}{H_\xi}\frac{\partial u_\phi}{\partial \xi} 
				- \frac{u_\xi}{H_\xi H_\phi} \dfrac{\partial H_\xi}{\partial \phi} 
				\right)
				\biggr\}
				\\
				&\bigg(-\dfrac{1}{H_\xi}\dfrac{\partial \eta}{\partial \xi}\bigg)
				\frac{\partial \eta}{\partial z}
				+ \frac{\mu \mu_{12}}{\mu_1} \biggl\{
				\frac{\partial u_\xi}{\partial z} 
				+ \frac{1}{H_\xi}
				\frac{\partial u_z}{\partial \xi}
				\biggr\}
				\bigg(-\dfrac{1}{H_\xi}\dfrac{\partial \eta}{\partial \xi}\bigg)
				\\
				&+ 2 \frac{\mu \mu_{12}}{\mu_1} \biggl\{
				\frac{1}{H_\phi}\frac{\partial u_\phi}{\partial \phi}		
				+ \frac{u_\xi}{H_\xi H_\phi} \dfrac{\partial H_\phi}{\partial \xi} 
				\biggr\}
				\frac{1}{H_\phi}
				\frac{\partial \eta}{\partial z}
				\\
				&+ \frac{\mu \mu_{12}}{\mu_1} \biggl\{
				\frac{\partial u_\phi}{\partial z} 
				+ \frac{1}{H_\phi}
				\frac{\partial u_z}{\partial \phi}
				\biggr\}
				\bigg[\bigg(- \frac{\partial \eta}{\partial z}\bigg)
				\frac{\partial \eta}{\partial z}
				+ \frac{1}{H_\phi} \frac{1}{H_\phi}\bigg]
				+ 2 \frac{\mu \mu_{12}}{\mu_1} \frac{\partial u_z}{\partial z}
				\bigg(- \frac{\partial \eta}{\partial z}\bigg)
				\frac{1}{H_\phi}
				\Biggr\rrbracket_{\phi = \phi_\eta} 
				\\
				&= 0,
			\end{aligned}
		\end{align}
		\begin{align}
			&\begin{aligned}
				&\Biggl \llbracket
				-p
				+ \bigg[\left(\dfrac{1}{H_\xi}\dfrac{\partial \eta}{\partial \xi}\right)^2
				+ \left(\dfrac{1}{H_\phi}\right)^2
				+ \bigg(\dfrac{\partial \eta}{\partial z}\bigg)^2\bigg] ^{-1}
				\bigg[2 \frac{\mu \mu_{12}}{\mu_1} \left(\frac{1}{H_\xi}\frac{\partial u_\xi}{\partial \xi} 
				+ \dfrac{u_\phi}{H_\xi H_\phi} \dfrac{\partial H_\xi}{\partial \phi}\right)
				\left(-\frac{H_\phi}{H_\xi}\frac{\partial \eta}{\partial \xi}\right)^2
				\\
				&+ 2 \frac{\mu \mu_{12}}{\mu_1} \biggl\{\left(
				\frac{1}{H_\phi}\frac{\partial u_\xi}{\partial \phi} 			
				- \dfrac{u_\phi}{H_\xi H_\phi} \dfrac{\partial H_\phi}{\partial \xi}\right) 
				+ \left(
				\frac{1}{H_\xi}\frac{\partial u_\phi}{\partial \xi} 
				- \frac{u_\xi}{H_\xi H_\phi} \dfrac{\partial H_\xi}{\partial \phi} 
				\right)
				\biggr\}
				\left(-\frac{H_\phi}{H_\xi}\frac{\partial \eta}{\partial \xi}\right)
				\\
				&+ 2 \frac{\mu \mu_{12}}{\mu_1} \biggl\{
				\frac{\partial u_\xi}{\partial z} 
				+ \frac{1}{H_\xi}
				\frac{\partial u_z}{\partial \xi}
				\biggr\}
				\left(-\frac{H_\phi}{H_\xi}\frac{\partial \eta}{\partial \xi}\right)
				\left(-H_\phi\frac{\partial \eta}{\partial z}\right)
				+ 2 \frac{\mu \mu_{12}}{\mu_1} \left(
				\frac{1}{H_\phi}\frac{\partial u_\phi}{\partial \phi}		
				+ \frac{u_\xi}{H_\xi H_\phi} \dfrac{\partial H_\phi}{\partial \xi} 			
				\right)
				\\
				&+ 2 \frac{\mu \mu_{12}}{\mu_1} \biggl\{
				\frac{\partial u_\phi}{\partial z} 
				+ \frac{1}{H_\phi}
				\frac{\partial u_z}{\partial \phi}
				\biggr\}
				\left(-H_\phi\frac{\partial \eta}{\partial z}\right)
				+ 2 \frac{\mu \mu_{12}}{\mu_1} \frac{\partial u_z}{\partial z} 
				\left(-H_\phi\frac{\partial \eta}{\partial z}\right)^2\bigg]
				\Biggr\rrbracket_{\phi = \phi_\eta} 
				\\
				&= \frac{1}{\We}
				\frac{1}{H_\xi H_\phi} 
				\bigg[\bigg(\dfrac{H_\phi}{H_\xi}\dfrac{\partial \eta}{\partial \xi}\bigg)^2 
				+ 1
				+ \bigg(H_\phi \frac{\partial \eta}{\partial z}\bigg)^2\bigg]^{-\frac{1}{2}}
				\bigg[\frac{\partial}{\partial \xi}
				\bigg(\frac{H_\phi^2}{H_\xi}
				\frac{\partial \eta}{\partial \xi}\bigg)
				- 2 H_\xi^2 \frac{\sin\phi}{\sin\phi_0}
				+ H_\xi H_\phi^2 \frac{\partial^2 \eta}{\partial z^2}\bigg]
				\\
				&- \frac{1}{2 H_\xi H_\phi}
				\bigg[\bigg(\dfrac{H_\phi}{H_\xi}\dfrac{\partial \eta}{\partial \xi}\bigg)^2 
				+ 1
				+ \bigg(H_\phi \frac{\partial \eta}{\partial z}\bigg)^2\bigg]^{-\frac{3}{2}}
				\Bigg[
				- \frac{H_\phi^2}{H_\xi}
				\frac{\partial \eta}{\partial \xi} 
				\biggl\{2 \frac{H_\phi}{H_\xi}\frac{\partial \eta}{\partial \xi} 
				\frac{\partial}{\partial \xi} \bigg(\frac{H_\phi}{H_\xi}\frac{\partial \eta}{\partial \xi}\bigg)
				\\
				&+ 2 H_\phi \frac{\partial \eta}{\partial z} \frac{\partial}{\partial \xi} 
				\bigg(H_\phi \frac{\partial \eta}{\partial z}\bigg)\biggr\}
				+ \frac{H_\xi}{H_\phi} 
				\biggl\{2 \frac{H_\phi}{H_\xi}\frac{\partial \eta}{\partial \xi} 
				\frac{\partial}{\partial \phi} 
				\bigg(\frac{1}{H_\xi}\frac{\partial \eta}{\partial \xi}\bigg)
				+ 2 H_\phi \frac{\partial \eta}{\partial z} \frac{\partial}{\partial \phi} 
				\bigg(H_\phi \frac{\partial \eta}{\partial z}\bigg)
				\biggr\}
				\\
				&- H_\xi H_\phi \frac{\partial \eta}{\partial z}
				\biggl\{2 \frac{H_\phi}{H_\xi}
				\frac{\partial \eta}{\partial \xi} 
				\frac{H_\phi}{H_\xi}
				\frac{\partial^2 \eta}{\partial \xi \partial z}
				+ 2 H_\phi^2 \frac{\partial \eta}{\partial z} 
				\frac{\partial^2 \eta}{\partial z^2}
				\biggr\}
				\Bigg].
			\end{aligned}
		\end{align}
	\end{subequations}

\end{appendices}
%%%%%%%%%%%%%%%%%%%%%%%%%%%%%%%%%%%%%%%%%%%%%%%%%%%%%%%%%%%%%%%%%%%%%%%%%%%%%%%%
% Bibliography
\bibliographystyle{../apalike-onesort}
\bibliography{/home/ilya/Documents/bibliography/ilya_bibliography}
%%%%%%%%%%%%%%%%%%%%%%%%%%%%%%%%%%%%%%%%%%%%%%%%%%%%%%%%%%%%%%%%%%%%%%%%%%%%%%%%

\end{document}